\documentclass[aps,twocolumn,showpacs]{revtex4}
\usepackage{graphicx}
\usepackage{textcomp}
\usepackage{psfrag}

\begin{document}

\title{Intermittent  Peel Front Dynamics and the Crackling Noise in an Adhesive Tape }
\author{Jagadish Kumar$^{1}$}
\author{Rumi De$^{2}$}  
\author{G. Ananthakrishna$^{1}$}
\affiliation{{$^1$Materials Research Centre, Indian Institute of Science, Bangalore 560012, India\\} 
$^2$ Department of Materials and Interfaces, Weizmann Institute of Science, Rehovot 76100, Israel \\}

\begin{abstract}

We report a comprehensive  investigation of a model for peeling of an adhesive tape along with a nonlinear  time series analysis of experimental acoustic emission signals in an effort to understand the origin of intermittent peeling of an adhesive tape and its connection to acoustic emission. The model represents the acoustic energy dissipated in terms of  Rayleigh dissipation functional that depends on the local strain rate. We show that the nature of the peel front exhibits rich spatiotemporal patterns ranging from smooth, rugged and stuck-peeled configurations that depend on three parameters, namely, the ratio of inertial time scale of the tape mass to that of the roller, the dissipation coefficient and the pull velocity. The stuck-peeled configurations are reminiscent of fibrillar peel front patterns observed in experiments.  We show that while the intermittent peeling is controlled by the peel force function, the model acoustic energy dissipated depends on the nature of the peel front and its dynamical evolution. Even though the acoustic energy is a fully dynamical quantity, it can be quite noisy for a certain set of parameter values suggesting the deterministic origin of acoustic emission in experiments. To verify this suggestion, we have carried out a dynamical analysis of experimental acoustic emission time series for a wide range of traction velocities. Our analysis shows an unambiguous presence of chaotic dynamics within a subinterval of pull speeds within the intermittent regime. Time series analysis of the model acoustic energy signals is also found to be chaotic within a subinterval of pull speeds. Further, the model provides insight into several statistical and dynamical features of the experimental AE signals including the transition from burst type acoustic emission to continuous type with increasing pull velocity  and the connection between acoustic emission and stick-slip dynamics. Finally, the model also offers an explanation for the recently observed feature that the duration of the slip phase can be less than that of the stick phase.
\end{abstract}
 
\pacs{83.60.Df, 05.45.-a, 05.45.Tp, 62.20.Mk}

\maketitle

\section{Introduction}

Adhesive tapes are routinely used in a variety of situations including daily usage as stickers, in packing and sealing. Yet, day-to-day experiences like intermittent peeling of an adhesive tape and the origin of the accompanying audible noise have remained ill understood. This may be partly attributed to the fact that adhesion is a highly interdisciplinary subject involving diverse but interrelated physical phenomena such as intermolecular forces of attraction at the interface, mechanics of contact, debonding and rupture,  visco-plastic deformation and fracture \cite{Kendall00}, and frictional dissipation  which operates during peeling \cite{Kendall00,Urbakh04,Persson}. Yet another reason is that most information on adhesion is obtained from quasistatic or near stationary conditions. Apart from scientific interest, understanding the intermittent peel or the stick-slip process   has relevance to industrial applications as well. For example, optimizing production schedules that involve pasting or peeling of an adhesive tapes at a rapid pace in an assembly line requires a good understanding of stick-slip dynamics. Moreover, insight into the time dependent and dynamical aspects of adhesion is expected to be important in design of adhesives with  versatile properties required in variety of  applications, in understanding the mechanisms leading to the failure of adhesive joints as also in understanding biologically relevant systems such as the gecko\cite{Jagota07} or reorientation dynamics of cells  \cite{Rumi07}.

Adhesion tests are essentially fracture tests designed to study adherence of solids and generally involve normal pulling off and peeling. Such experiments can be performed under quasistatic or near-stationary and nonequilibrium conditions as well. The latter kind of experiments demonstrate the rate dependence of adhesive properties. It is this rate dependence and the inherent nonlinearity that leads to a variety of instabilities. These kinds of peeling experiments are comparatively easy to setup in a laboratory. Moreover, the set up also allows one to record unusually long force waveforms and AE signals that should be helpful in extracting useful information on the nonlinear features of the system.

One type of peeling experiment that yields dynamical information is carried out with an adhesive tape mounted on a roller subjected to a constant pull velocity \cite{MB,CGVB04}. Peeling experiments have also been performed under constant load conditions \cite{BC97,CGVB04}. At low pull velocities, the velocity of the contact point $v$ keeps pace with the imposed velocity $V$. The same is true at high velocities as well. However, there is an intermediate regime of traction velocities where the peeling is intermittent. Peeling in this regime is accompanied by a characteristic audible noise \cite{MB,BC97,CGVB04}. It must be stressed that these two stable dissipative branches refer to stationary branches. Even so, the  stick-slip dynamics observed in the intermediate region of pull velocities has been attempted by assuming an unstable branch connecting the two stable branches. The strain energy release rate shows a power law for low velocities with an exponent around 0.3. The high velocity branch also shows a power law but with a much higher exponent value of about 5.5 \cite{MB}.  The low velocity branch is known to arise from viscous dissipation and that at high velocity corresponds to fracture. These studies report a range of wave forms starting from saw tooth, sinusoidal or even irregular wave form that has been termed 'chaotic' \cite{MB}. More recently, the dynamics of the peel point has been imaged as well \cite{Cortet}.

Stick-slip processes  are usually observed in systems subjected to a constant response where-in the force developed in the system is measured by dynamically coupling the system to a measuring device.  The phenomenon  is experienced routinely, for example, while writing with chalk piece on a black board, playing violin or walking down a staircase with the hand placed on the hand-rail. A large number of studies on stick-slip dynamics have been reported in systems ranging from atomic length scales, for instance,  stick-slip observed using atomic force microscope \cite{soc04} to geological length scales  like the stick-slip of tectonic plates causing  earthquakes ~\cite{BK,BKC}.  A few well known laboratory scale systems are - sliding friction \cite{Heslot,Urbakh04,Persson} and the Portevin-Le Chatelier (PLC) effect \cite{PLC,GA07}, a kind of plastic instability observed during tensile deformation of dilute alloys \cite{GA07,Anan04}, to name only two. Most stick-slip processes are characterized by the system spending a large part of the time in the stuck state and a short time in the slip state. This feature is observed both in experiment and in models. See for instance \cite{Heslot,soc04,BK}. A counter example where the time spent in the stuck is less than that in the slip state (observed at high applied strain rates) is the PLC effect \cite{PLC}. These studies show that while the physical mechanisms that operate in different situations can be quite varied \cite{Persson}, in general, stick-slip results from a competition among the inherent internal relaxational time scales \cite{GA07,Anan04} and the applied time scale. In the case of peeling, one identifiable internal relaxation time scale is the viscoelastic time scale of the adhesive. Other relevant time scales that may be operative need to be included for a proper description of the dynamics.  All stick-slip systems are governed by deterministic nonlinear dynamics.

Models that attempt to explain the dynamical features of stick-slip systems  use the macroscopic phenomenological negative force response (NFR) feature as an input although the unstable region is not accessible. This is also true for models dealing with the dynamics of the adhesive tape including the present work. In this context, it must be stated that there is no microscopic theory that predicts the negative force-velocity relation in most stick-slip situations except in the case of the PLC effect where we have provided  a dynamical interpretation of the negative strain rate sensitivity of the flow stress \cite{Anan82,Rajesh} (see below).
There are several theoretical attempts to model the stick-slip process observed during peeling of an adhesive tape.  Maugis and Barquin \cite{MB}, were the first to write down a model set of equations suitable for the experimental situation and to carry out approximate dynamical analysis. These equations were later modified and a dynamical analysis of these equations was reported \cite{HY1}. However, the stick-slip oscillations were {\it not} obtained as a natural consequence of the equations of motion \cite{HY2,CGB}.  Indeed, these equations are singular\cite{Rumi04}. Subsequently, we devised a special algorithm to solve these differential algebraic equations (DAE) \cite{HLR,Rumi04}. This algorithm allows for dynamical jumps across the two stable branches of the peel force function. This was followed by converting the DAE into a set of nonlinear ordinary differential equations (ODE) by including the missing kinetic energy of the stretched tape thereby lifting the singular nature of the DAE \cite{Rumi04,Rumi05}. Apart from supporting  dynamical jumps, the ODE model exhibits rich dynamical features. However, all these studies discuss only  contact point dynamics while  the tape has a finite width. The ODE model  has been extended to include the spatial degrees of freedom that is crucial for describing the dynamics of the peel front as also for  understanding the origin of acoustic emission \cite{Rumiprl,Jagdish08}.

Acoustic emission is commonly observed in an unusually large number of systems such as seismologically relevant fracture studies of rock samples \cite{Scholz68a,Lockner96,Sam92}, martensite transformation \cite{vives,rajeevprl,kalaprl}, micro-fracturing process \cite{Petri94}, volcanic activity \cite{Diodati}, collective dislocation motion \cite{Miguel,Weiss} etc.  The general mechanism attributed to AE is the abrupt release of the stored potential energy although the underlying mechanisms triggering AE are system specific. The nondestructive nature of the AE technique has been useful in tracking the microstructural changes during the course of deformation by monitoring the AE signals. For instance, it is used in   fracture studies of rock samples \cite{Lockner96} and more recently, a similar approach has been used in understanding collective behavior of dislocations \cite {Weiss}.  In both these cases, multiple transducers are used to locate the hypocenters through an inversion process of arrival times \cite{Lockner96,Weiss}. In the latter case, by analysing the dislocations sources generating AE signals, the study establishes the fractal nature of the collective motion of dislocations. (In contrast to these dynamical studies, most studies on AE \cite{Petri94,Diodati,Miguel} are limited to compiling the statistics of the AE signals in an effort to find experimental realizations of self-organized criticality \cite{Bak}.) However, in the case of peeling, using multiple transducers is far from easy and only a single transducer is used leading to scalar time series. In such situations, dynamical information is traditionally recovered using nonlinear time series analysis \cite{GP,HKS,KS}.  However, a major difficulty arises in the present case due to a high degree of noise present  and the associated difficulties involved in curing the noise content. 

Despite large number of experimental investigations and to a lesser extent model studies, several issues related to intermittent peeling and the associated acoustic emission remain ill understood. For instance, there are no models (even in the general area of stick-slip) which show that  the duration of the stick phase can be equal to or even less than that of the slip phase \cite{Cortet}, a feature  which is quite unlike conventional  stick-slip dynamics. From a dynamical point of view, this is also suggestive of the existence of at least three time scales.  The model represents the acoustic energy in terms of the Rayleigh dissipation functional that depends on the local strain rate of the peel front and thus is sensitive to the nature of the peel front dynamics. While  preliminary results of the model \cite{Rumiprl,Jagdish08}  based on a  small domain of parameters were encouraging, no systematic study of the influence of all the relevant time scales on the dynamics of the peel front was carried out.  In particular, while the nature of experimental AE signals changes with the traction velocity, the study of the influence of pull speed on internal relaxational mechanisms, the consequent peel front dynamics and its relationship with the acoustic energy  was not studied either.

The principal objective of the present study is to understand the  various contributing mechanisms to the intermittent peel process and its connection to acoustic emission. The objective is accomplished by carrying out a systematic  study of the influence of the three internal relaxational time scales namely the two inertial time scales of the tape mass and the roller inertia, and dissipative time scale of the peel front.  In particular, we report the influence of the experimentally relevant pull velocity (covering the entire range) on the peel front dynamics.  These studies show that the model exhibits rich spatiotemporal peel front patterns (including the stuck-peeled configurations that mimic fibrillar patterns seen in experiments) arising due to the interplay of the  three time scales. Consequently, varied patterns of model acoustic signals are seen. Another consequence of the inclusion of the three  time scales is that it explains the recent observation that the duration of the slip phase can be larger than that of the stick-phase  \cite{Cortet}.  Interestingly, the model studies show that it is possible to establish a correspondence between the various types of model acoustic energy profiles with certain peel front patterns.    More importantly, the study shows that even as the acoustic energy dissipated is the spatial average of the local strain rate, it can be noisy suggesting the possible deterministic origin of the experimental acoustic signals.   Here,  we report a detailed analysis of the statistical and dynamical analysis of the experimental AE signals. The study shows that while the intermittent peeling is controlled by the peel force function, acoustic emission is controlled by the dynamics of the peel front patterns that determine the local strain rate. This coupled with a comparative study of  a comprehensive nonlinear time series analysis (TSA) of the experimental AE signals for a wide range of traction velocities supplemented by a similar study on the model acoustic energy time series provides additional insights into the connection between AE signals and stick-slip dynamics. In particular, the model displays the recently observed experimental feature that the duration of the slip phase can be more than that of the stick phase with increase in the pull velocity. Finally, the model studies together with the dynamical analysis of the model acoustic signal provide a dynamical explanation for the changes in the nature of the experimental AE signal in terms of the changes in the peel front patterns.  

\begin{figure}[!t]
\hbox{
\centering
\includegraphics[height=2.5cm,width=4.8cm]{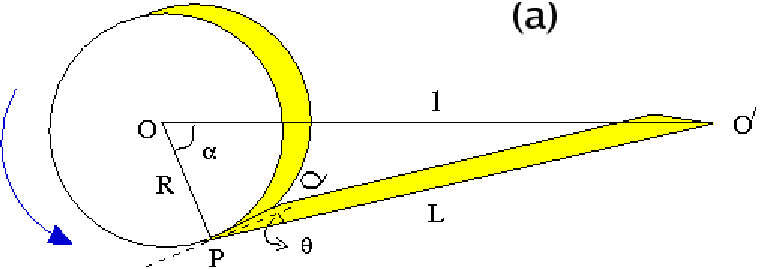}
\includegraphics[height=3.0cm,width=3.5cm]{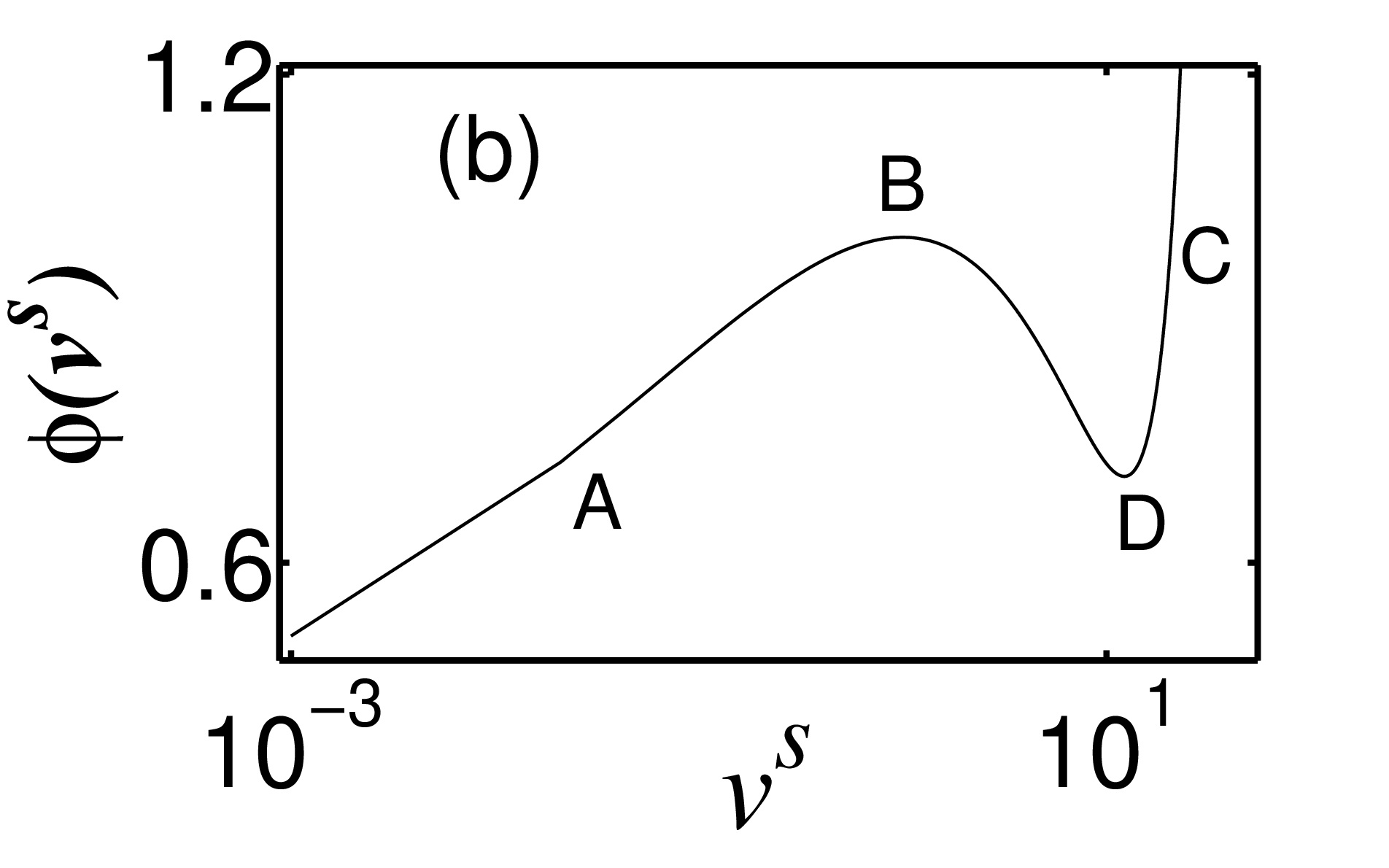}
}

\caption{ (a) (Color online) A schematic representation of the experimental setup. (b) Plot of $\phi(v^s)$ as a function of $v^s$.
}
\label{tapewidth}
\end{figure}

\section{The model}

A typical experimental set up consists of  an adhesive tape mounted on a roller. The tape is pulled at a constant pull velocity using a motor. A schematic representation of the set up is shown in Fig. \ref{tapewidth}(a). The axis of the roller passes through the point O into the plane of the paper. The drive motor is positioned at O$^{\prime}$. Let the distance between $O$ and $O^{\prime}$ be denoted by $l$. $P$ is the contact point on the peel front  $PQ$. Let the peeled length of the tape $PO'$ be denoted by $L$. Several geometrical features can be discussed using a projection on to the plane of the paper.   Let the angle between the tangent to the contact point $P$ and $PO^{\prime}$ be denoted by $\theta$ and the angle $\angle{POO'}$ by $\alpha$.   Then, from the geometry of the Fig. \ref{tapewidth}(a), we get $L\ {cos}\, \theta = -l\ {sin}\,\alpha$ and  $L\ {sin}\,\theta = l\ { cos}\,\alpha - R$ where $R$ is the diameter of the roller tape.  Let the local velocity of the peel point be denoted by $v$ and the displacement (from a uniform stuck state) of the peel front by $u$. Then,  the pull velocity has to satisfy  
\begin{equation}
V= v + \dot u + R \  {\rm cos}\ \theta  \ \dot \alpha. 
\label{Localconstr} 
\end{equation}
As the peel front has a finite width, we define the corresponding quantities along the peel front coordinate $y$ (i.e., along the contact line) by $v(y),\theta(y)$ and $\alpha(y)$. Then as the entire tape width is pulled  a constant  velocity, the above constraint generalizes to  
\begin{eqnarray}
{1\over b} \int^b_0 \big[V- v(y) -  \dot u(y)  -  R \ \ \dot\alpha(y) \ \ {\rm cos} \ \theta(y) \big]dy =0, 
\label{Vconstraint}
\end{eqnarray}
where $b$ is the width of the tape. However, we are interested in the deformation of the peel front of the adhesive, which is a soft visco-elastic material. For the purpose of modeling, while we shall ignore the viscoelastic nature of the adhesive, we recognize its low modulus, i.e., we assume an effective spring constant $k_g$ (along the  contact line) whose value is  much smaller than the spring constant of the tape material $k_t$. This also implies that the force along $PO'$ equilibrates fast and therefore the integrand in  Eq. (\ref{Vconstraint}) can be assumed to vanish for all $y$. Thus, the above equation  reduces to Eq. (\ref{Localconstr}).

The present model is an extension of the ODE model for the contact point dynamics \cite{Rumi05}. The ODE model already contains information on the inertial time scale of the tape mass that allows for dynamical jumps across the two branches of the peel force function. The extension involves introducing the Rayleigh dissipation functional to deal with acoustic emission apart from introducing the spatial degrees of freedom.
 The equations of motion  for  the contact line dynamics are derived by writing down the relevant energy terms consisting of the kinetic energy, potential energy and the energy dissipated during the peel process.  The total kinetic energy $U_k$ is the sum of the rotational kinetic energy of the roller tape and the kinetic energy of the stretched part of the  tape. This is given by 
\begin{equation}
U_K={1\over2}\int^b_0 \xi \big[\dot \alpha(y) +{v(y)\over R} \big]^2 dy + {1\over2}\int^b_0 \rho \big[\dot u(y) \big]^2 dy.
\label{KE}
\end{equation}
Here, $\xi$ is the moment of inertia per unit width of the roller tape and $\rho$  the mass per unit width of the tape. The total potential energy $U_p$ consists of the contribution from the displacement of the peel front due to stretching of the peeled tape and possible inhomogeneous nature of the peel front. This is given by

\begin{equation}
U_P={1\over2}\int^b_0 {k_t\over b} \Big[u(y) \Big]^2 dy + {1\over2}\int^b_0 {k_g b} \Big[{\partial u(y) \over \partial y} \Big]^2 dy. 
\label{PE}
\end{equation}
The peel process always involves dissipation. Indeed,  the peel force function with the two stable branches, one corresponding to low velocities and another at high velocity arises from two different dissipative mechanisms. Apart from this, there is an additional dissipation that arises from the rapid rupture of the peel front which in turn  results in the accelerated motion of local regions of the peel front. We consider this accelerated motion of the local slip as the source responsible for the generation of acoustic signals \cite{Rumiprl}. Any rapid movement also prevents the system from attaining a quasistatic equilibrium which in turn generates dissipative forces that resist the motion of the slip. Such dissipative forces  are modeled by the Rayleigh dissipation functional that depends on the gradient of the local displacement rate \cite{Land}. Indeed, such a dissipative term has proved useful in explaining the power law statistics of the AE signals during martensitic transformation \cite{vives,rajeevprl,kalaprl} as also in explaining certain AE features in fracture studies of rock sample \cite{Rumiepl}.  Then, the total dissipation can be written as the sum of these two contributions

\begin{equation}
{\cal R}={1\over b} \int^b_0 \int f(v(y)) dv dy + {1\over2}\int^b_0 {\Gamma_u\over b} \Big[{\partial \dot u(y) \over \partial y} \Big]^2 dy,
\label{diss}
\end{equation}
where $f(v)$ physically represents the peel force function assumed to be derivable from a potential function  $\Phi(v) = \int f(v)dv$ (see Ref. \cite{Rumi05}). We denote the second term in Eq. (\ref{diss}) by  ${\cal R}_{ae}$ which is identified with the energy dissipated in the form of AE. In the context to plastic deformation, the acoustic energy arising from the abrupt motion of dislocations is given by  ${\cal R}_{ae} \propto \dot \epsilon^2(r)$, where $\dot \epsilon(r)$ is  the local plastic strain rate \cite{Rumiepl}. Following this,  we interpret  ${\cal R}_{ae}$ as the energy dissipated in the form of AE signals. Note that $\frac{\partial \dot u}{\partial y}$ is the local strain rate of the peel front.  As for the first term in Eq. (\ref{diss}), the form of the peel force function we use is given by
\begin{equation}
f(v) = 402v^{0.34}+171v^{0.16}+68e^{(v/7.7)}-369.65v^{0.5}-2.
\label{f}
\end{equation}
We stress here that as we are interested in the generic properties of the peeling process,   the exact form of the peel force function used here is not important as long as major experimental features like the magnitude of the jump in the velocity across the two branches, the range of  values of the measured peel force function, in particular the values  at the maximum and minimum, are captured. 

As can be seen from Eq.  (\ref{KE}), there are two time scales; one corresponding to the inertia of the tape mass and the other due to the roller inertia. In addition, there is a third time scale, namely the dissipative time scale in Eq. (\ref{diss}) (second term). Thus, there are three internal relaxational time scales in the model. Apart from this, there is also a time scale due to the pull speed. Then the nature of the dynamics is determined by an interplay among all these time scales.

It is more convenient to deal with scaled quantities. Consider introducing  basic length  and time scales which will be used to rewrite all the energy terms in scaled form. A natural choice for a time like variable is $\tau = \omega_{u} t$ with $\omega_{u}^2={k_t/(b \ \rho)}$. In a similar way, we introduce a basic length scale defined by $d=f_{max}/k_t$, where  $f_{max}$ is the value of $f(v)$ at $v_{max}$ on the left  stable branch.  We define scaled variables  by $u = X d = X (f_{max}/k_t) $, $l =  l^s d$, $L =  L^s d$   and $R =  R^s d$. The peel force function $f$ can be written as  $\phi(v^s) = f(v(v^s))/f_{max}$. Here $v^s=v/v_c\omega_u d$   and $V^s=V/v_c\omega_u d$ are the dimensionless peel and pull velocities  respectively with $v_c = v_{max}/ \omega_u d$ representing the dimensionless critical  velocity at which the unstable branch starts. Using this we can define  a few relevant  scaled parameters $C_f=(f_{max}/k_t)^2(\rho/\xi)$,   $k_0=k_g b^2/(k_t a^2)$,   $\gamma_u = \Gamma_u \omega_u/(k_t a^2)$, and $y = ar$, where $a$ is a  unit length variable along the peel front. The parameter $C_f$ is a measure of the relative strengths of the inertial time scale of the stretched tape to that of the roller, $k_0$ the relative strengths of the effective elastic constant of the adhesive to that of the tape material and $\gamma_u$ the strength of the dissipation coefficient. Then, the scaled local form of  Eq. (\ref{Localconstr}) takes the form 

\begin{equation}
\dot X = (V^s - v^s)v_c + R^s \ {l^s \over L^s} \ ({sin}\ \alpha)\ \dot \alpha. 
\label{localconstraint}
\end{equation}  

In terms of the scaled variables, the scaled kinetic energy $U^s_K$ and scaled potential energy $U^s_P$ can be
respectively written as 

\begin{eqnarray}
U^s_K &= &{1\over2 C_f}\int^{b/a}_0 \Big[ \Big(\dot \alpha(r) +{v_c v^s(r)\over R^s} \Big)^2  + C_f{\dot X}^2(r)\Big] dr, \\
\label{ScKE}
U^s_P &=& {1\over2}\int^{b/a}_0 \Big[ X^2(r) + k_0 \Big({\partial X(r) \over \partial r} \Big)^2 \Big]dr. 
\label{ScPE}
\end{eqnarray} 

The total dissipation in the scaled form is
\begin{equation}
{\cal R}^s = R_f^s +  R_{ae} = {1\over b} \int^{b/a}_0 \Big[\int \phi(v^s(r)) dv^s  + {\gamma_u\over2}\Big({\partial \dot X(r) \over \partial r} \Big)^2 \Big]dr.
\label{ScDiss}
\end{equation}

The first term on the right hand side is the frictional dissipation arising from  the peel force function. The scaled peel force function, $\phi(v^s)$, can be obtained by using the scaled velocities in Eq. (\ref{f}). The nature of $\phi(v^s)$ is shown in Fig. \ref{tapewidth}(b). Note that the maximum occurs at $v^s =1$. We shall refer to the left branch AB as the `stuck state' and  the high velocity branch CD as the 'peeled state'. The second term on the right hand side denotes the scaled form of the acoustic energy dissipated.

The Lagrange equations of motion in terms of the generalized coordinates $\alpha(r), \dot\alpha(r), X(r) $ and $\dot X(r)$ are
\begin{eqnarray}
{d\over d\tau}\left({\partial{\cal L} \over {\partial\dot\alpha(r)}}\right)-
{\partial{\cal L} \over {\partial\alpha(r)}}+
{\partial{\cal R}^s \over {\partial\dot\alpha(r)}}&=0&,\\
{d\over d\tau}\left({\partial{\cal L} \over {\partial \dot X(r)}}\right)-
{\partial{\cal L} \over {\partial X}(r)}+
{\partial{\cal R}^s \over {\partial \dot X(r)}}&=&0.
\end{eqnarray}

Using this, we get the equations of motion as
\begin{eqnarray}
 \ddot \alpha &=& - {v_c \dot v^s \over R^s}  - C_f R^s {{l^s \over L^s} \, {sin}\, \alpha \over (1 + {l^s \over L^s} \, {sin}\, \alpha)} \phi(v^s),\label{seqalpha} \\
\ddot X &=& - X + k_0 {\partial^2 X \over \partial r^2}  + {\phi(v^s)\over (1+ {l^s\over L^s} \, {sin}\, \alpha)} + \gamma_u {\partial^2 \dot X \over \partial r^2}.  
\label{sequ}
\end{eqnarray}

However, Eqs. (\ref{seqalpha}, \ref{sequ}) should  satisfy  the constraint Eq. (\ref{localconstraint}). This consistency can be imposed by using the theory of  mechanical  systems with constraints \cite{ECG}. This leads to an  equation for the acceleration  variable $\dot v^s (r)$  obtained by differentiating Eq. (\ref{localconstraint}) and using Eqs. (\ref{sequ}), 

\begin{eqnarray}
\nonumber
 \dot v^s &=& \big[- {\ddot  X} + {R^sl^s\over L^s} \big(\dot \alpha^2 (cos \alpha -{R^s l^s}({sin \alpha \over L^s})^2  \big) \\
 & + & sin \alpha {\ddot \alpha} \big) \big]/v_c.
\label{sdotv}
\end{eqnarray}

These Eqs. (\ref{localconstraint}, \ref{seqalpha}) and (\ref{sdotv}) constitute a set of nonlinear partial differential equations that determine the dynamics of the peel front. They have been solved by discretizing the peel front on a grid of  N points and using an adaptive step size stiff differential equations solver (MATLAB package).  We have used open boundary conditions appropriate for the problem. The initial conditions were drawn from the stuck configuration (i.e., the values are from the left branch of $\phi(v^s)$) with a small spatial inhomogeneity in $X$ such that  they satisfy Eq. (\ref{localconstraint}) approximately. The system is evolved till a steady state is reached before the data is accumulated.

The nature of the dynamics depends on the pull velocity $V^s$, the dissipation coefficient $\gamma_u$ and $C_f$. We have carried out detailed studies of the dynamics of the model over a wide range of values of these parameters keeping other parameters fixed at $R^s=0.35$, $l^s=3.5 $, $k_0=0.1$ ($k_t=1000$ N/m) and $N=50$ (in units of the grid size). Larger system size $N=100$ is used whenever necessary.

\section{Time series analysis of experimental AE signals}

One of the objectives is to carry out statistical and nonlinear time series analysis of experimental AE signals associated with the jerky peel process with a view to understand the results on the basis of model studies. Acoustic emission data files were obtained from peel experiments  under  constant traction velocity conditions that  cover a wide range of values from $0.2$ to $7.6$ cm/s \cite{Ciccotti07}.   Signals were recorded at the standard audio sampling frequency of $44.1$ kHz (having $6$ kHz band width) using a high quality microphone. They were digitized and stored as $16$ bit signals in raw binary files.  There are 38 data files each containing approximately $1.2 \times 10^6$ points. The AE signals are noisy as in most experiments on AE.

Two characteristic features of low dimensional chaos are the existence of a strange attractor with self similar properties quantified by a fractal dimension (or equivalently the correlation dimension) and sensitivity to initial conditions quantified by the existence of a positive Lyapunov exponent.  Given the equations of motion, these quantities can be directly calculated. However, when a scalar time series is suspected to be a projection from a higher dimensional dynamics, they are traditionally analyzed by using embedding methods that attempt to recover the underlying dynamics.  The basic idea is to unfold the dynamics through a phase space reconstruction of the attractor by embedding the time series in a higher dimensional space using a suitable time delay\cite{Packard,GP}.  Consider a scalar time series measured in units of sampling time $\Delta t$ defined by $[x(k),k=1,2,3,\cdots ,N]$. Then, we can construct $d-$dimensional vectors defined  by   $\vec{\xi}_{k}=[x(k),x(k+\tau),\cdots,x(k+(d-1)\tau)]; \,\, k=1,\cdots ,[N-(d-1)\tau]$. The delay time $\tau$ suitable for the purpose is either obtained from the autocorrelation function or from mutual information \cite{HKS}. Once the reconstructed attractor is obtained, the existence of converged values of correlation dimension and a positive exponent is taken to be a signature of  the underlying chaotic dynamics.  In real systems, most experimental signals contain noise which in this case is high. There are several methods designed to cure the noise component \cite{HKS,KS,KS1,GHKSS,CH}. Usually, the cured data sets are then subjected to further analysis.

The correlation integral defined as the fraction of pairs of points $\vec{\xi}_{i}$ and $\vec{\xi}_{j}$ whose distance is less than $r$, is given by
\begin{equation}
C(r)=\frac{1}{N_p}\sum_{i,j} \Theta(r-|\vec{\xi}_i-\vec{\xi}_j|),
\end{equation}
where $\Theta(\cdots)$ is the step function and $N_p$ the number of vector pairs summed. A window is imposed to exclude temporally correlated points \cite{HKS}.  The  method provides equivalence between the reconstructed attractor and the original  attractor. It has been shown that a proper equivalence is possible if the time series is noise free and long \cite{Ding}. For a self similar attractor $C(r)\sim r^{\nu}$, where $\nu$ is the correlation dimension \cite{GP}. Then, as $d$ is increased, one expects to find a convergence of the slope $dln C(r)/d ln r$ to a finite value in the limit of small $r$. However, in practice,  the scaling regime is found at  intermediate length scales due to the presence of noise.

The existence of a positive Lyapunov exponent is considered as an unambiguous quantifier of chaotic dynamics. However, the presence of superposed noise component, which in the present case is high, poses problems.  In principal the noise component can be cured and then the Lyapunov exponent calculated \cite{HKS,KS,KS1,GHKSS,CH}.  Here, we use an algorithm that does not require preprocessing of the data;  it is designed to average out the influence of superposed noise. The algorithm, which is an extension of Eckmann's algorithm, has been shown to work well for reasonably high levels of noise in model systems as well as for short time series. The method has  been used to analyze experimental time series as well (for details, see Ref. \cite{Anan99,Noro01}).

In the conventional Eckmann's algorithm \cite{Eckmann}, a sequence of tangent matrices are constructed that connect the initial small difference vector $\vec{\xi}_i - \vec{\xi}_j$ to evolved difference vectors $\vec{\xi}_{i+k} - \vec{\xi}_{j+k}$, where $k$ is the propagation time.  In the algorithm, the number of neighbors used is small typically min$[2d,d+4]$ contained in a spherical shell of size $\epsilon_s$. A simple modification of this is to use those neighbors falling between an inner and outer radii $\epsilon_i$ and $\epsilon_0$ respectively. Then, the inner shell $\epsilon_i$ is expected to act as a noise filter. However, so few neighbors will not be adequate to average out the noise  component superposed on the signal. Thus, the modification we effect is to allow more number of neighbors so that the noise statistics is sampled properly. (See for details \cite{Anan99,Noro01}.) As  the sum of the exponents should be negative for a dissipative system, we impose this as a constraint. In addition, we also demand the existence of stable positive and zero exponents (a necessary requirement for continuous time systems like AE) over a finite range of shell sizes $\epsilon_s$.  As a cross check, we have also calculated the correlation integral and Lyapunov spectrum using the TISEAN package as well \cite{HKS}. 

\section{Dynamics of the peel front}

A systematic study of the dynamics of the model is essential to understand the influence of the various parameters on the spatiotemporal dynamics of the peel front, its connection to intermittent peeling and to the accompanying acoustic emission.  From Eq. (\ref{ScDiss}), it is clear that the acoustic energy $R_{ae}$ is the spatial average of the local strain  rate. As the peel front patterns determine the nature of acoustic energy,  a detailed study of the dependence of the patterns on the relevant parameters and on the pull velocity should help us to get insight into AE generation process during peeling.

\subsection{General considerations on time scales and parameter values} 

We begin by making some general observation about the various parameters and their influences. The dynamics of the model is sensitive to the three time scales (reduced from four due to scaling) determined by the parameters $C_f$, $\gamma_u$ and $V^s$. $C_f$ is related to the ratio of inertial time of the tape mass to that of roller inertia (see below). The dissipation parameter $\gamma_u$ reflects the rate at which the local strain rate relaxes.  The pull velocity $V^s$ determines the duration over which all the internal relaxations are allowed to occur. The range of $C_f$ is determined by the allowed values of the tape mass $m$ and the roller inertia $I$. Following our earlier studies,  we vary $I$ from $10^{-5} $ to $10^{-2}$ and $m$ from  0.001 to 0.1.  Thus, $C_f$ can be varied over a few orders of magnitude keeping one of them fixed. For model calculations, the dissipation parameter is varied from 0.001 to 1. (However, an order of magnitude estimate shows that $\gamma_u << 1$, see below.)  The range of $V^s$ of interest is determined by the instability domain which is from 1 to $\sim 12$ as shown in Fig. \ref{tapewidth}(b).

To appreciate the influence of inertial time scale of tape mass parameterized by $C_f$, consider the low mass limit of the ODE model \cite{Rumi05} which has been shown to lead to the DAE model equations \cite{Rumi04}. In this limit, the velocity jumps across the two branches of the peel force function are abrupt with infinite acceleration. However, finite tape mass introduces an additional time scale that leads to jumps in $v^s$ to occur over a finite time scale which in turn the magnitude of the velocity jumps. Indeed, the phase space trajectory need not jump to the high velocity branch of $\phi(v^s)$, as we shall see. This can be better appreciated by considering the ODE model (that ignores the spatial degrees of freedom). Consider the relevant ODE model equations \cite{Rumi05} (in unscaled form).
\begin{eqnarray}
\label{Ieqn}
{\ddot \alpha} &=& -\frac{\dot v}{R} + \frac{R}{I}\frac{cos \, \theta}{(1-cos \, \theta)}f(v),\\
m{\ddot u}  &=& \frac{1}{(1-cos \, \theta)}[ f(v) -ku(1-cos \, \theta)],
\label{Meqn}
\end{eqnarray}
where $\alpha$ is shown in Fig. \ref{tapewidth}(a) and $u$ the displacement of the contact point.  $m$ is mass of the tape and $k$ the spring constant of the tape. From, Eqs. ( \ref{Ieqn}, \ref{Meqn}), two inertial time scales can be identified, one corresponding to the roller inertia  $\Omega_{\alpha} = (Rf/I)^{1/2}$ and another to that of the tape mass  $\Omega_u = (k/m)^{1/2}$. (Note that $k$ in Eq. (\ref{Meqn}) of the ODE model corresponds to $k_t$ in the present model.) Thus, $C_f$ in present model is directly related to the ratio of these two inertial time scales. Differentiating Eq. (\ref{Localconstr}), we get
\begin{equation}
{\dot v} + {\ddot u} + R \, {\ddot\alpha} \, cos \, \theta = R \, {\dot \alpha} \, {\dot \theta} \,  sin \theta.
\label{veqn}
\end{equation}   
Eq. (\ref{Meqn}) is the force balance equation. In the limit $m \rightarrow 0$, we have the algebraic constraint $f(v) = F(t)(1- cos \, \theta(t))$. Differentiating this equation shows that $\dot v$ diverges at points of maximum and minimum of the peel force function $f(v)$.  This demonstrates that in the low mass limit, the orbits jump to the high velocity branch abruptly. Now consider Eq. (\ref{veqn}) that relates the acceleration of the peel point ($\dot v$), acceleration of the displacement $u$ i.e., $\ddot u$  and $\ddot \alpha$.  This again is basically a force balance equation as can be seen by multiplying the equation by  the tape mass $m$. As the right hand side is small, any increase in one of these acceleration variables implies a decrease in the other variables. As low mass limit implies infinite acceleration of the peel front ($\dot v$) across the peel force function, finite mass implies the velocity jumps across the peel force function is reduced. It is worthwhile to note that  the effect of inertial time scale causing jumps across the unstable branch to occur at a finite time scale is a general feature. This has been recognized and demonstrated experimentally in the context of the PLC effect \cite{SF85}.

\begin{figure}[!b]
\vbox{
\includegraphics[height=3.2cm,width=7.5cm]{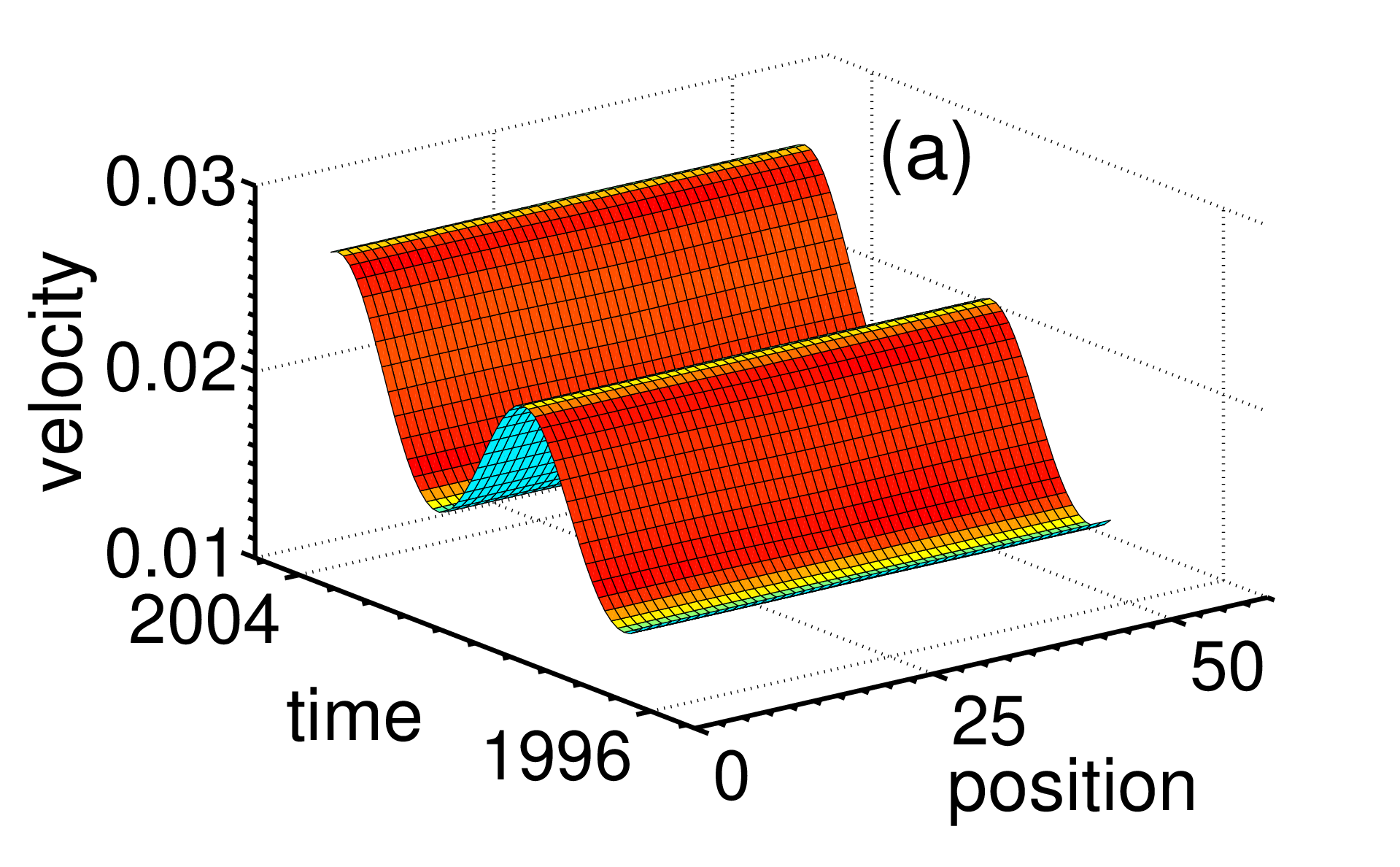}
\includegraphics[height=3.2cm,width=7.5cm]{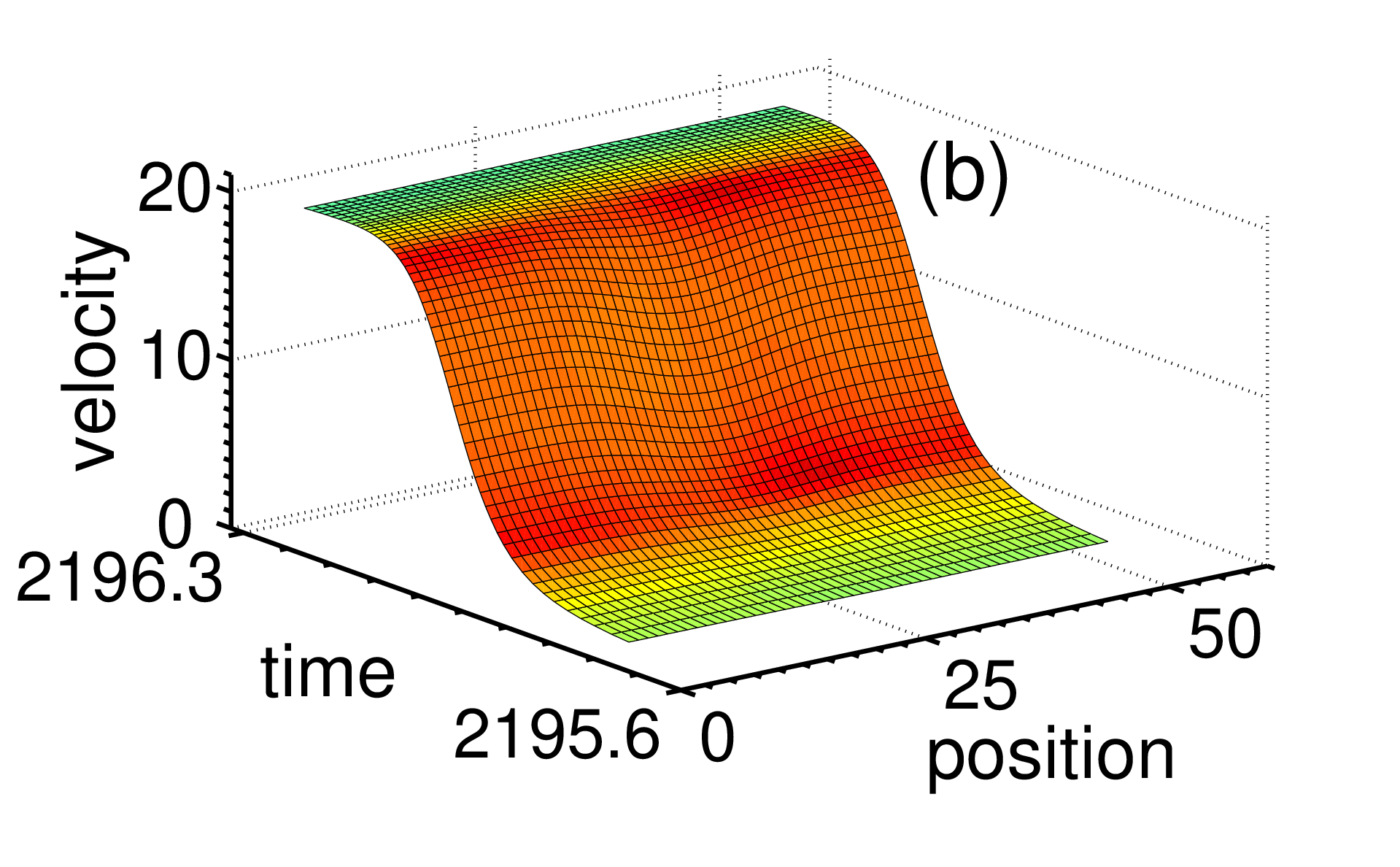}
\includegraphics[height=3.0cm,width=6.0cm]{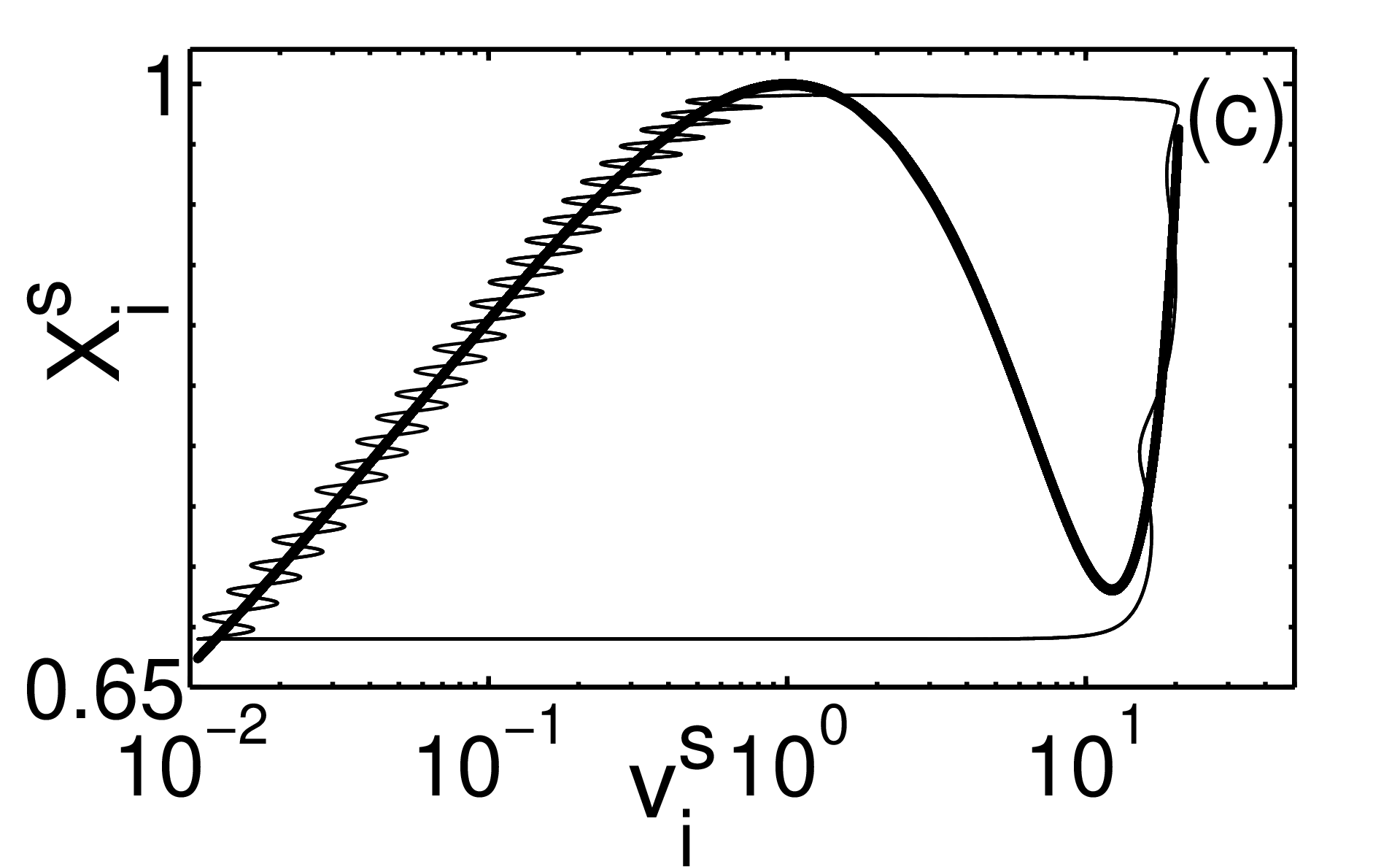}
\includegraphics[height=3.0cm,width=6.0cm]{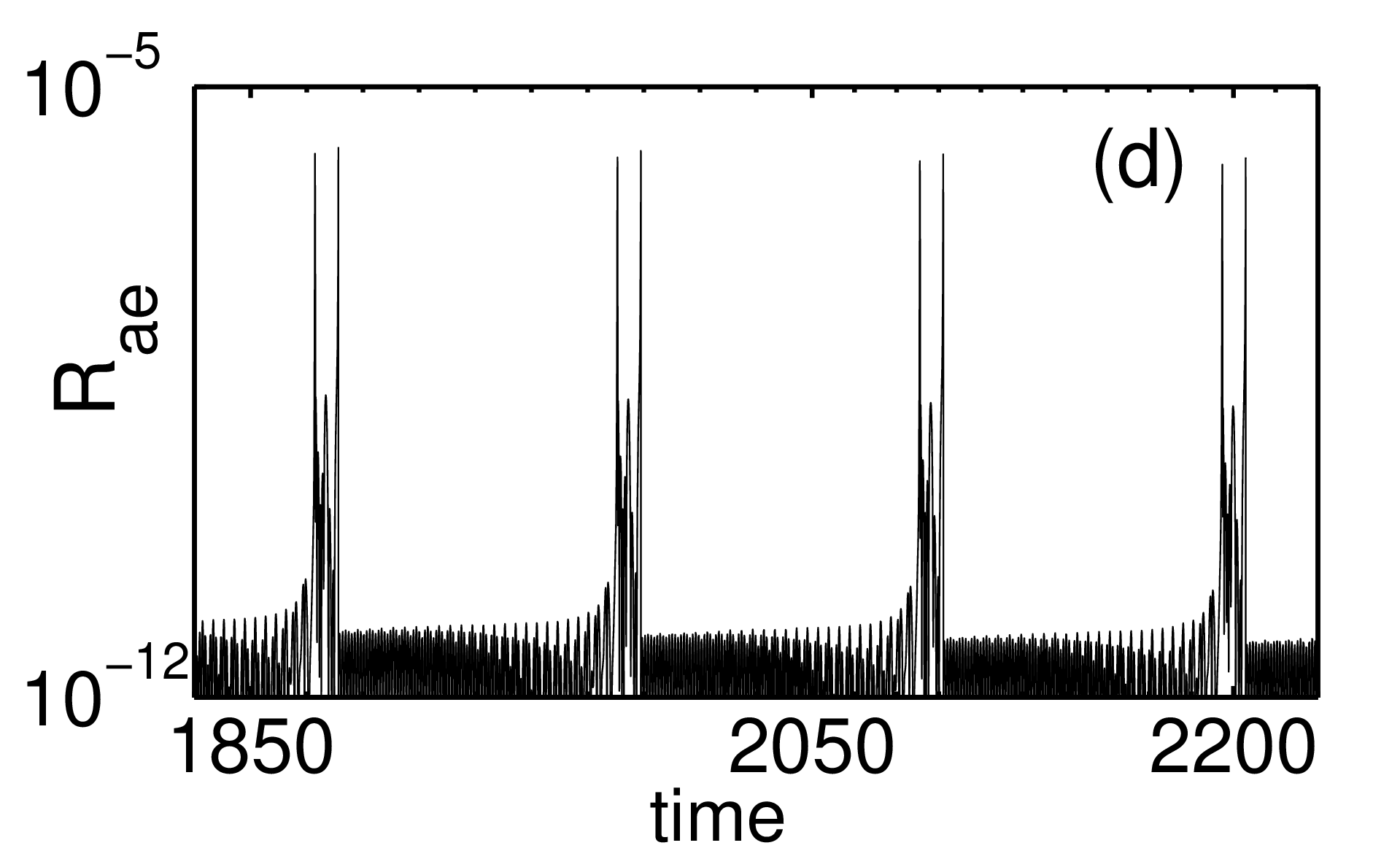}
\includegraphics[height=3.44cm,width=7.5cm]{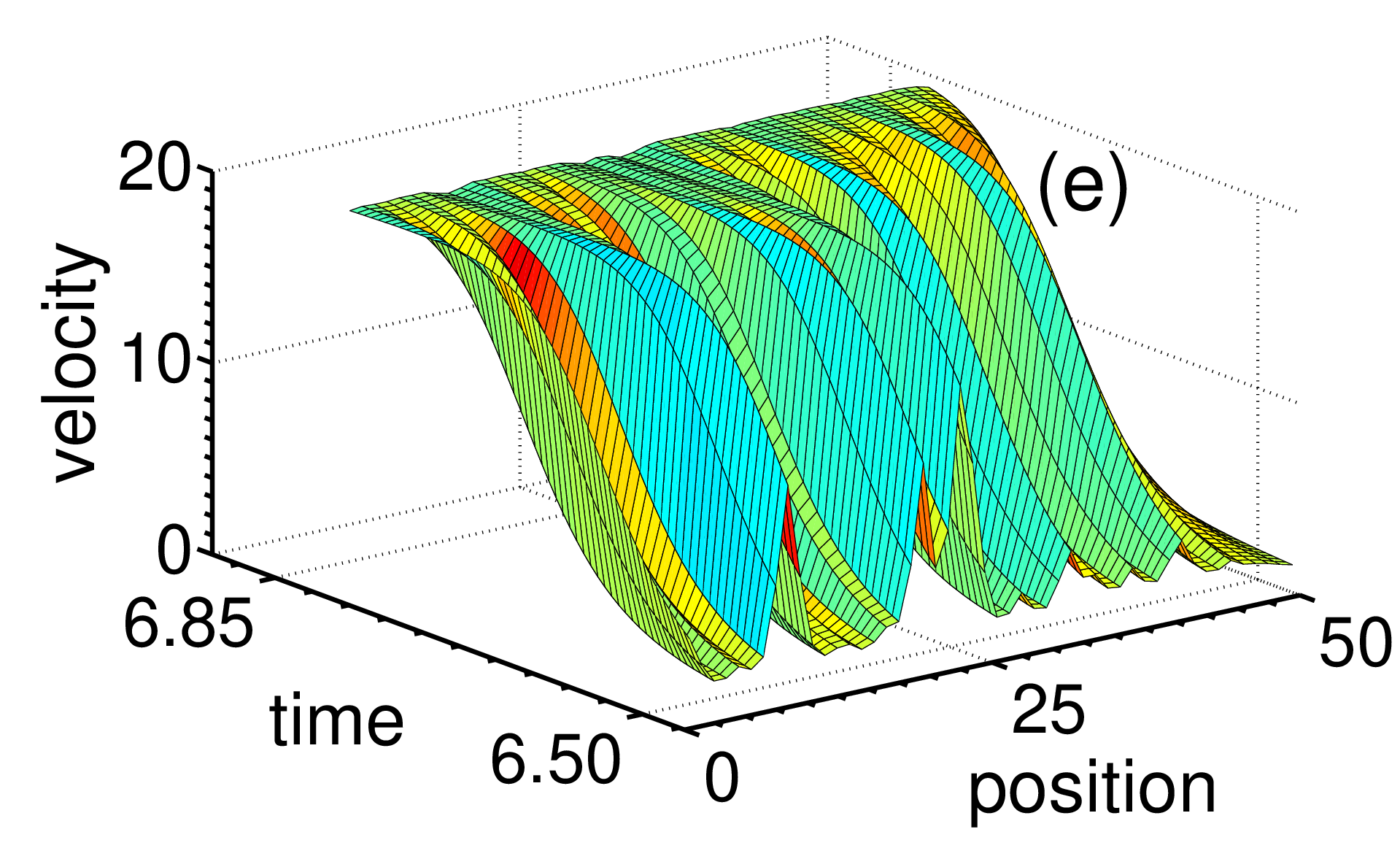}
\includegraphics[height=3.44cm,width=7.5cm]{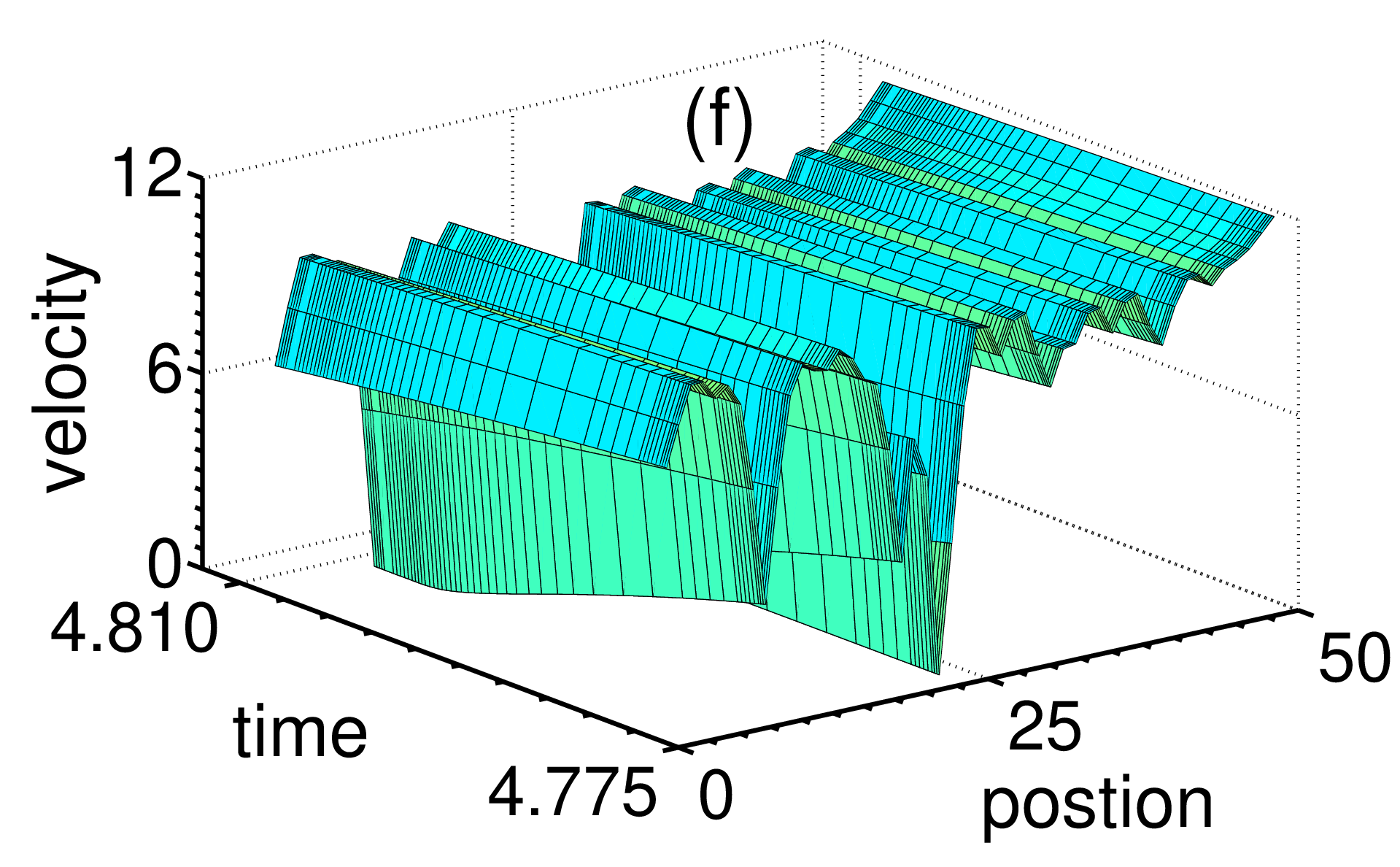}
}
\caption{(Color online a, b, e, f) (a, b) Snapshots for $C_f = 7.88$, $V^s=1.48$, and $\gamma_u=1.0$. (a) a smooth peel front  with small amplitude high frequency oscillations due to finite roller inertia and (b) a smooth peel front. (c) Phase plot for an arbitrary spatial point on the  peel front. Bold line represents $\phi(v^s)$. (d) Model acoustic energy plot. (e, f) Snapshots  for $C_f = 7.88$, $V^s=1.48$,  and $\gamma_u=0.1$. (Rugged peel front and the onset of stuck-peeled configuration.) }
\label{I5m3V1gu1}
\end{figure}

Now consider estimating the order of magnitude of the dissipative time scale. The unscaled dissipation parameter $\Gamma_u$ is related to the fluid shear viscosity $\eta$ \cite{Land} and thus an order of magnitude estimate can be obtained. Typical values of $\eta$ for adhesives at low shear rates is $\sim 1000 -10000$ Pa.s. As stress is directly related to shear viscosity, $\Gamma_u$ can be estimated using typical dimensions of the peel front.   It has been shown that  deformed peel front dimension is about 100 $\mu m$ \cite{Dickinson,Yamazaki}, the thickness of the adhesive is $ \sim$ 50 $\mu m$ and the width of the peel front $\sim 20$ mm (width of the tape). It is easy to show that $\Gamma_u \sim 10^{-3} -10^{-2}$ J.s.  Thus, the range of $\gamma_u$ is $\sim 10^{-3} -10^{-4}$ taking $\eta \sim 1000$ Pa.s.  As some of the numbers used are material dependent, this is  just an order of magnitude estimate. For model studies, the range of $\gamma_u$ is taken to be from 1 to 0.001. However, we will not discuss the results for  $\gamma_u =0.001$ as these are similar to 0.01.

Within the scope of the model, the model acoustic energy given by $ R_{ae}(\tau)={1\over 2}\gamma_u \sum_i(\dot X_{i+1}- \dot X_i)^2$ (in the discretized form) depends on the nature of the local displacement rate. Based on this relation, some general observations can be made on the nature of $R_{ae}$ and its dependence on the peel front dynamics.  From Eq. (\ref{sequ}), high $\gamma_u$ implies that the coupling between neighboring sites is strong and hence the local dynamics at one spatial location has no freedom to  deviate from that of its neighbor. Thus, the displacement rate at a point on the peel front cannot differ from that of its neighbor. For the same reason, low $\gamma_u$ implies weak coupling between displacement rates on neighboring points on the peel front which therefore can differ substantially. This clearly should lead to significantly more inhomogeneous peel velocity profile. Based on the above arguments, high $\gamma_u$ should lead to smooth peel front and consequently sharp bursts in the model acoustic energy $R_{ae}$ that occurs during jumps between the two branches of $\phi(v^s)$. In contrast, when $\gamma_u$ is small, $R_{ae}$ should be high as also spread out in time. However, as the exact nature of the peel front pattern is sensitive to the values of  $C_f$, pull velocity $V^s$ and $\gamma_u$, the nature of $R_{ae}$  depends on all the three time scales.   Indeed, one should expect that the more rapidly the peel front patterns change with time, the noisier the model acoustic energy should be. This is one feature that we hope to compare with experimental acoustic signals.

\subsection{Results of the model}

We have carried out extensive studies on the  nature of the dynamics for a wide range of values of the parameters stated above. The peel front dynamics is analyzed by recording the velocity-space-time patterns of the peel front, the phase plots in the $X^s-v^s$ plane for an arbitrary spatial point on the peel front and the model acoustic energy dissipated $R_ {ae}$. (Unless otherwise stated, these plots refer to steady state dynamics after all the transients have died out.)  Here, we present a few representative solutions for different sets of parameters within the range of interesting dynamics. Our analysis shows that while the nature of the dynamics results from  competing influences of the three time scales, the dissipation parameter $\gamma_u$ appears to have a significant influence on the spatiotemporal dynamics of the peel front. 

\subsubsection{ Case (i), $C_f = 7.88$ - high (low) tape mass, low (high) roller  inertia }

\begin{figure}[!h]
\vbox{
\includegraphics[height=3.6cm,width=7.5cm]{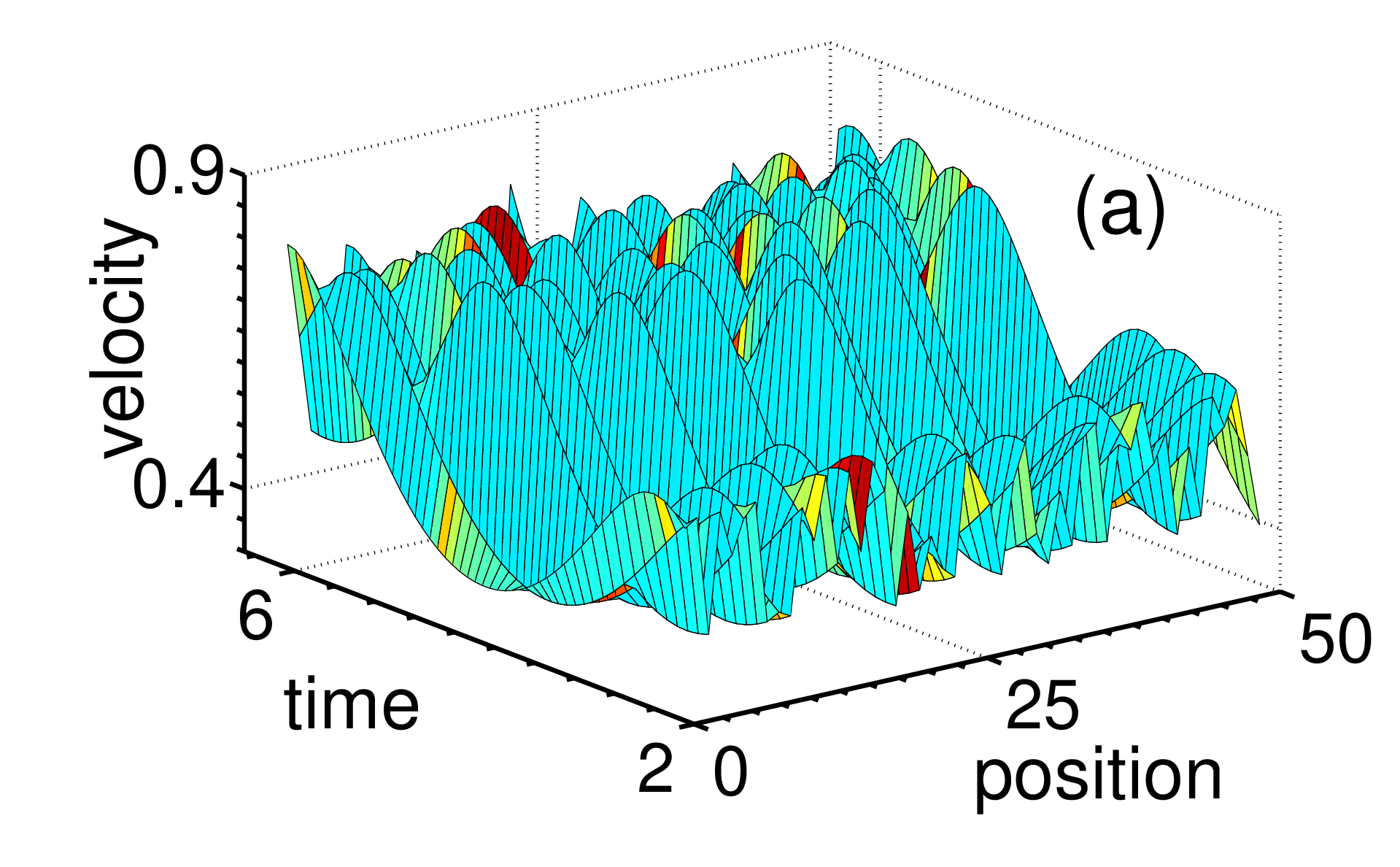}
\includegraphics[height=3.6cm,width=7.5cm]{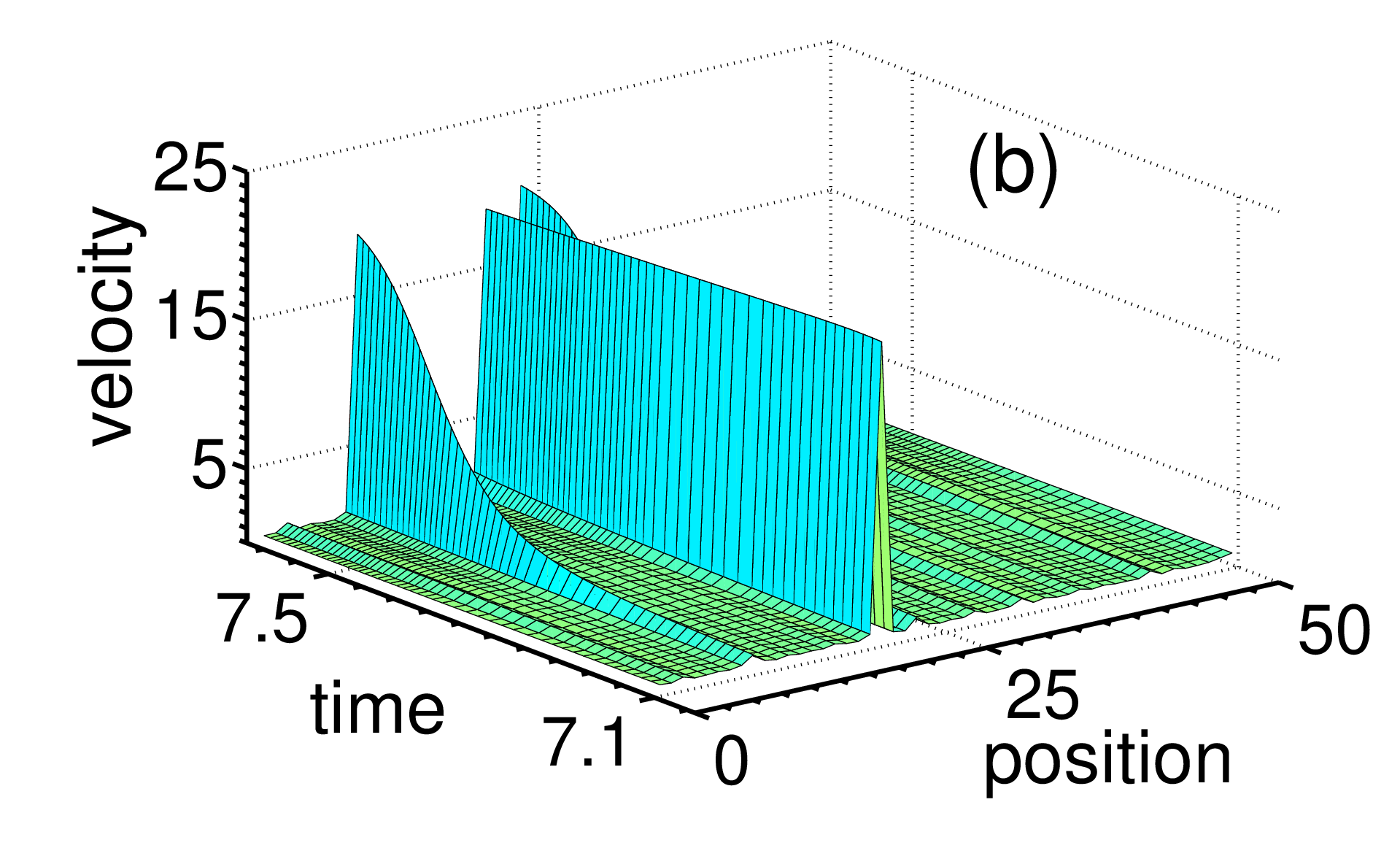}
\includegraphics[height=3.6cm,width=7.5cm]{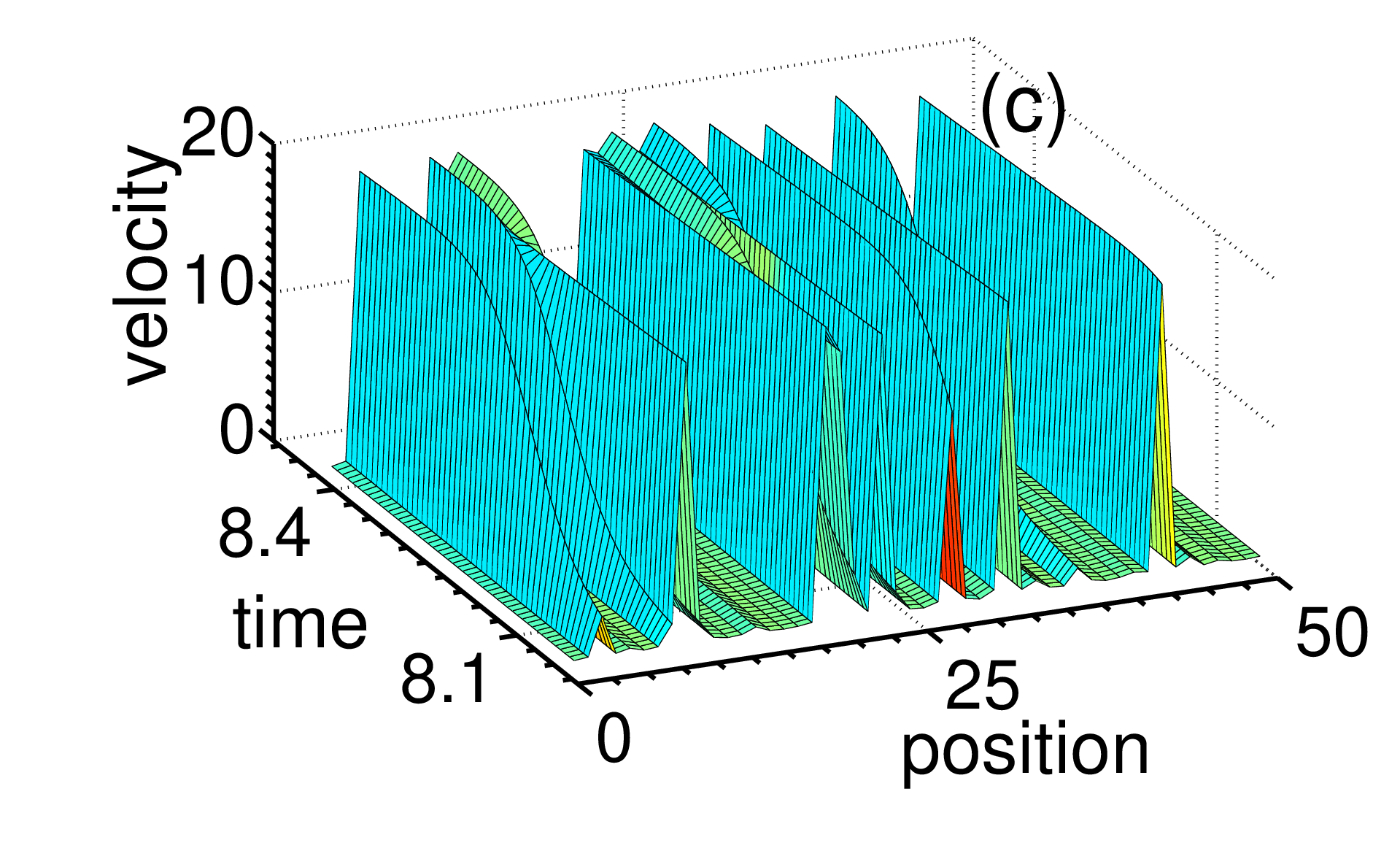}
\includegraphics[height=3.6cm,width=7.5cm]{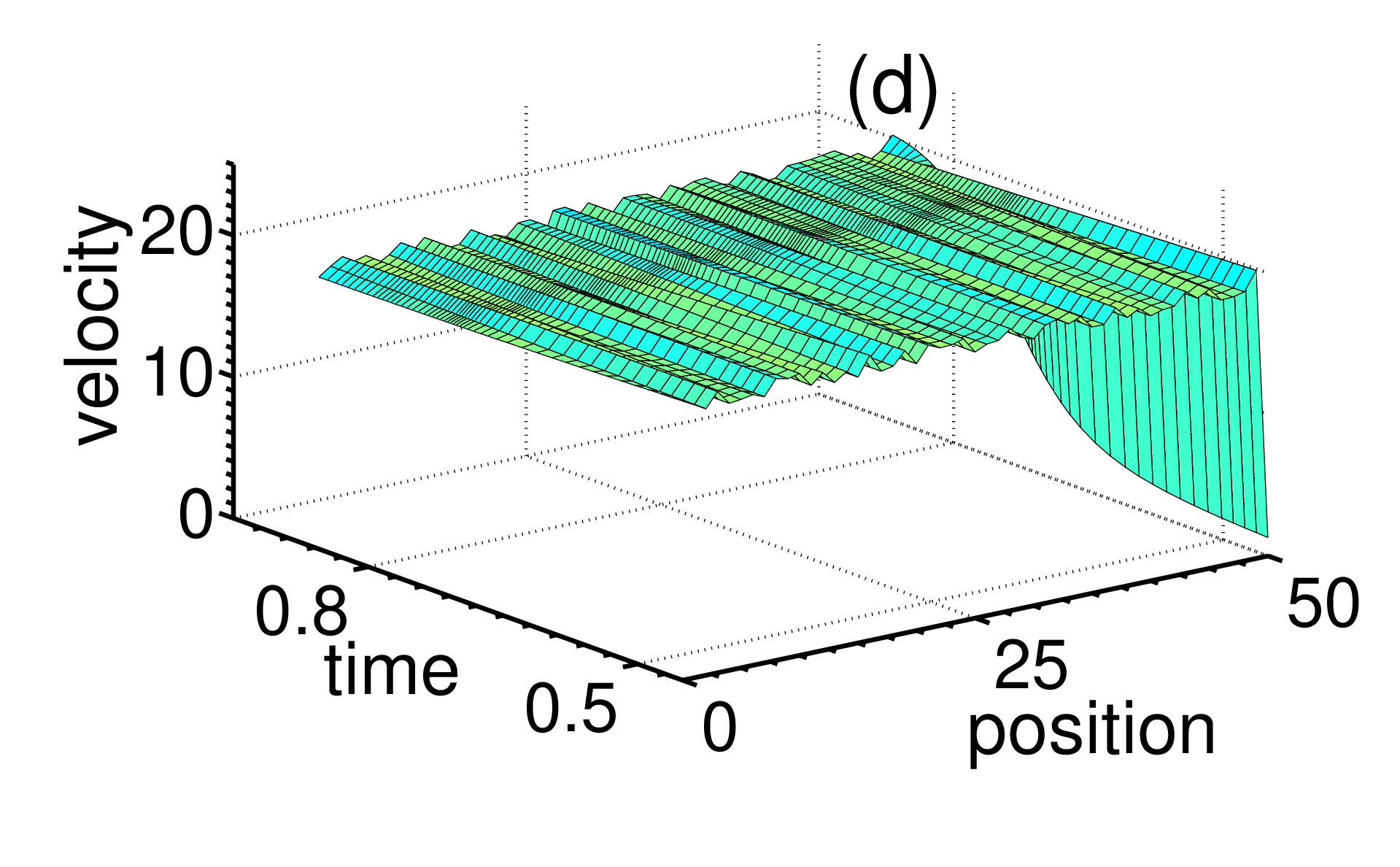}
}
\caption{(Color online) (a-d) Snapshots during the peeling process for $C_f = 7.88, V^s=1.48$,  and $\gamma_u=0.01$ : (a) highly rugged peel front when the system is on the left branch of $\phi(v^s)$,  (b) the onset of peel process, (c) stuck-peeled configuration and (d) resulting nearly uniform peeled state.}
\label{I5m3V1gu0_1}
\end{figure}

Given a  value of $C_f$ there is a range of values of $(m,I)$. In this case, the set of values are:   $(0.1,10^{-3}), (0.01,10^{-4})$ and  $(0.001,10^{-5})$. The dissipation coefficient is varied from $\gamma_u =1 $ to 0.01. For high $\gamma_u =1.0$, only smooth peeling is seen independent of the magnitude of the pull velocity. The peel front switches between the low and high velocity branches of the peel force function $\phi(v^s)$. Plots of the smooth nature of the entire peel front are shown in Figs. \ref{I5m3V1gu1}(a, b) for $V^s =1.48$. Figure \ref{I5m3V1gu1}(a) shows the nature of the peel front when the system is on the low velocity branch of $\phi(v^s)$,  i.e., the local velocities of all spatial elements follow the AB branch of $\phi(v^s)$. The small amplitude synchronous high frequency oscillation of the entire peel front results  from the roller inertia. (Compare the values of $v^s$ in the two figures.)  A phase plot in the $X^s - v^s$ plane for an arbitrary point on the peel front is shown in Fig. \ref{I5m3V1gu1}(c). The small amplitude oscillation of the peel front shown in Fig. \ref{I5m3V1gu1}(a) corresponds to the velocity oscillations in the phase plot (Fig. \ref{I5m3V1gu1}(c)). As high $C_f$ implies relatively low values of $I$, it can be shown (on lines similar to Ref. \cite{Rumi04}) that the orbit sticks  to the stationary branches (slow manifold) of the peel force function $\phi(v^s)$ jumping between the branches only at the limit of stability typical of relaxation oscillations.   
The corresponding  model acoustic energy $ R_{ae}(\tau)$ shows a sequence of small amplitude spikes corresponding to the small amplitude oscillations arising from the roller inertia [Fig. \ref{I5m3V1gu1}(d)] followed by large bursts that occur  at regular intervals. The bursts result from the peel front jumping from the stuck to the peeled state and back.  Note that the duration of the bursts are short compared to duration between them.

However, as we decrease $\gamma_{u}$ to $0.1$ keeping $V^s=1.48$, we observe rugged and stuck-peeled configurations.  The rugged pattern is seen when the system is on the AB branch of  $\phi(v^s)$. Even so, on reaching the limit of stability, the entire contact line peels nearly at the same time as shown in Fig. \ref{I5m3V1gu1}(e).  But once it jumps to the high velocity branch CD of $\phi(v^s)$, the peel front that has nearly uniform peel velocity commensurate with that of the right branch of $\phi(v^s)$  becomes unstable and breaks up into stuck and peeled segments as shown in Fig. \ref{I5m3V1gu1}(f). The width of these segments increases in time with a concomitant decrease in the magnitude of the velocity jumps of peeled segments, eventually the entire peel front goes into a stuck state. Then, the cycle restarts with the peel front switching between the rugged and stuck-peeled (SP) states. The phase plot is similar to that for $\gamma_u= 1$ again sticking to the slow manifold. The model acoustic energy dissipated $R_{ae}$ is also similar to that for $\gamma_u=1$ except that the large bursts are comparatively broader as should be expected due to presence of  stuck-peeled configurations that contribute to large changes in the local velocity.
\begin{figure}[!b]
\vbox{
\includegraphics[height=3.15cm,width=7.5cm]{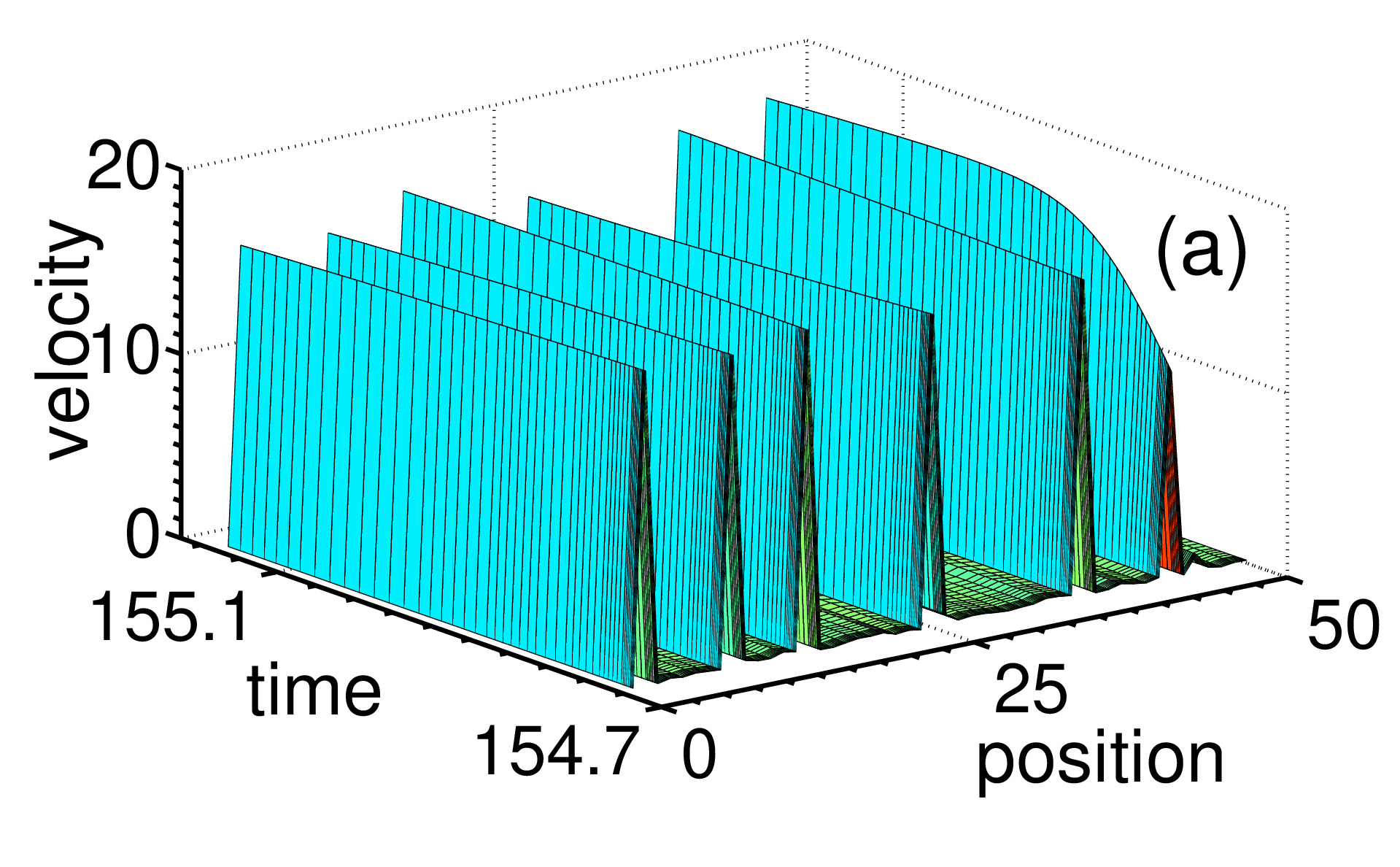}
\includegraphics[height=3.0cm,width=7.5cm]{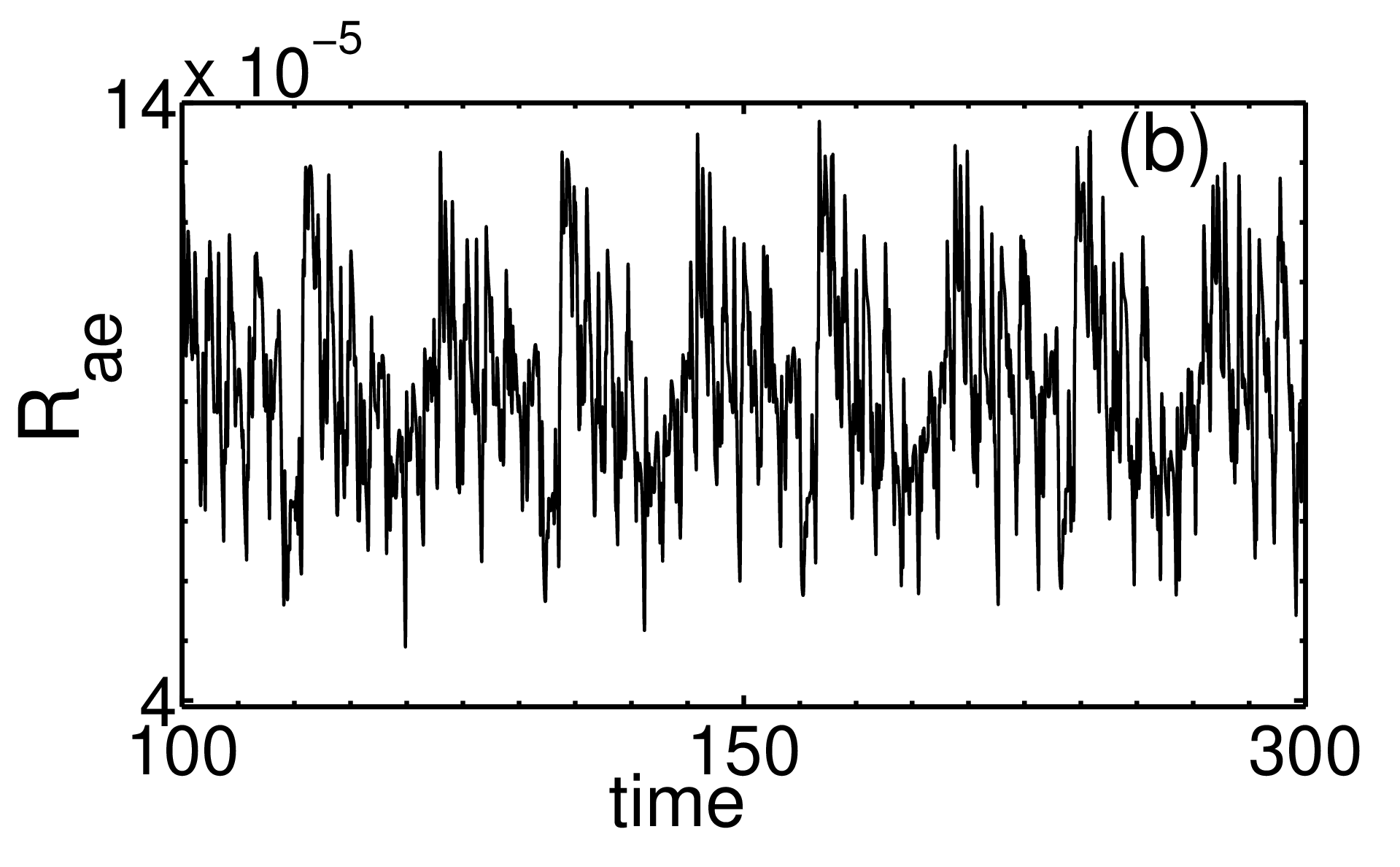}
\includegraphics[height=3.22cm,width=6.0cm]{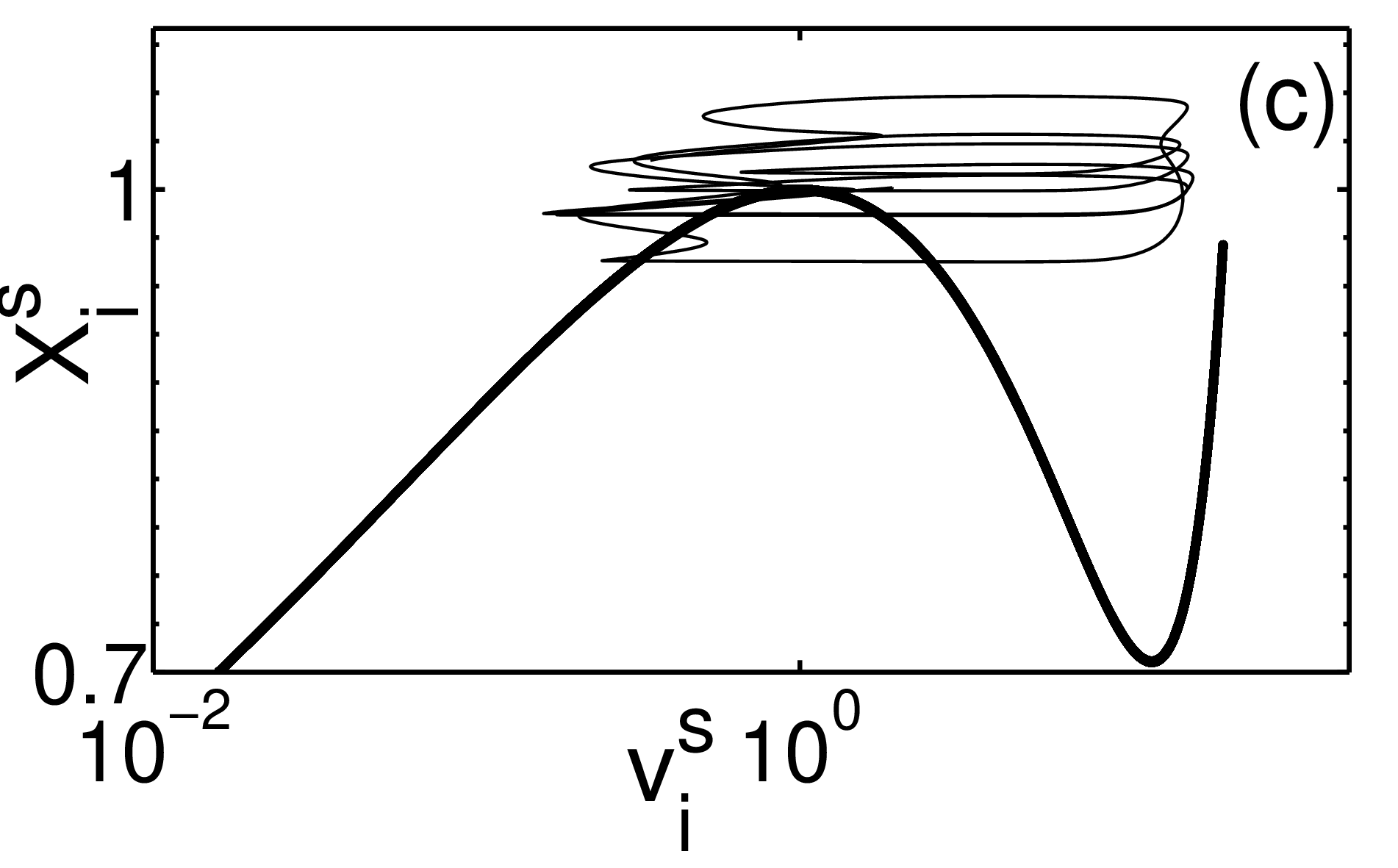}
\includegraphics[height=3.0cm,width=7.5cm]{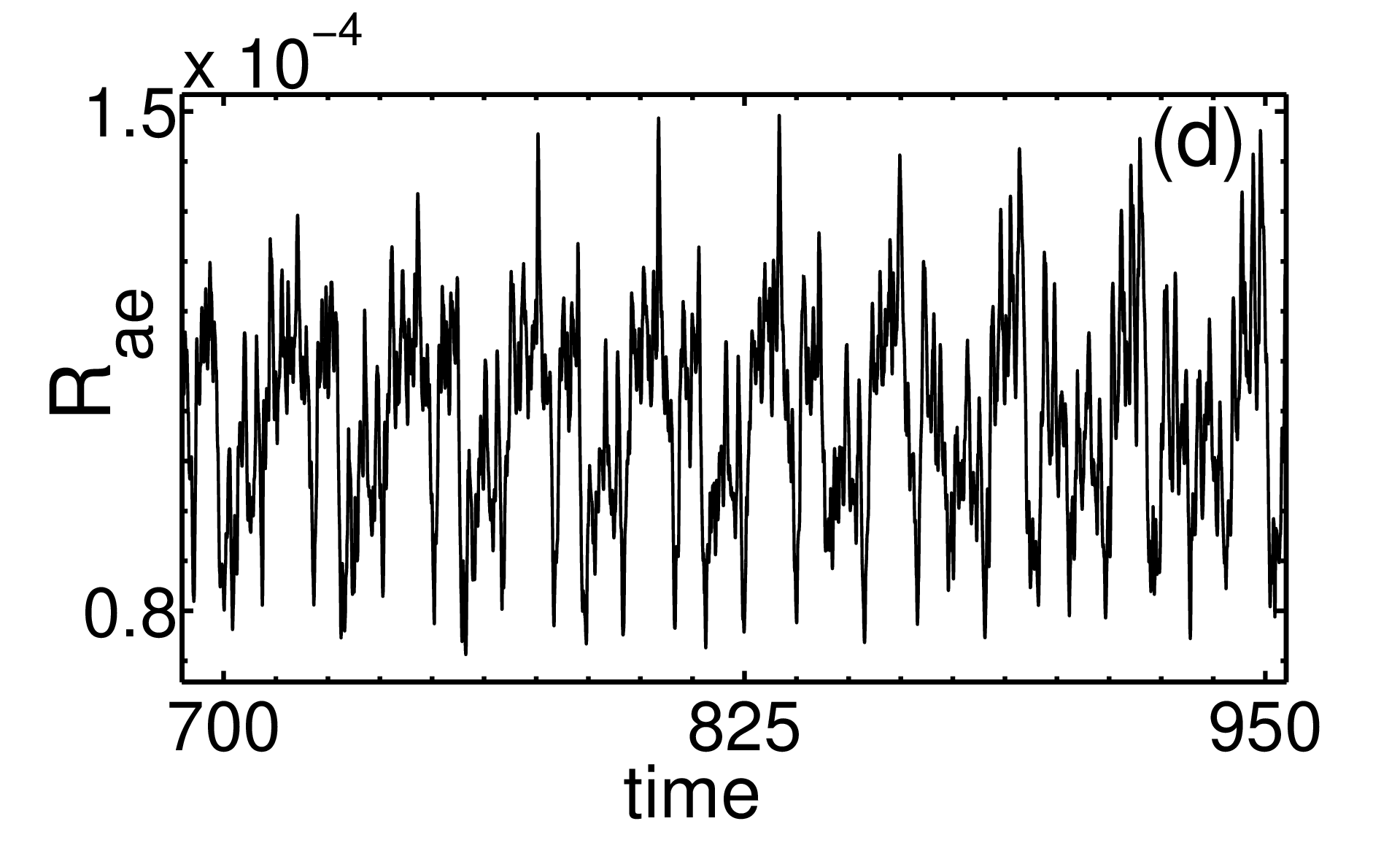}
\includegraphics[height=3.9cm,width=7.5cm]{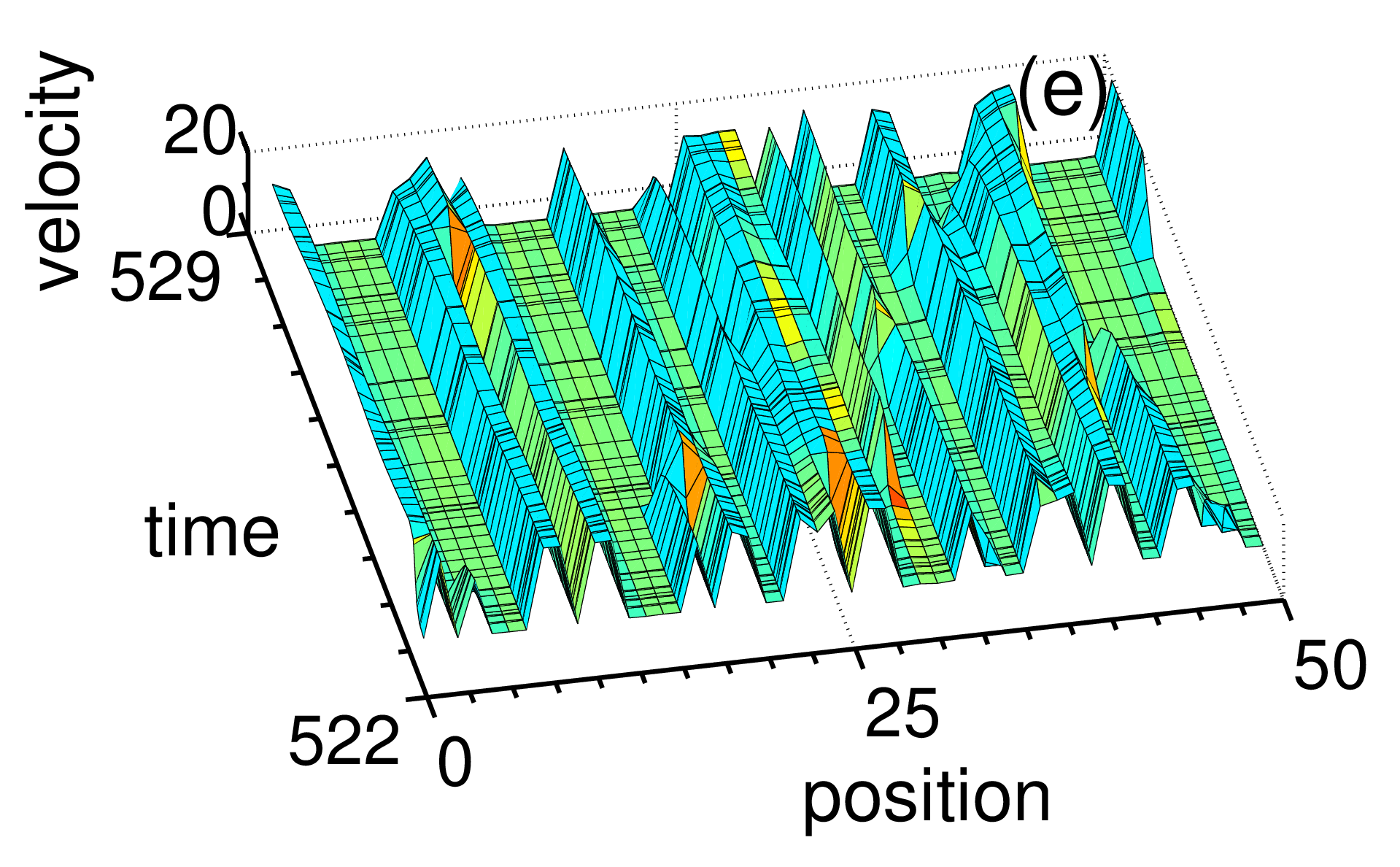}
\includegraphics[height=3.4cm,width=7.5cm]{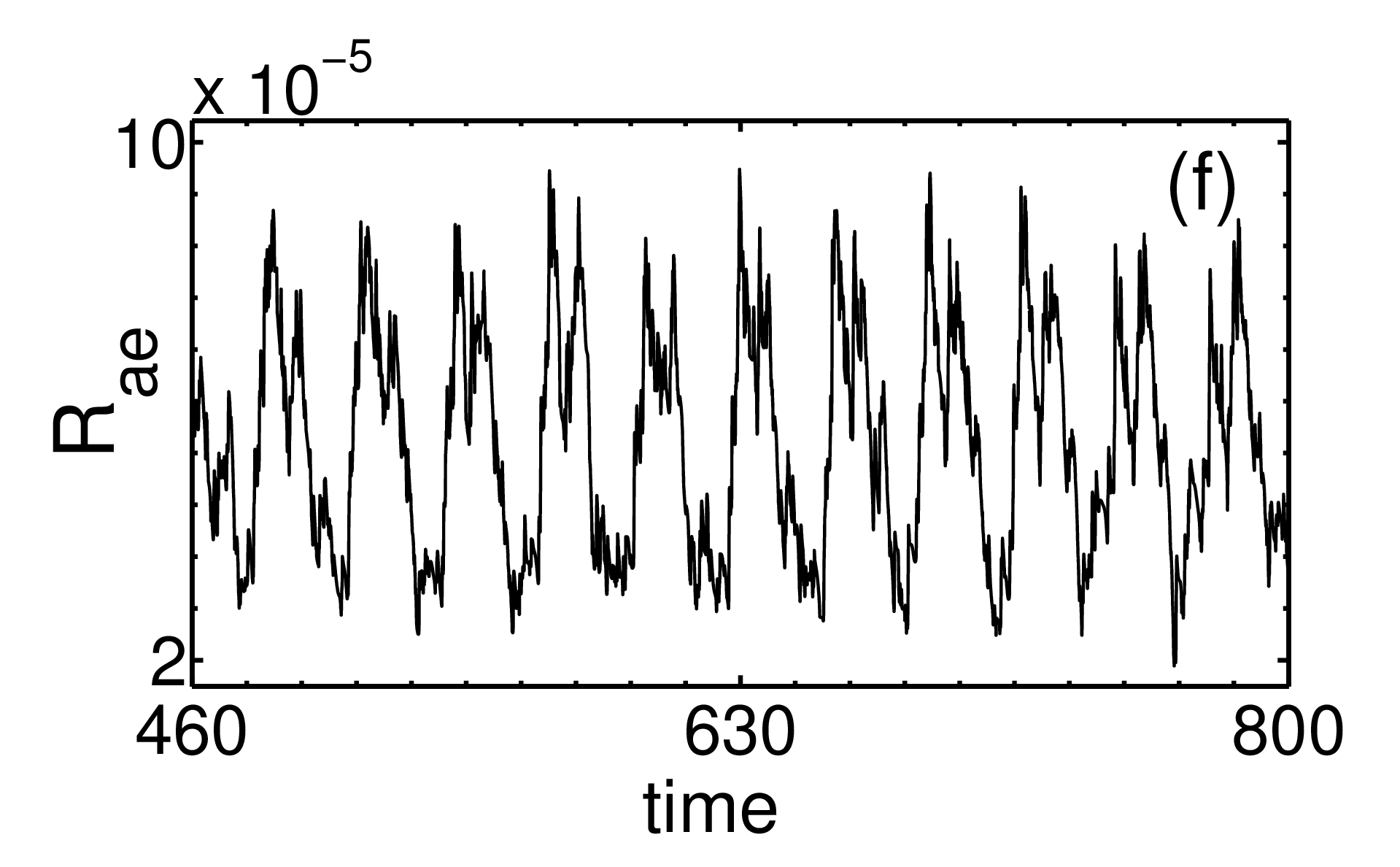}}
\caption{(Color online a,e) (a, b) Snapshot of a stuck-peeled configuration and model acoustic energy respectively for $C_f = 7.88$, $\gamma_u=0.01$ and $V^s=2.48$. (c, d) Phase plot and model acoustic energy plot respectively for $C_f = 7.88$, $V^s=4.48$, and $\gamma_u=0.01$. (e) Snapshot of long lived stuck-peeled configuration for $C_f = 7.88$, $V^s=5.48$, and $\gamma_u=0.01$ and (f) the associated model acoustic energy.}
\label{I5m3V2gu0_1_01}
\end{figure}

As we decrease $\gamma_u$ to 0.01, the observed patterns are similar to those for $\gamma_u=0.1$ but the sequence of the peel front patterns is different.  Starting with a low velocity configuration that is even more rugged compared to that for $\gamma_u=0.1$ as shown in Fig. \ref{I5m3V1gu0_1}(a), the peel process starts with a small stuck segment getting peeled [Fig. \ref{I5m3V1gu0_1}(b)]. There after, several stuck segments peel out leading to a  stuck-peeled pattern as shown in Fig. \ref{I5m3V1gu0_1}(c), eventually, the entire peel front peels-out leaving  a nearly uniform peeled state as shown in Fig. \ref{I5m3V1gu0_1}(d) (with a velocity commensurate with the high velocity branch of $\phi(v^s)$). This is again destabilized with some segments of the peel front getting stuck as in the case of $\gamma_u =0.1$ (similar to Fig. \ref{I5m3V1gu1}(f)). The number of such stuck segments increases with time, eventually the whole peel front goes into a stuck state. The cycle restarts.  The phase plot is similar to $\gamma_u =0.1$   and 1. Indeed, for a given $C_f$, independent of $\gamma_u $ the phase plot changes only when $V^s$ is increased. But $R_ {ae}$ shows broader bursts compared to $\gamma_u=0.1$ as the corresponding stuck-peeled configurations last longer.  Even so, the duration of the SP configurations in a cycle is short, i.e., the duration of the bursts is short compared to the duration between them. 

Now we consider the influence of increasing the pull velocity (keeping $C_f$ fixed at 7.88) which in turn should leave less time for internal relaxational mechanisms to operate. Intuitively one should expect that  some patterns observed for low $V^s$ may not be seen for higher values of $V^s$. $\gamma_u =1$ case is uninteresting for  the reasons stated above. But, reducing $\gamma_u$ to 0.1 does provide some degree of freedom for the local dynamics to operate at each point. Even so, for $V^s =2.48$, the peel front switches between a SP configuration with most segments momentarily in the stuck state (similar to Fig. \ref{I5m3V1gu0_1}(b)) and  a configuration that  has several stuck-peeled segments (similar to Fig. \ref{I5m3V1gu0_1}(c)). The corresponding $X^s-v^s$ phase plot shows that the orbit jumps moves slightly beyond the upper value of $\phi(v^s)$ and jumps back from the right branch even before reaching the minimum of $\phi(v^s)$ ( not shown). As we decrease $\gamma_u$ to 0.01, the rugged configuration seen for $V^s =1.48$ is no longer seen and only SP configurations are observed as shown in  Fig. \ref{I5m3V2gu0_1_01}(a). The SP configurations are dynamic in the sense, segments that are in stuck state at one time become unstuck at a later time and vice versa. For this case ($C_f=7.88,V^s=2.48,\gamma_u=0.01$), these rapid changes occur over a short time scale. Consequently, the model acoustic energy is quite noisy as shown in Fig. \ref{I5m3V2gu0_1_01}(b) but has a noticeable periodic component. The points of minima correspond to configurations that have fewer peeled segments compared to those near the peak of $R_{ae}$. The phase plot in the $X^s-v^s$ plane is limited to the upper part of  $\phi(v^s)$. Even as the phase plots for any two spatial points look similar, there is a phase difference. For instance, at any given time, the phase point of stuck segment will be on the left branch while that for peeled point will be on the right branch.   

As we increase $V^s$ to $4.48$ (keeping $C_f$ at $7.88$), there is even lesser time for peel front inhomogeneities to relax and thus, we observe a smooth peeling for $\gamma_u=1$ and as also for $0.1$. As we decrease $\gamma_u$ to $0.01$, we see only SP patterns (not shown but similar to Fig. \ref{I5m3V2gu0_1_01}(a)). The corresponding $X^s-v^s$ phase plot  for an arbitrary point on the peel front shown in Fig. \ref{I5m3V2gu0_1_01}(c) is confined to the top of $\phi(v^s)$. The corresponding $R_{ae}$ is noisy and irregular as shown in Fig. \ref{I5m3V2gu0_1_01}(d). However, when we increase $V^s$ to $5.48$, initially, one does observe the patterns switching between rugged and SP configurations. If we wait long enough, we observe only SP configurations that are different from those for lower $V^s$. In this case, the stuck and peeled segments are long lived. A top view of the SP pattern is shown in Fig. \ref{I5m3V2gu0_1_01}(e). The phase space orbit in the $X^s-v^s$ plot is pushed beyond the upper limit of $\phi(v^s)$. The energy dissipated is quite regular (but aperiodic) unlike that for lower pull velocities as shown in Fig. \ref {I5m3V2gu0_1_01}(f). This regularity is clearly due to the long lived nature of these SP configurations. The long lived nature of the SP configurations for high pull velocity is a general feature, i.e., the duration over which the stuck segments remain stuck (peeled segments remain peeled) increases as we increase the pull velocity. The dynamics is no longer interesting beyond $V^s=6.48$ as only smooth peeling is seen.
\begin{figure}[!b]
\vbox{
\includegraphics[height=3.6cm,width=7.5cm]{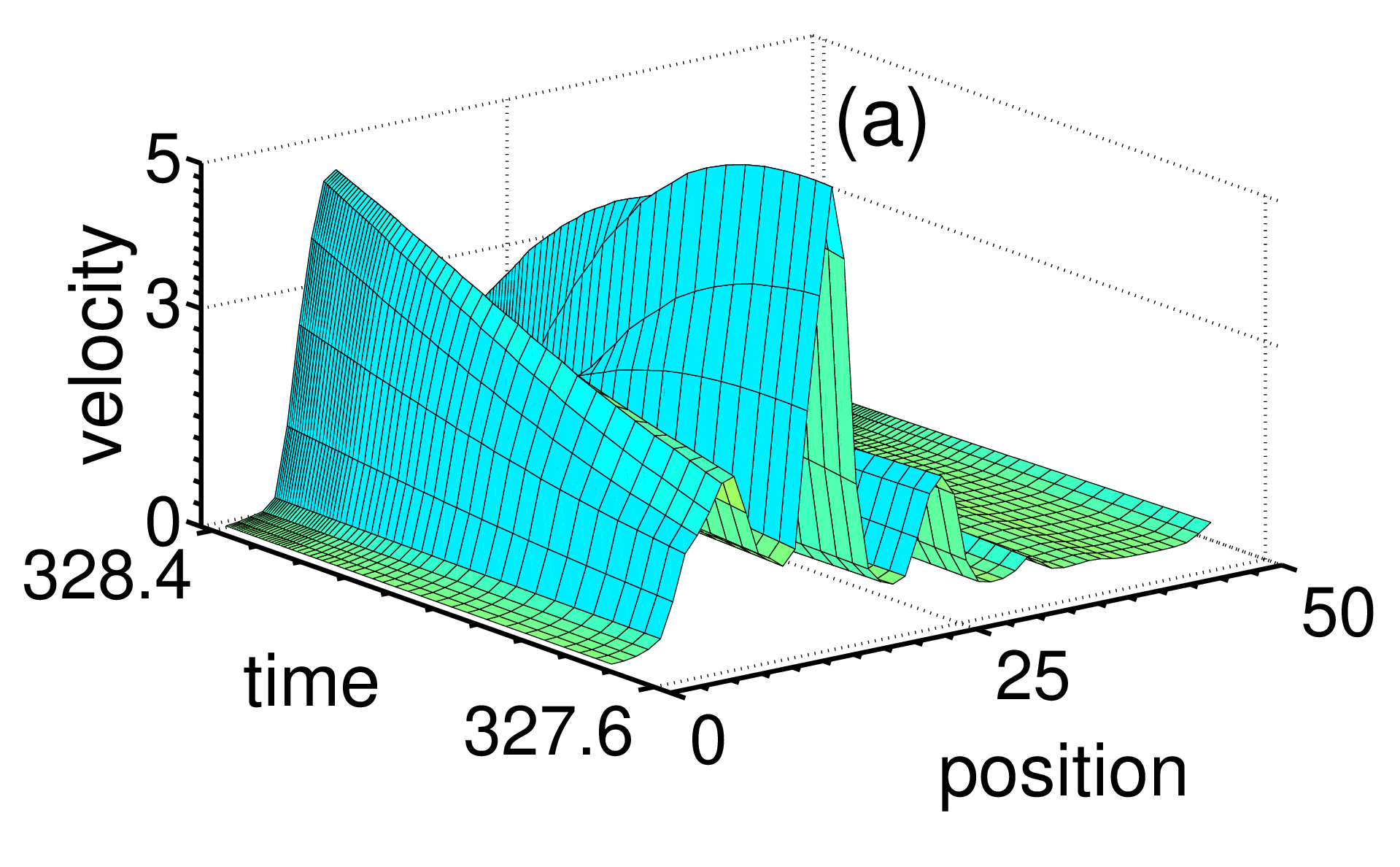}
\includegraphics[height=3.6cm,width=7.5cm]{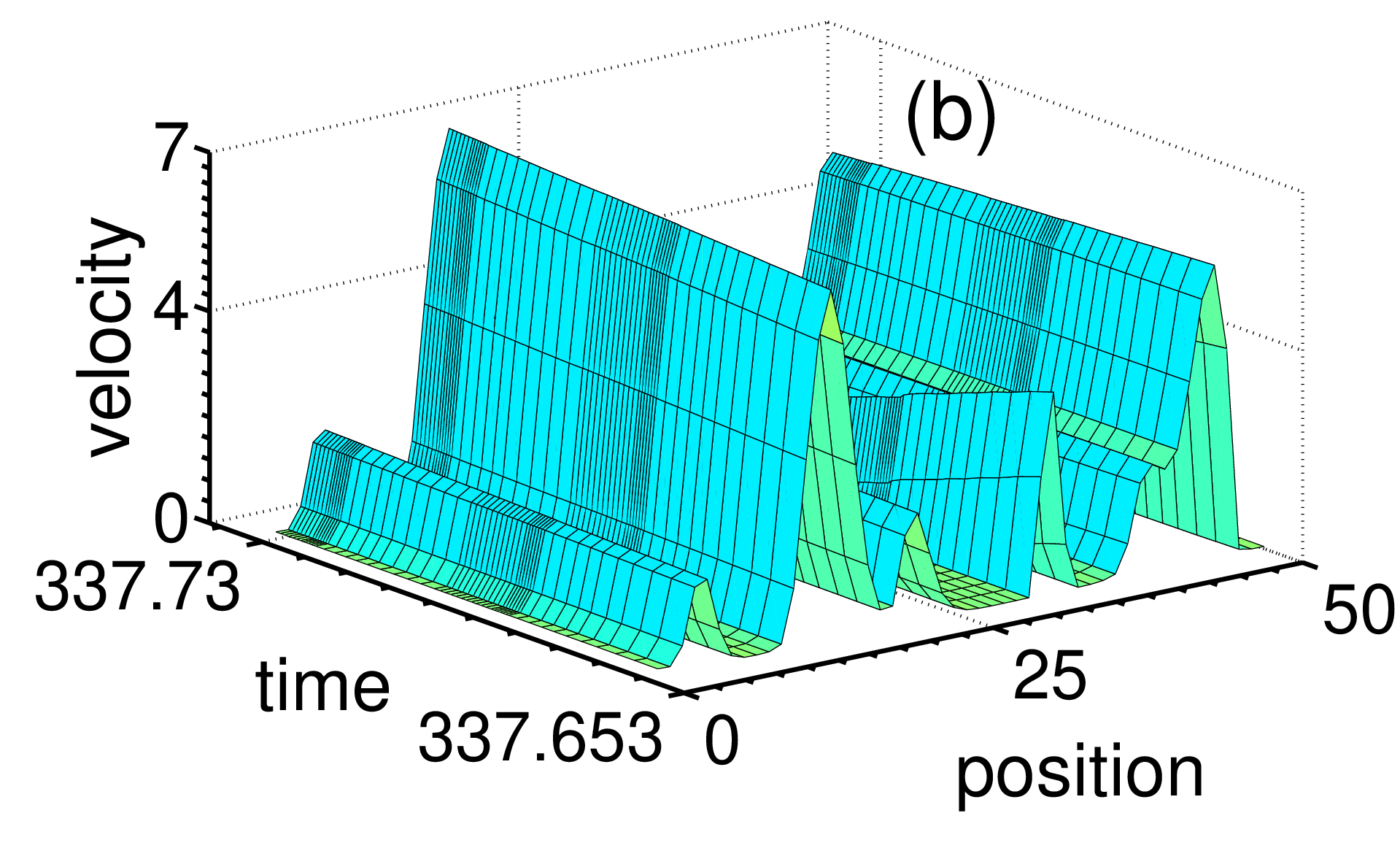}
\includegraphics[height=3.6cm, width=6.0cm]{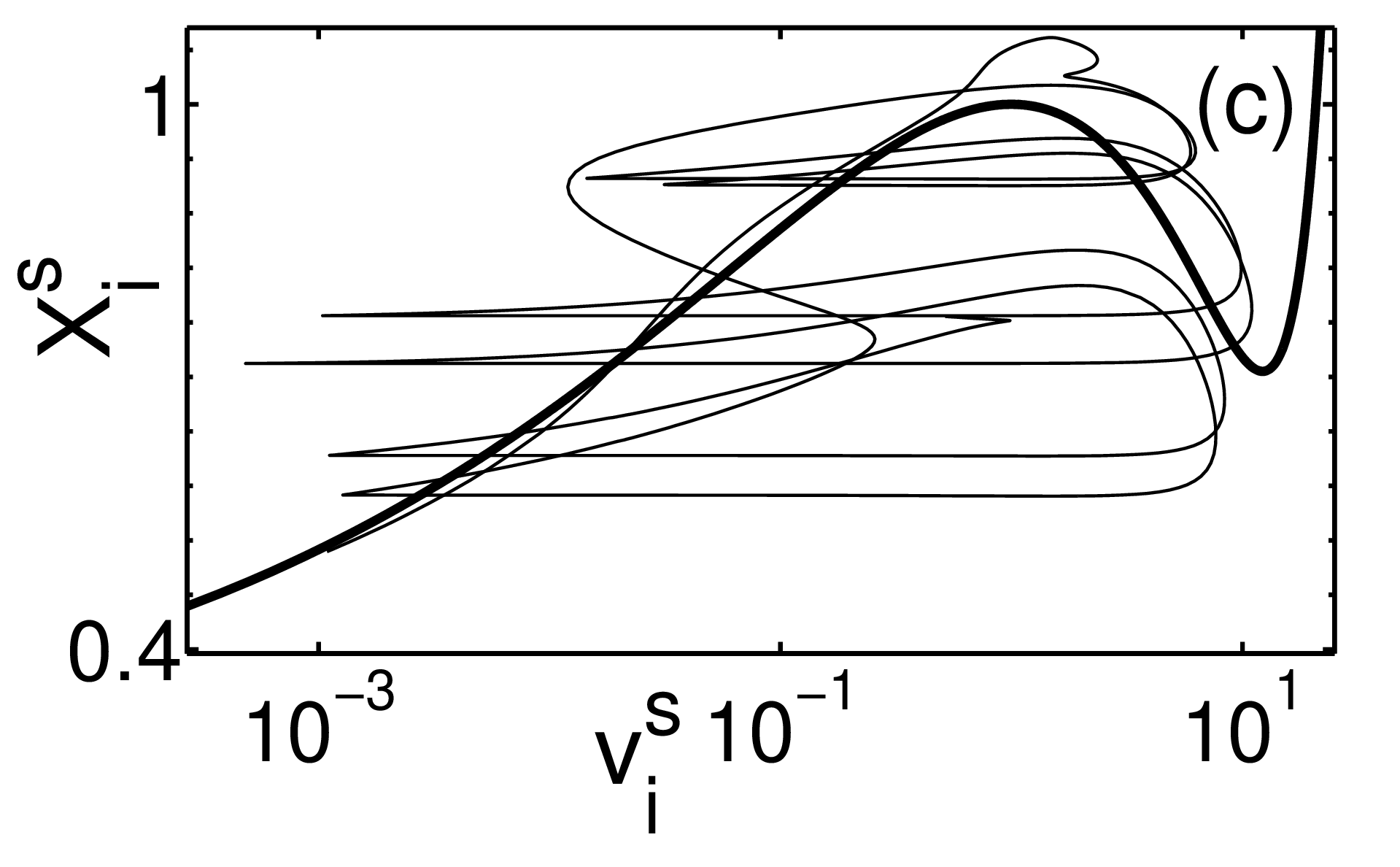}
\includegraphics[height=3.6cm,width=7.5cm]{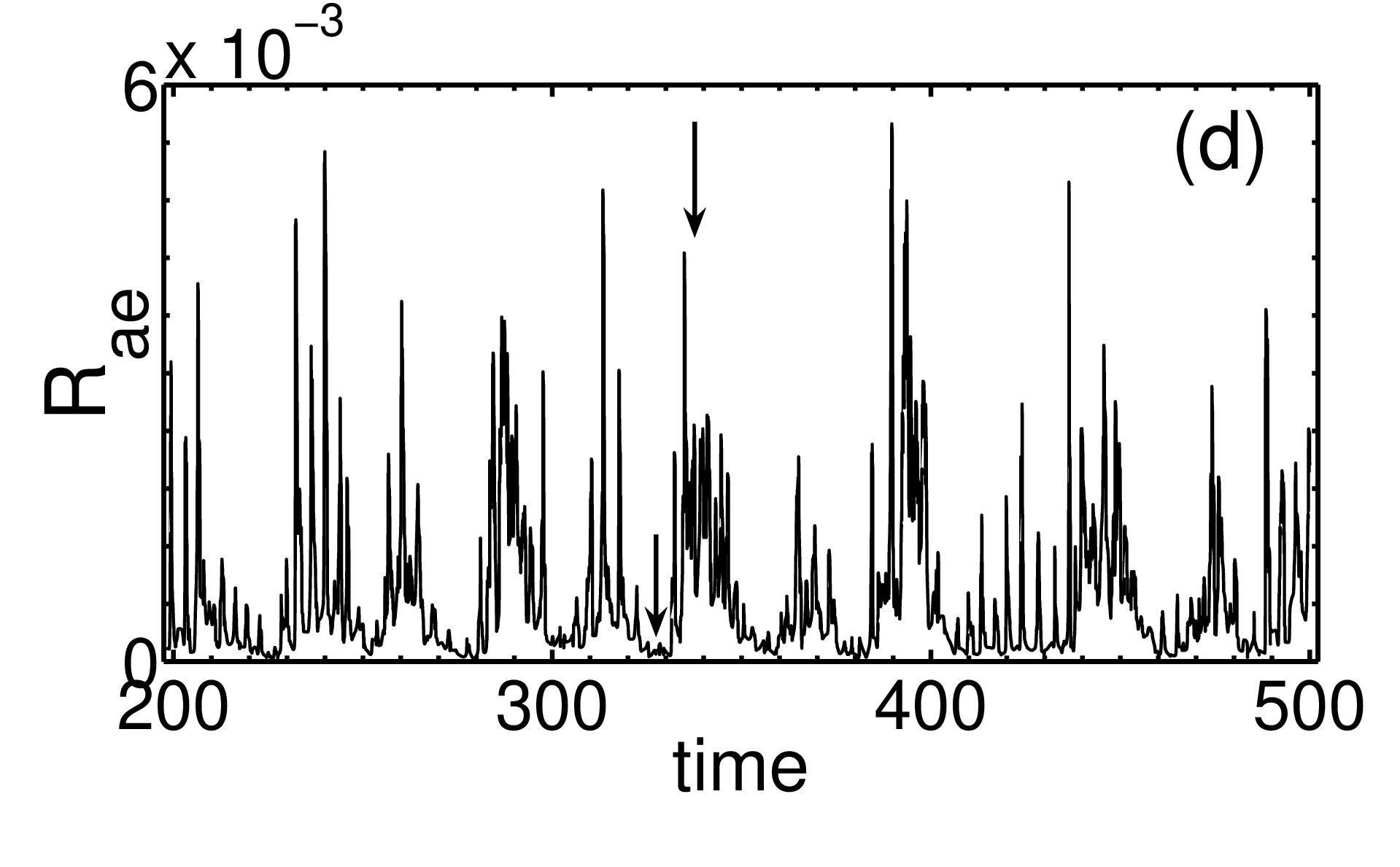}
}
\caption{(Color online a, b) (a, b) Snapshots of stuck-peeled configurations for $C_f = 0.788, V^s = 1.48$, and $\gamma_u =1.0$.  Note that (a) has fewer peeled segments compared to (b). (c) Phase plot for an arbitrary point on the peel front, and (d) model acoustic energy plot.  }
\label{I2m1V1gup1}
\end{figure}
\begin{figure}[!t]
\vbox{
\includegraphics[height=3.6cm,width=7.5cm]{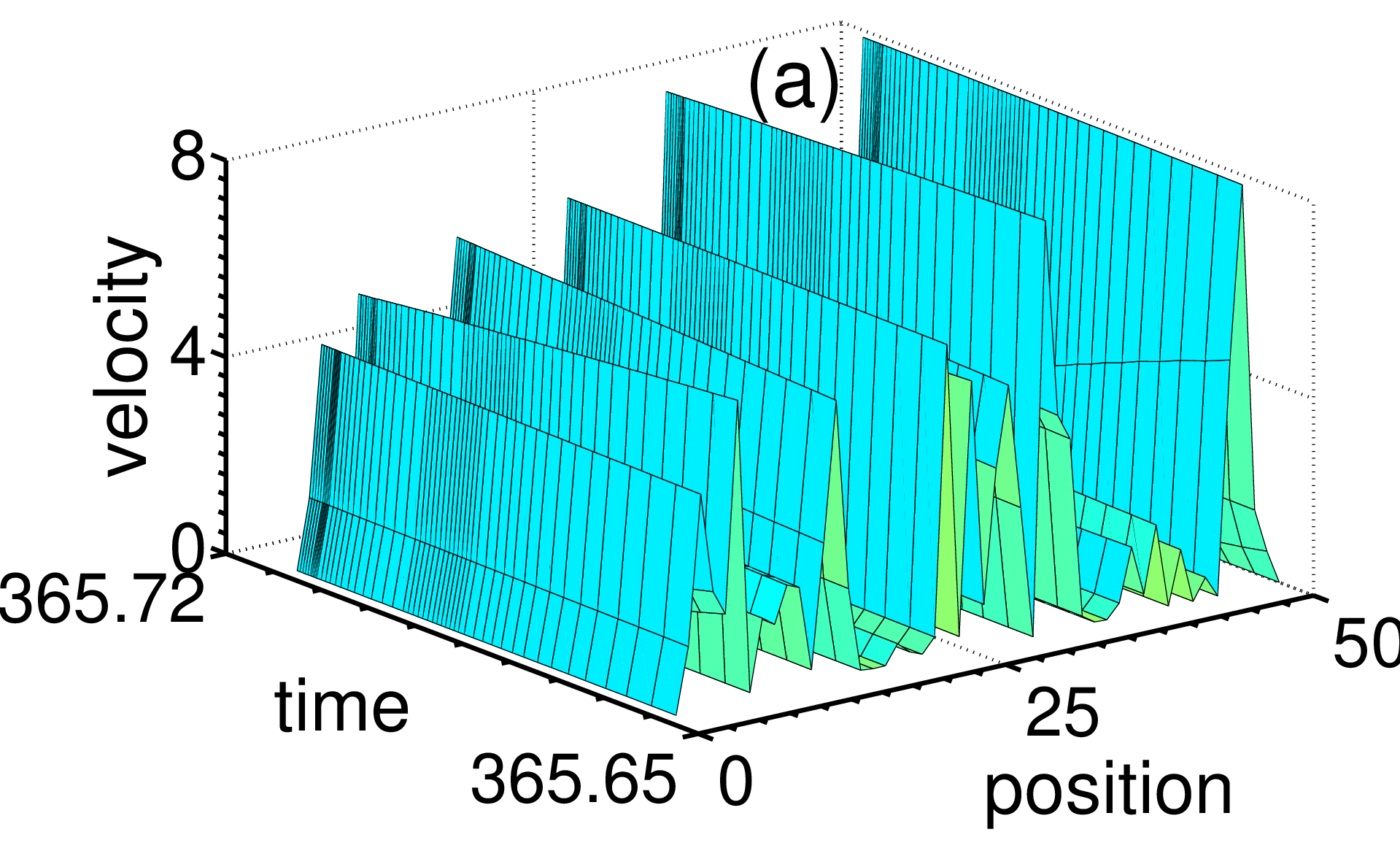}
\includegraphics[height=3.6cm,width=7.5cm]{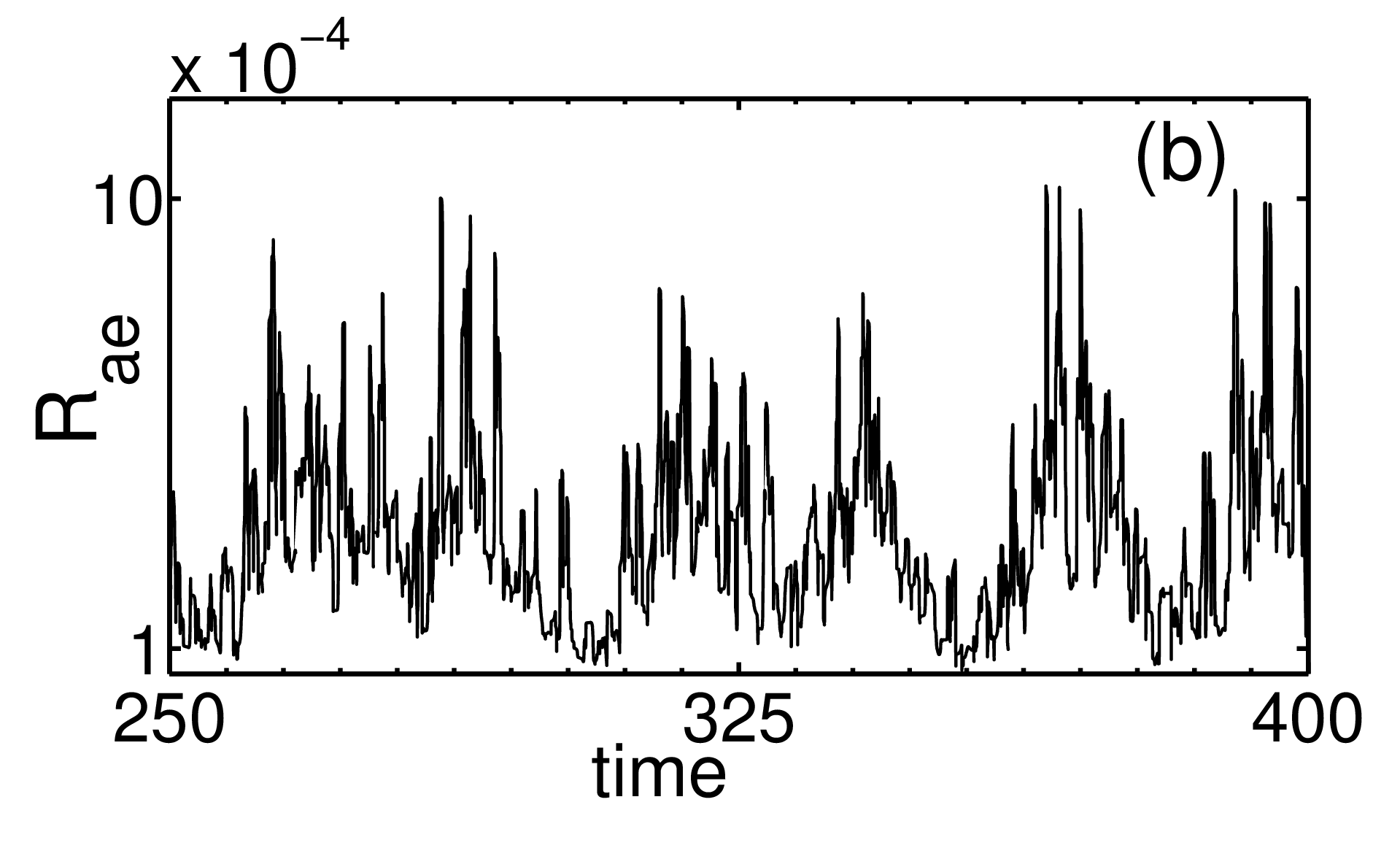}
\includegraphics[height=3.6cm, width=7.5cm]{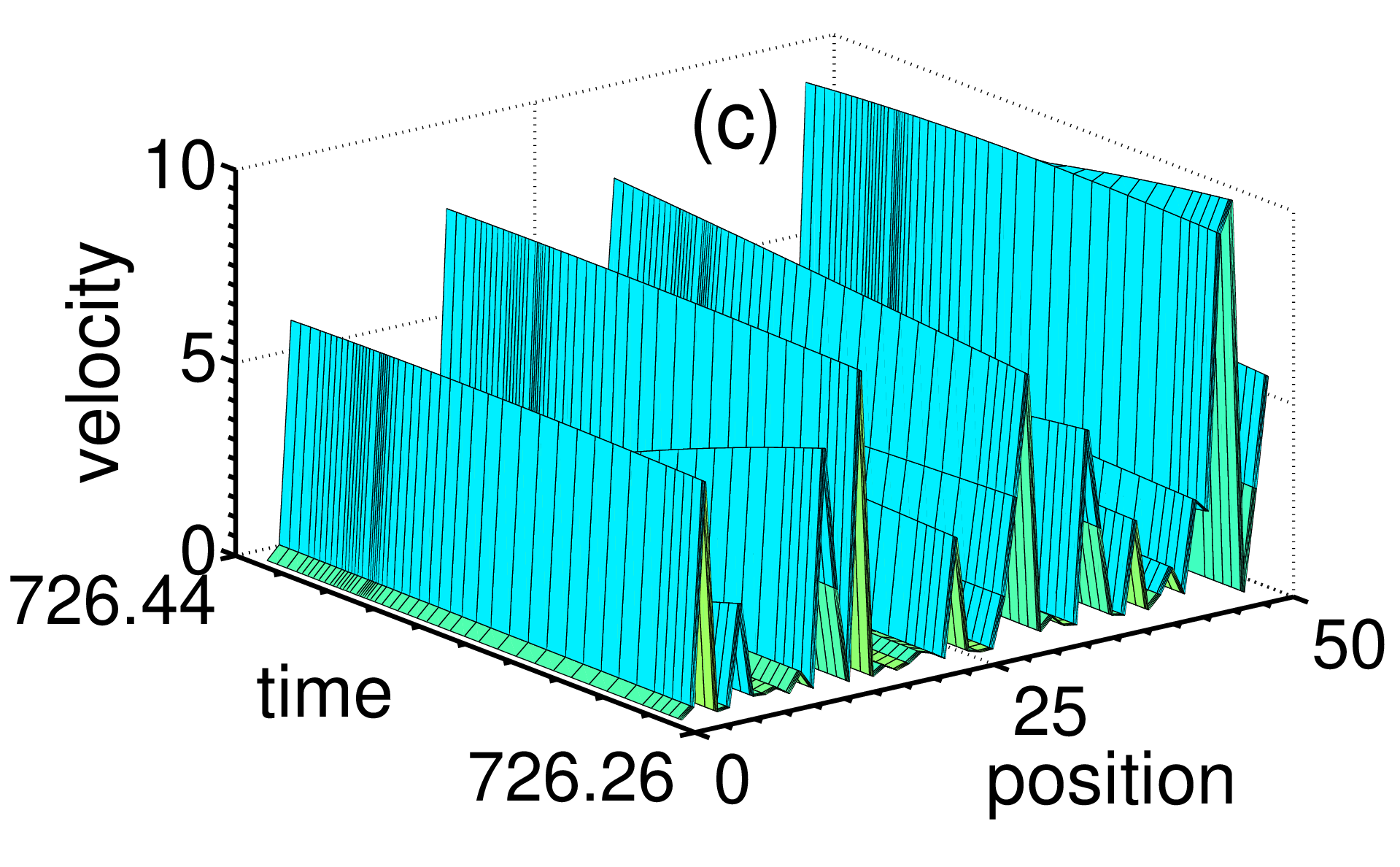}
\includegraphics[height=3.6cm,width=7.5cm]{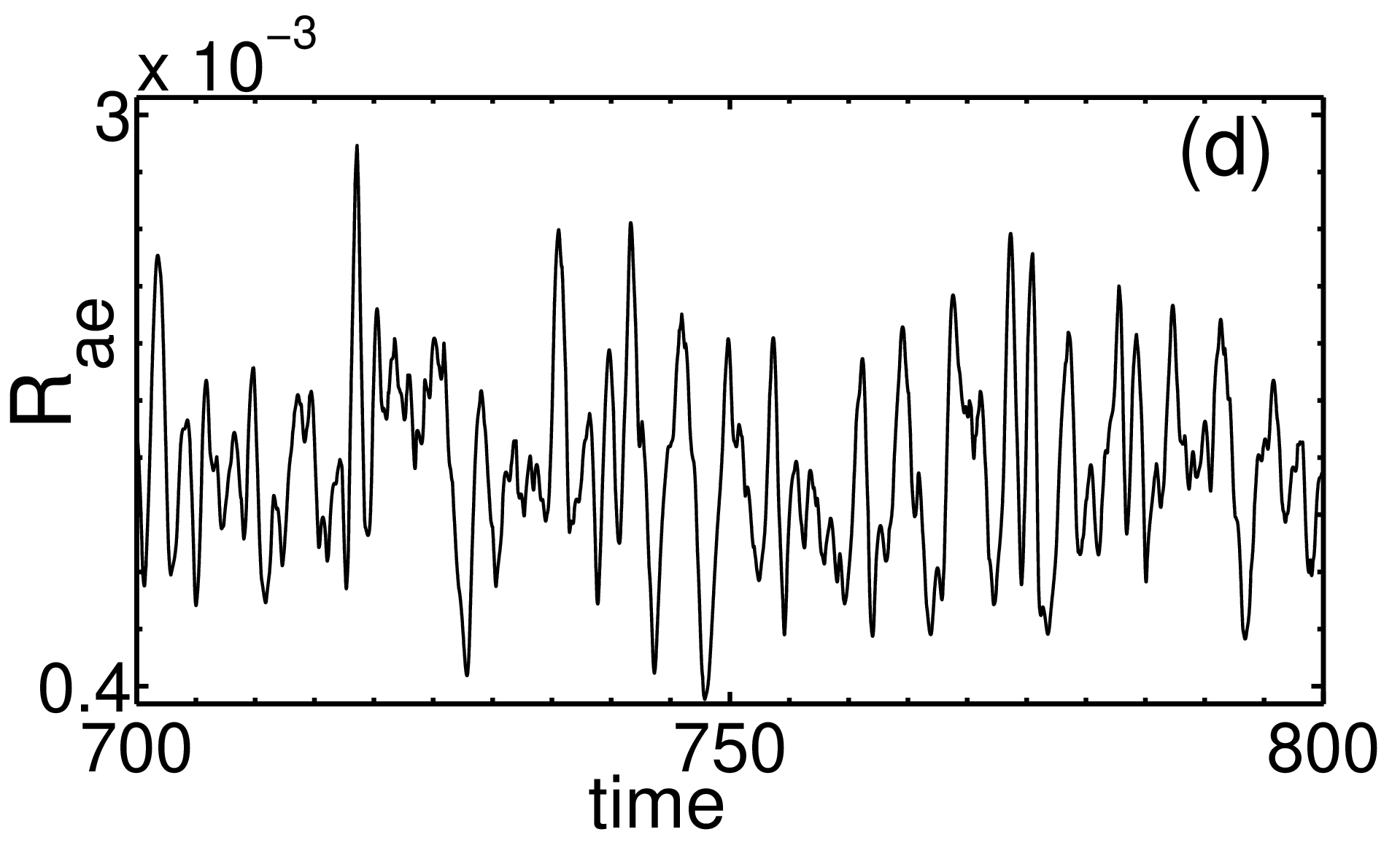}
}
\caption{(Color online a, c) (a, b) Snapshot of a stuck-peeled configuration for  $C_f = 0.788, V^s = 1.48$, and $\gamma_u =0.1$ and model acoustic energy plot respectively. (c, d) Plot of `an edge of peeling' configuration for $C_f = 0.788, V^s = 1.48$, and $\gamma_u =0.01$ and the corresponding model acoustic energy.}
\label{I2m1V1gup0_1}
\end{figure}
\begin{figure}[!b]
\vbox{
\includegraphics[height=3.6cm, width=7.5cm]{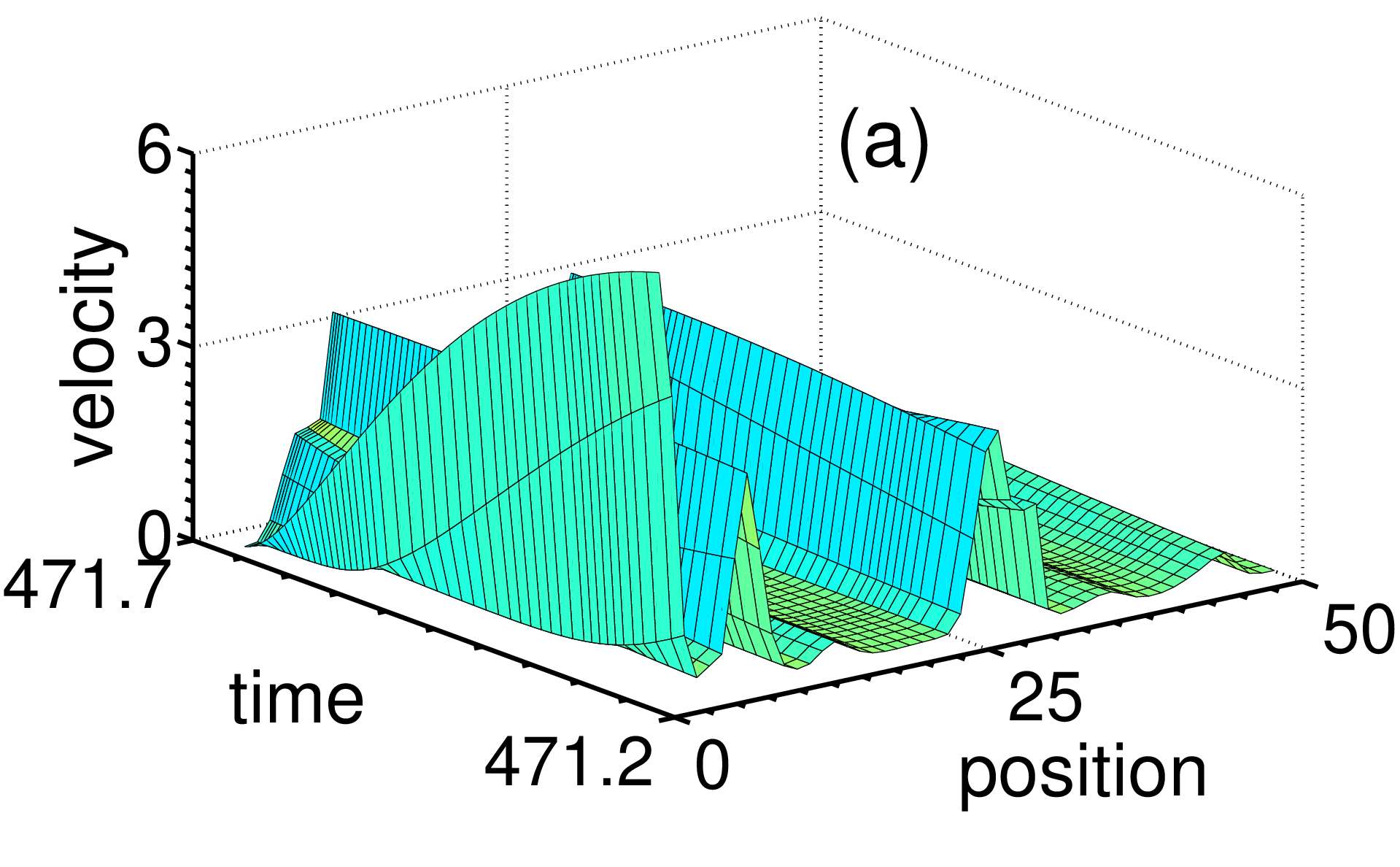}
\includegraphics[height=3.6cm,width=7.5cm]{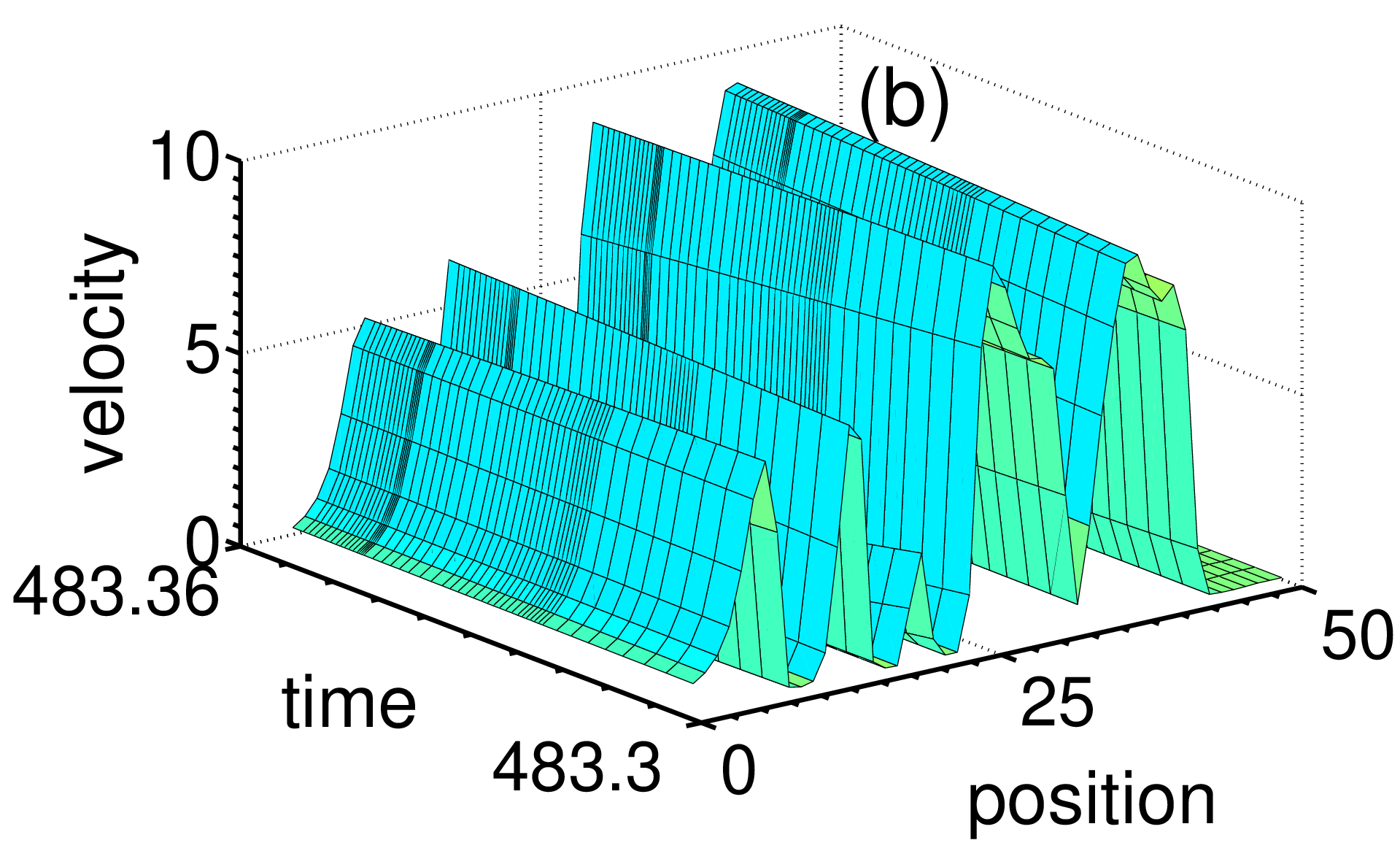}
\includegraphics[height=3.6cm,width=7.5cm]{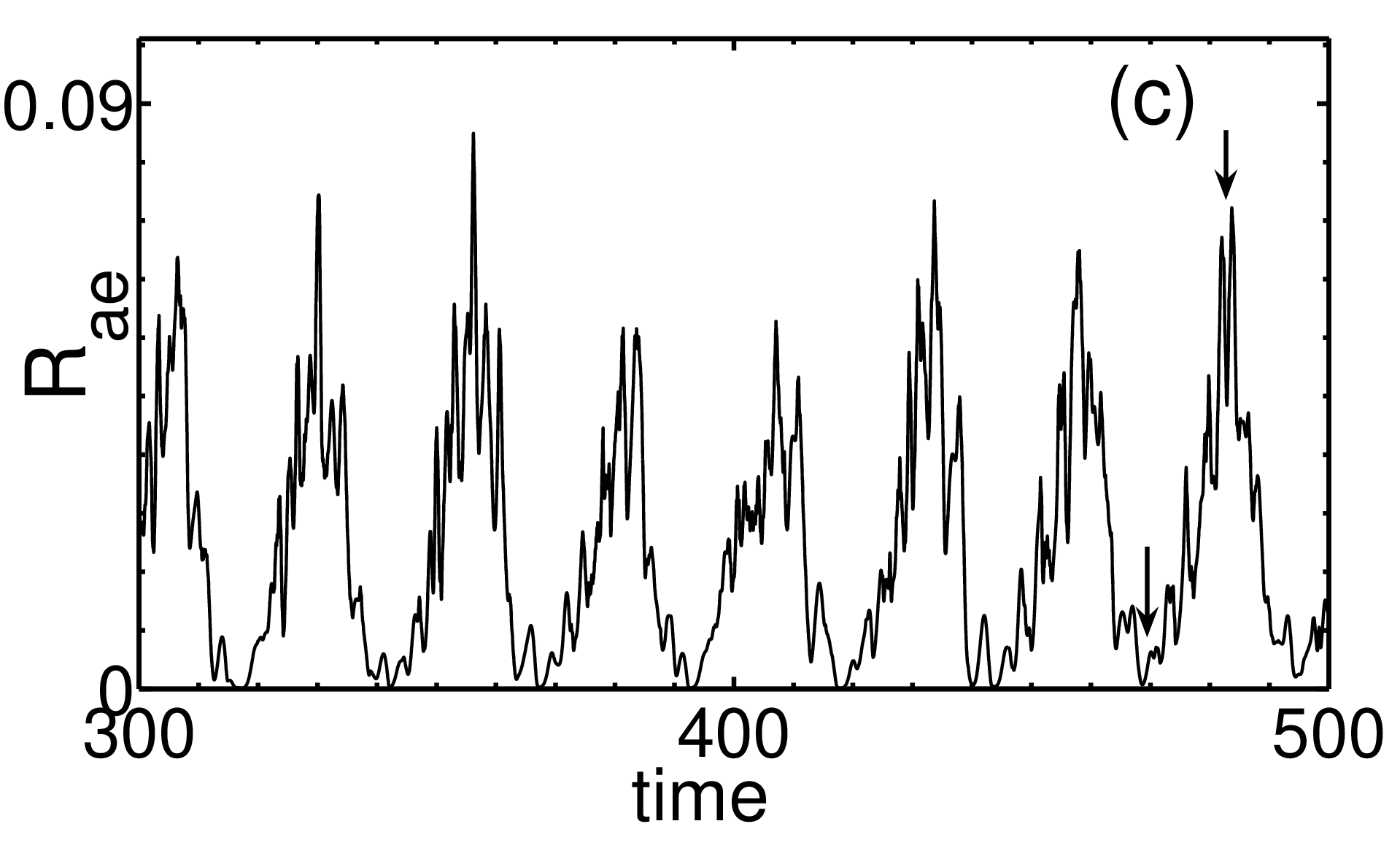}
\includegraphics[height=3.6cm,width=6.0cm]{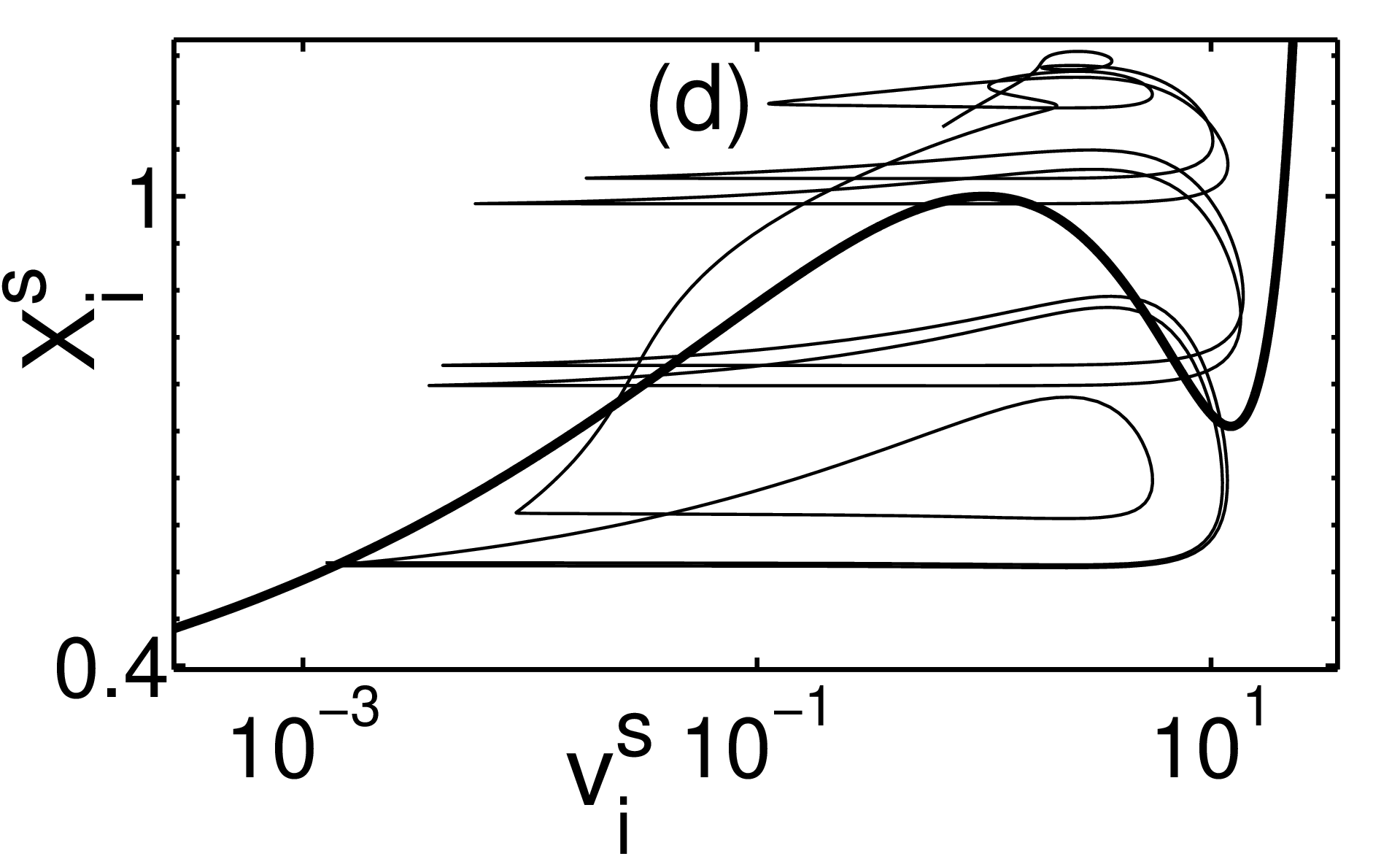}
}
\caption{(Color online a, b) 
(a, b) Snapshots during peel process for $C_f = 0.788, V^s = 2.48$, and $\gamma_u =1.0$. (c) Model acoustic energy plot. (d) Phase plot for an arbitrary spatial point on the  peel front. Bold line represents $\phi(v^s)$. }
\label{I2m1V2gup1}
\end{figure}

\subsubsection{ Case (ii), $C_f = 0.788$ - high (and low) tape mass, high (and intermediate) roller inertia }

For this  value of $C_f$, the allowed set of values of $(m,I)$ are $(0.1,10^{-2}), (0.01, 10^{-3})$ and $(0.001,10^{-4})$.  The dynamics is more interesting for this case as  there is a scope for competition among the three time scales.

We first study the dynamics keeping $V^s =1.48$ and varying the dissipation parameter. For $\gamma_u=$ 1.0, the uniform nature of the  peel front seen for $C_f= 7.88$ disappears and even for short times, stuck-peeled configurations are seen.  The peel front patterns stabilize to stuck-peeled configurations as shown in Figs. \ref{I2m1V1gup1}(a, b). As can be seen these SP patterns have only a few stuck or peeled segments with moderate velocity jumps and smooth variation along the peel front unlike the  SP configurations discussed earlier. (Note that the SP configuration in Fig. \ref{I2m1V1gup1}(b) has more stuck segments compared to Fig. \ref{I2m1V1gup1}(a).)  The moderate velocity jumps can be understood by noting that the phase  space orbit never visits the high velocity branch of $\phi(v^s)$ as can be seen from Fig. \ref{I2m1V1gup1}(c). It is interesting to note that the trajectory stays close to  the unstable branch of $\phi(v^s)$ even after attempting to jump from the low velocity branch. Such orbits are reminiscent of canard type solutions \cite{Canard}. The trajectory is irregular and is suggestive of spatiotemporal chaotic nature of the peel front. 
The energy dissipated $R_{ae}$  shown in Fig. \ref{I2m1V1gup1}(d) is continuous and irregular due to the dynamic SP pattern as should be expected, but there is a noticeable periodic component.  The rough periodicity of $R_{ae}$ can be traced to fact that the peel front configurations switch between patterns with more stuck segments and less stuck segments. (From the number shown on $x$-axis, Fig. \ref{I2m1V1gup1}(a) can be identified with  minimum and Fig. \ref{I2m1V1gup1}(b) with the peak of $R_{ae}$ in Fig. \ref{I2m1V1gup1}(d). See the marked arrows as well.)

As we decrease $\gamma_u$ to 0.1, the SP configurations observed have more  stuck and peeled segments compared to $\gamma_u=1$ (compare Fig. \ref{I2m1V1gup0_1}(a) with Fig. \ref{I2m1V1gup1} (a)). However, the magnitude of the velocity jumps remains moderate as in the previous case. This is again due to the fact that the orbit never visits the high velocity branch of $\phi(v^s)$. (Recall that given a value of $C_f$, the phase plot remains the same for different $\gamma_u$ values as long as $V^s$ is fixed). Indeed, for this value of $C_f=0.788$, the orbit never jumping  to the high velocity branch is a consequence of finite inertia of the tape mass compared to that of the roller inertia as discussed earlier. For this case, the model acoustic energy $R_{ae}$ is also irregular and continuous as shown in Fig. \ref{I2m1V1gup0_1}(b) with a noticeable periodic component. Now, if we decrease $\gamma_u$ further to 0.01, the peel front pattern displays increased  number of stuck  and peeled segments with each stuck segment having only a few contiguous  stuck points as can be seen from  Fig. \ref{I2m1V1gup0_1}(c).  Note also that  there is a large dispersion in the magnitudes of the velocity jumps of the peeled segments  even as the largest one is significantly smaller than the value of CD branch of $\phi(v^s)$. As can be seen from the  Fig. \ref{I2m1V1gup0_1}(c), even though the pattern is dynamic, the segments that are stuck are barely so. Thus, the configuration shown in  Fig. \ref{I2m1V1gup0_1}(c) gives the  feeling of a critically poised state. The corresponding $X^s-v^s$ phase plot (similar to that shown in  Fig. \ref{I2m1V1gup1}(c)) is irregular and possibly suggestive of spatiotemporal  chaotic nature of the peel front.  The acoustic energy $R_{ae}$ is very irregular without any trace of periodicity as shown in Fig. \ref{I2m1V1gup0_1}(d).

\begin{figure}[!t]
\vbox{
\includegraphics[height=3.6cm, width=7.5cm]{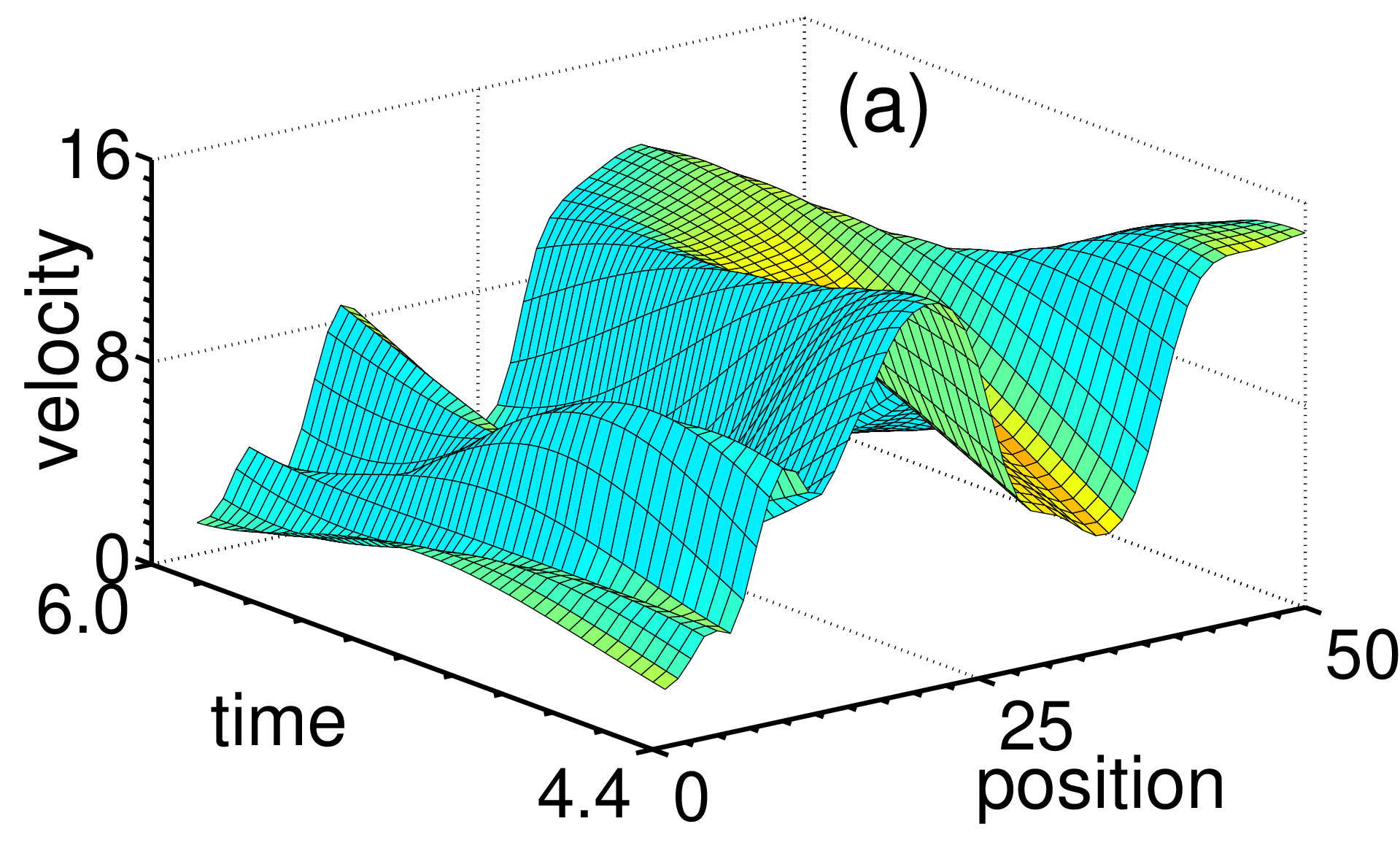}
\includegraphics[height=3.6cm,width=7.5cm]{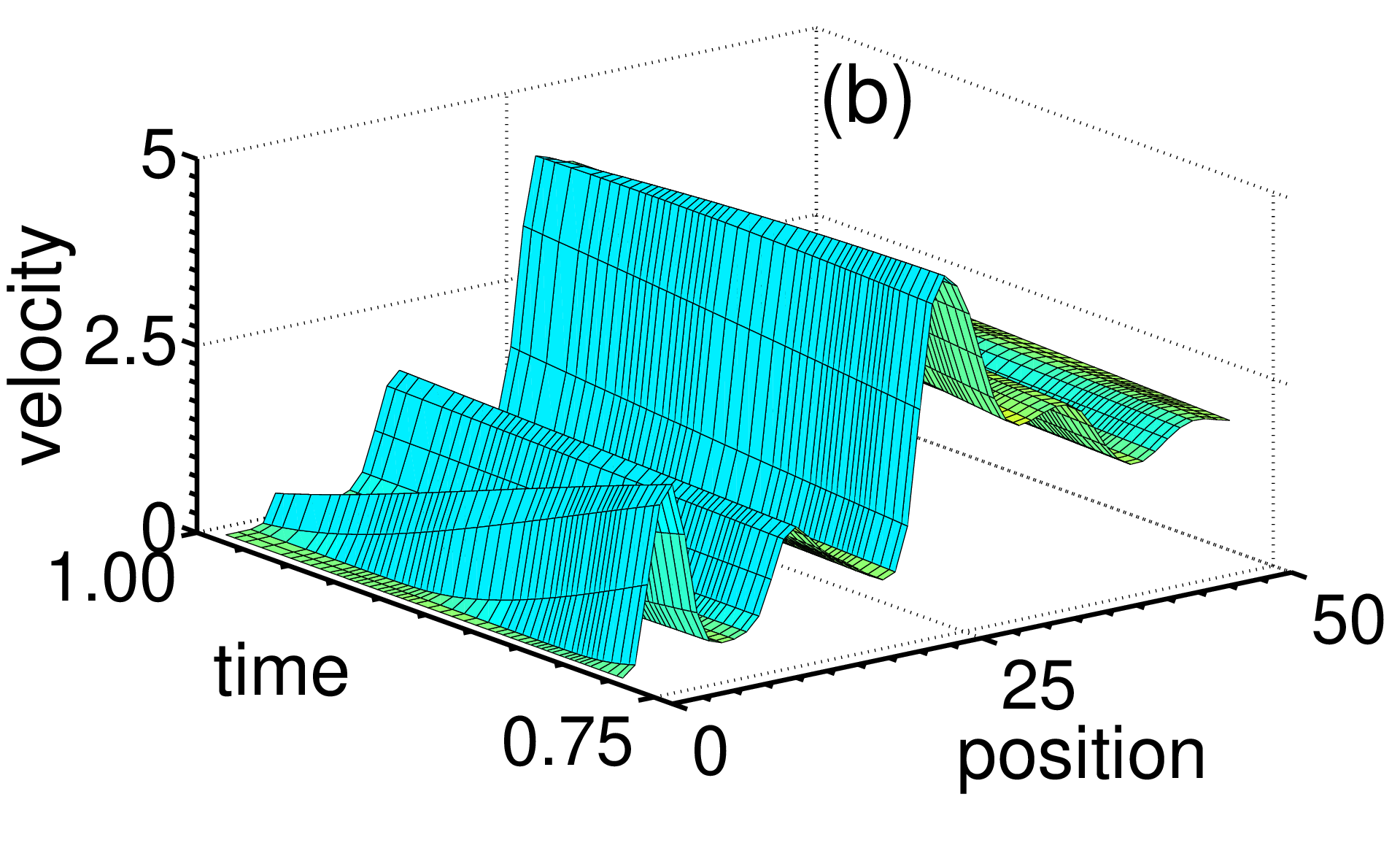}
\includegraphics[height=3.6cm, width=6.0cm]{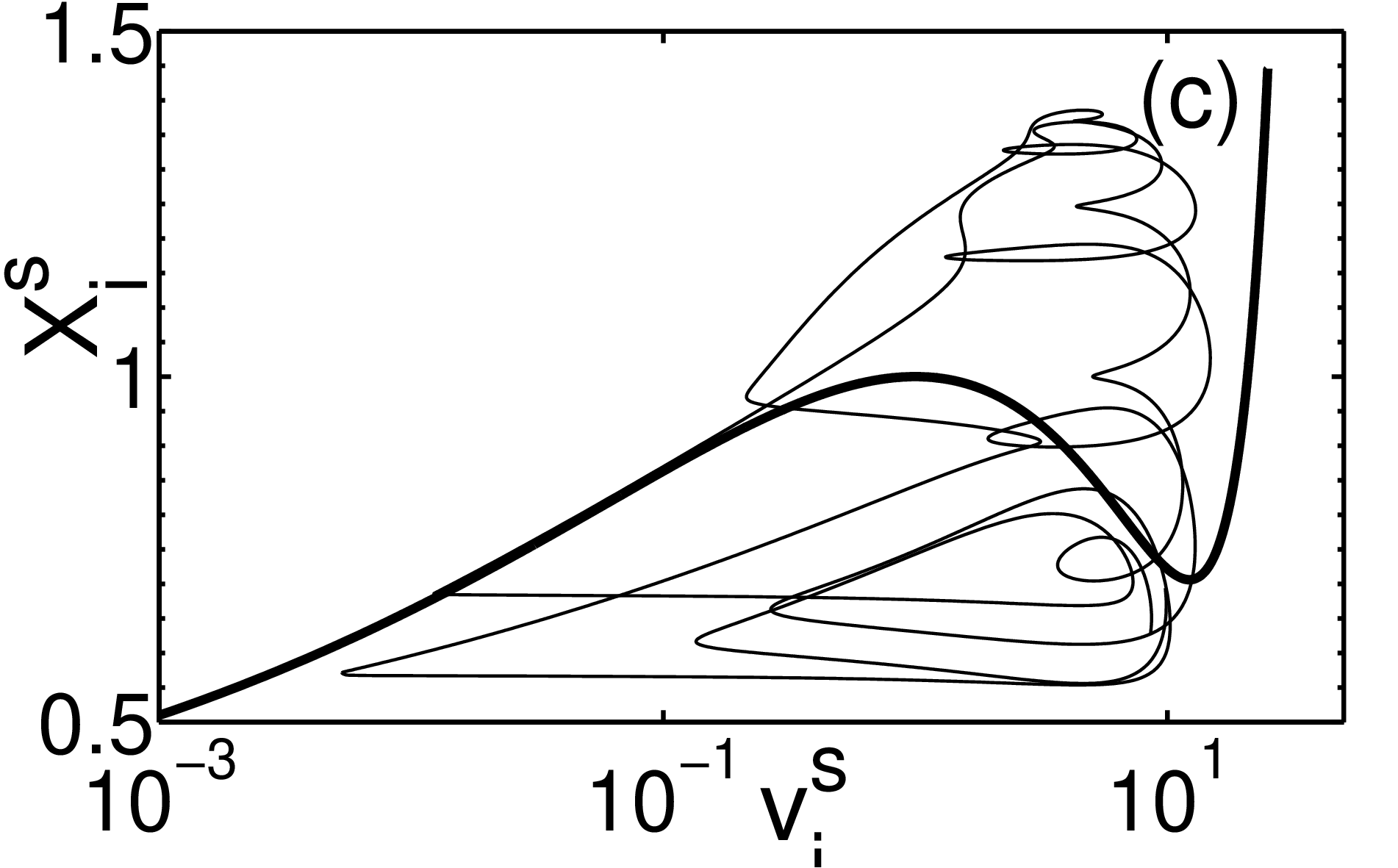}
\includegraphics[height=3.6cm,width=7.5cm]{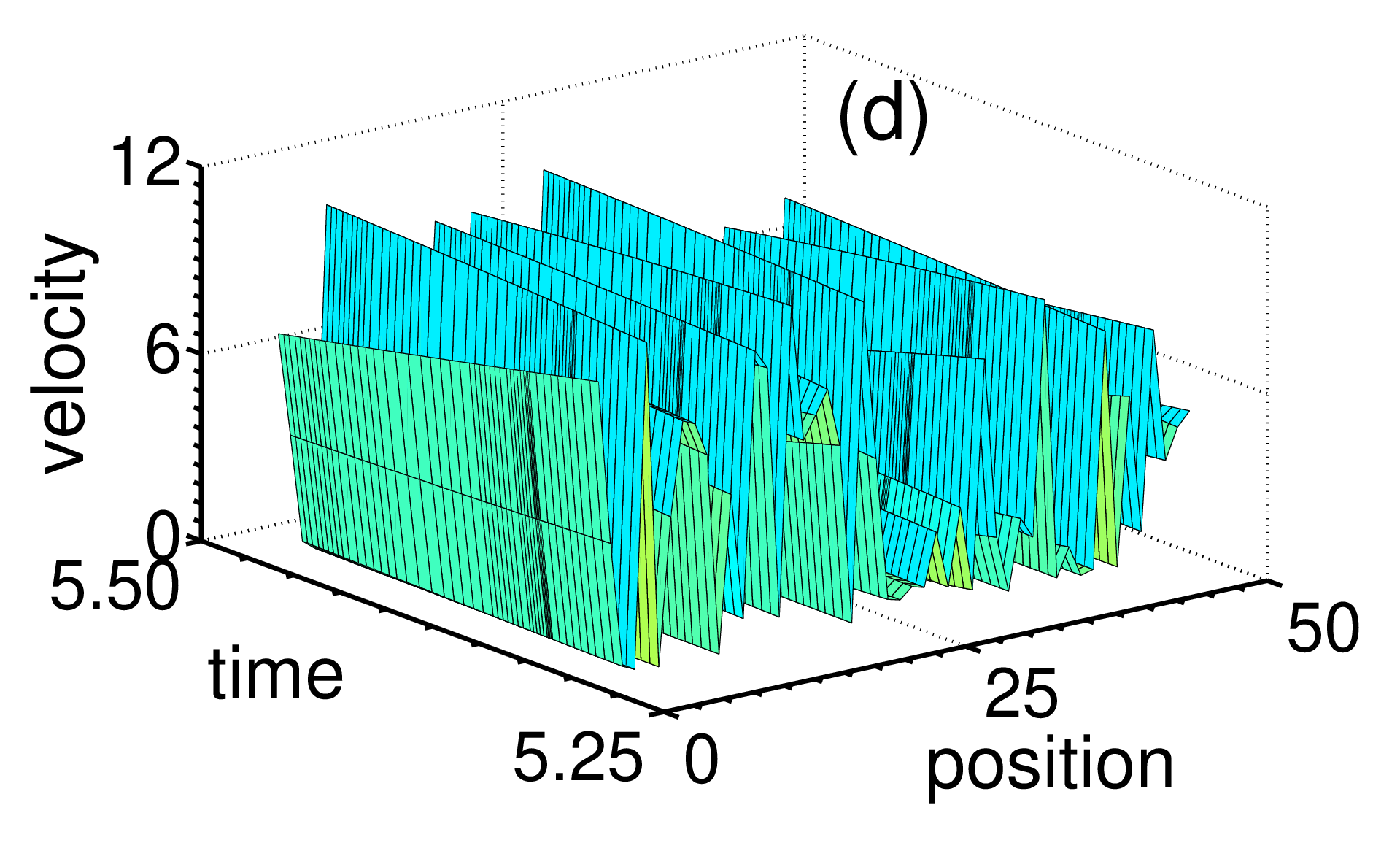}
\includegraphics[height=3.6cm,width=7.5cm]{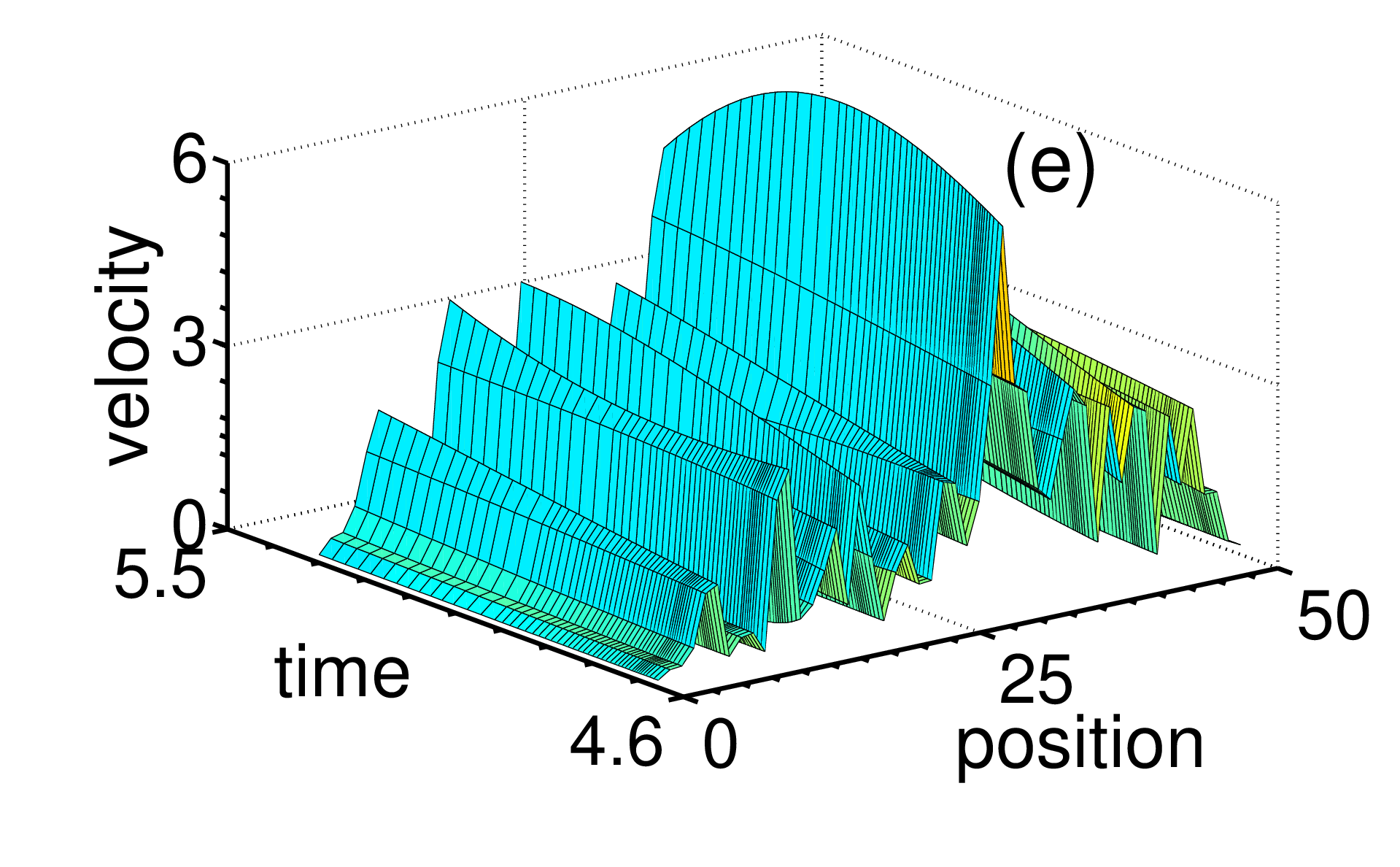}
}
\caption{(Color online a, b, d, e) 
(a, b) Snapshots during peel process $C_f = 0.788, V^s = 4.48$, and $\gamma_u =1.0$. (c) Phase plot for an arbitrary spatial point on the  peel front. Bold line represents $\phi(v^s)$. (d, e) Snapshots during peel process $C_f = 0.788, V^s = 4.48$, and $\gamma_u =0.01$.}
\label{I2m1V4gup1}
\end{figure}

We now consider  the influence of increasing the pull velocity $V^s$. As we increase $V^s$ to 2.48, the spatiotemporal patterns  seen for $\gamma_u= 1.0$, 0.1 and 0.01 are slightly different from those for $V^s =1.48$.  For $\gamma_u=1.0$, the peel process goes through a cycle of configurations shown in Figs. \ref{I2m1V2gup1}(a, b). It is clear that Fig. \ref{I2m1V2gup1}(a)  has more segments in the stuck state while Fig. \ref{I2m1V2gup1}(b) is the usual kind of SP configuration except that the stuck and peel segments are fewer. For this case, the stuck and peeled segments last longer than those for $V^s=1.48$.    The corresponding  $R_{ae}$ for each $\gamma_u$ exhibits noisy bursts overriding a periodic component. A typical plot for $\gamma_u=1$ is  shown in Fig. \ref{I2m1V2gup1}(c). From the time labels as also the arrows shown, the minima and maxima in $R_{ae}$ can be identified with Figs. \ref{I2m1V2gup1}(a, b) respectively. The orbit in the $X^s-v^s$ plane  moves into regions much beyond the values  allowed by $\phi(v^s)$ as is clear from  Fig. \ref{I2m1V2gup1}(d). The phase plots for $\gamma_u=0.1 $ and 0.01 are similar to this case.

For $\gamma_u=0.1$ also, the peel front pattern goes through a cycle of stuck-peeled configurations (with more stuck and peeled segments than for $\gamma_u=1.0$) and  stuck segments (similar to Fig. \ref{I2m1V2gup1}(a)). Yet, the energy dissipated $R_{ae}$ is similar to Fig. \ref{I2m1V2gup1}(c) for $\gamma_u=1.0$ which is surprising considering that there are more stuck and peeled segments compared to $\gamma_u=1$ case. This can be traced long lived of the stuck or peeled configurations that hardly change over a cycle (as in the case of $C_f=7.88,V^s=5.48,\gamma_u=0.01$, see Fig. \ref{I5m3V2gu0_1_01}(e)). The peel process is similar even for $\gamma_u=0.01$.

As we increase the peel velocity to 4.48, the influence of this time scale on the peel front pattern is discernable even for $\gamma_u=1.0$. The spatiotemporal patterns of the peel front switches sequentially from nowhere stuck  configuration shown in Fig. \ref{I2m1V4gup1}(a) to stuck-peeled configuration with few stuck and peeled segments  shown in Fig. \ref{I2m1V4gup1} (b). Note that there are very  few stuck and peeled segments. The corresponding $R_{ae}$ exhibits  noisy periodic pattern similar to Fig. \ref{I2m1V2gup1}(c) for  $V^s =2.48$. 
The $X^s-v^s$ phase plot in Fig. \ref{I2m1V4gup1}(c) shows that the orbit can move much beyond the values  allowed by $\phi(v^s)$. As we decrease $\gamma_u$ to 0.1, the nowhere stuck configuration [Fig. \ref{I2m1V4gup1}(a)] is replaced by  a partly stuck, partly peeled configuration and a SP configuration. For $\gamma_u=1.0$ case ( Fig. \ref{I2m1V4gup1}(c)), the  $X^s-v^s$ phase plot is slightly different as the orbit makes several loops before it jumps to low velocity branch without visiting the high velocity branch of $\phi(v^s)$. The nature of $R_{ae}$ is still noisy and periodic similar to Fig. \ref{I2m1V2gup1}(c). As we decrease $\gamma_u$ to 0.01, the peel process goes through SP configurations shown in Figures. \ref{I2m1V4gup1}(d, e). Note that Fig. \ref{I2m1V4gup1}(e) has large dispersion in the magnitude of velocity jumps of the peeled segments compared to that in Fig. \ref{I2m1V4gup1}(d). It is worth emphasizing that the increase in the number of stuck and peeled segments with decrease in $\gamma_u$ is a general feature. Despite the higher number of stuck and peeled segments,  $R_{ae}$ for $\gamma=0.01$ is similar to that for $\gamma_u=1.0$ as  these peel front configurations are long lived which again is a general feature observed at high pull velocities. Finally, it should be stated that for $C_f=0.788$, in general the velocity variation along the peel front is much more smooth compared to other values of $C_f$. The dynamics is uninteresting beyond $V^s =7.48$ as only smooth peeling is seen.
\begin{figure}[!h]
\vbox{
\centerline{\includegraphics[height=3.1cm, width=7.5cm]{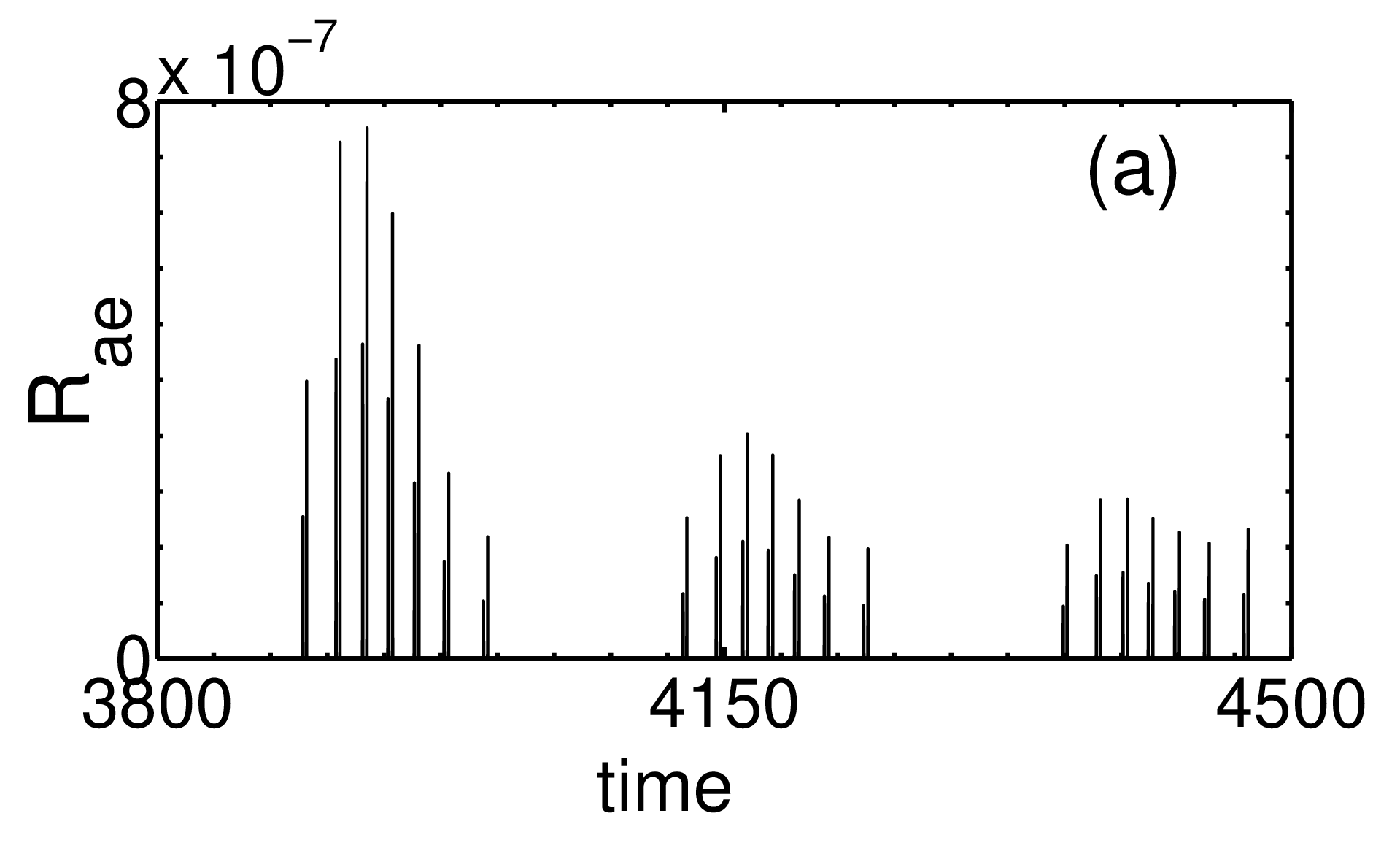}}
\centerline{\includegraphics[height=3.3cm, width=7.5cm]{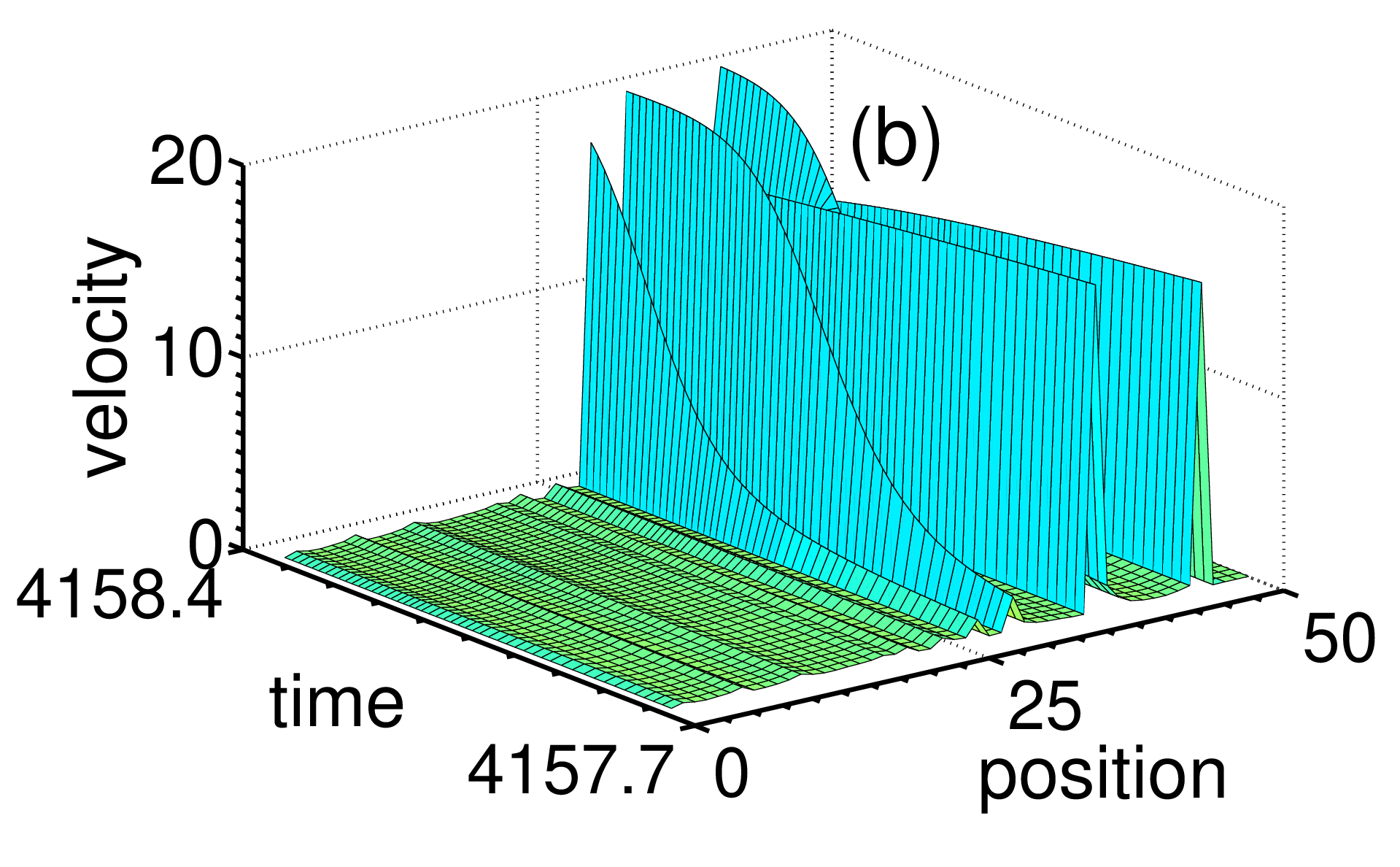}}
\centerline{\includegraphics[height=3.3cm, width=7.5cm]{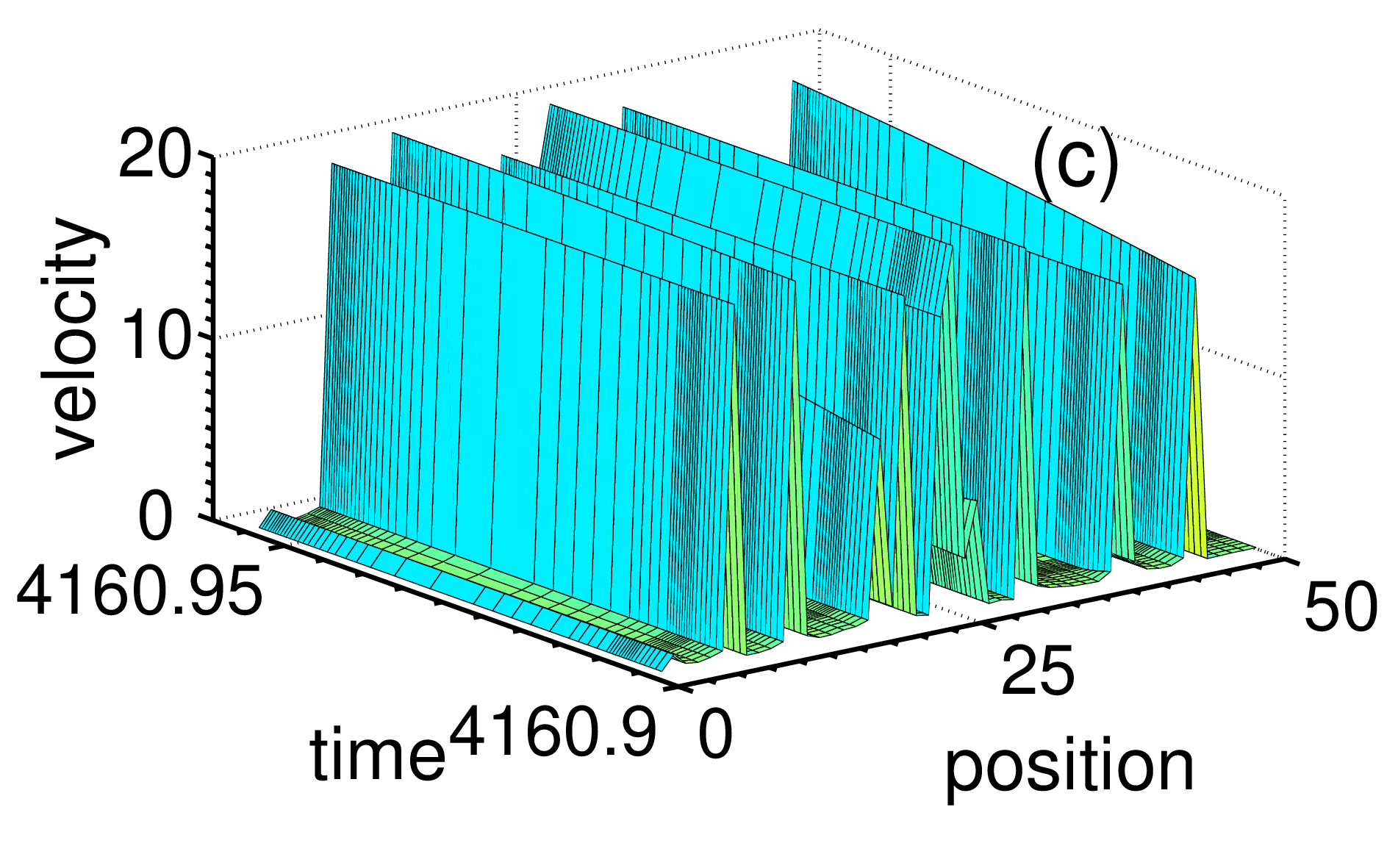}}
\centerline{\includegraphics[height=3.5cm, width=6.0cm]{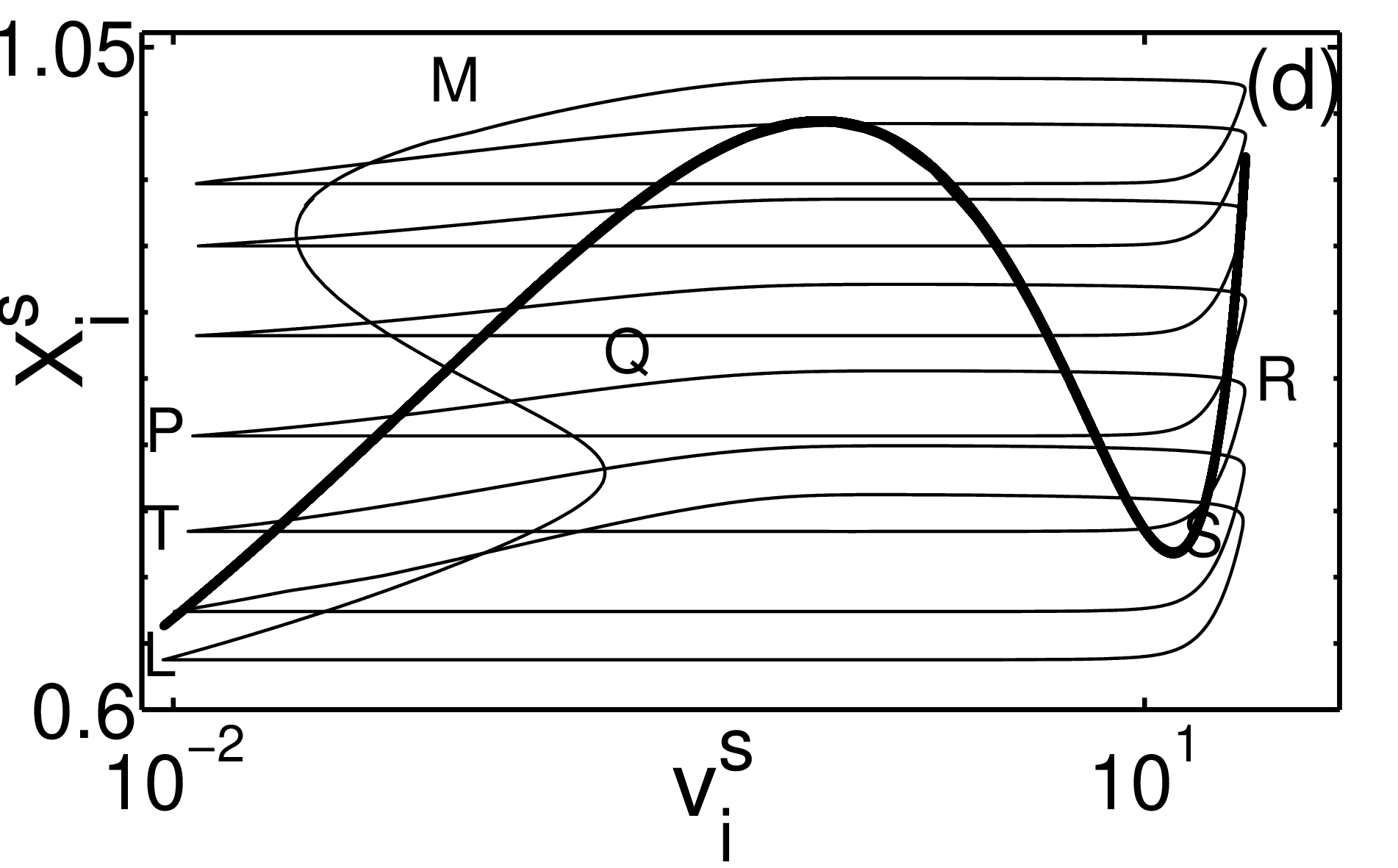}}
\centerline{\includegraphics[height=3.0cm, width=7.5cm]{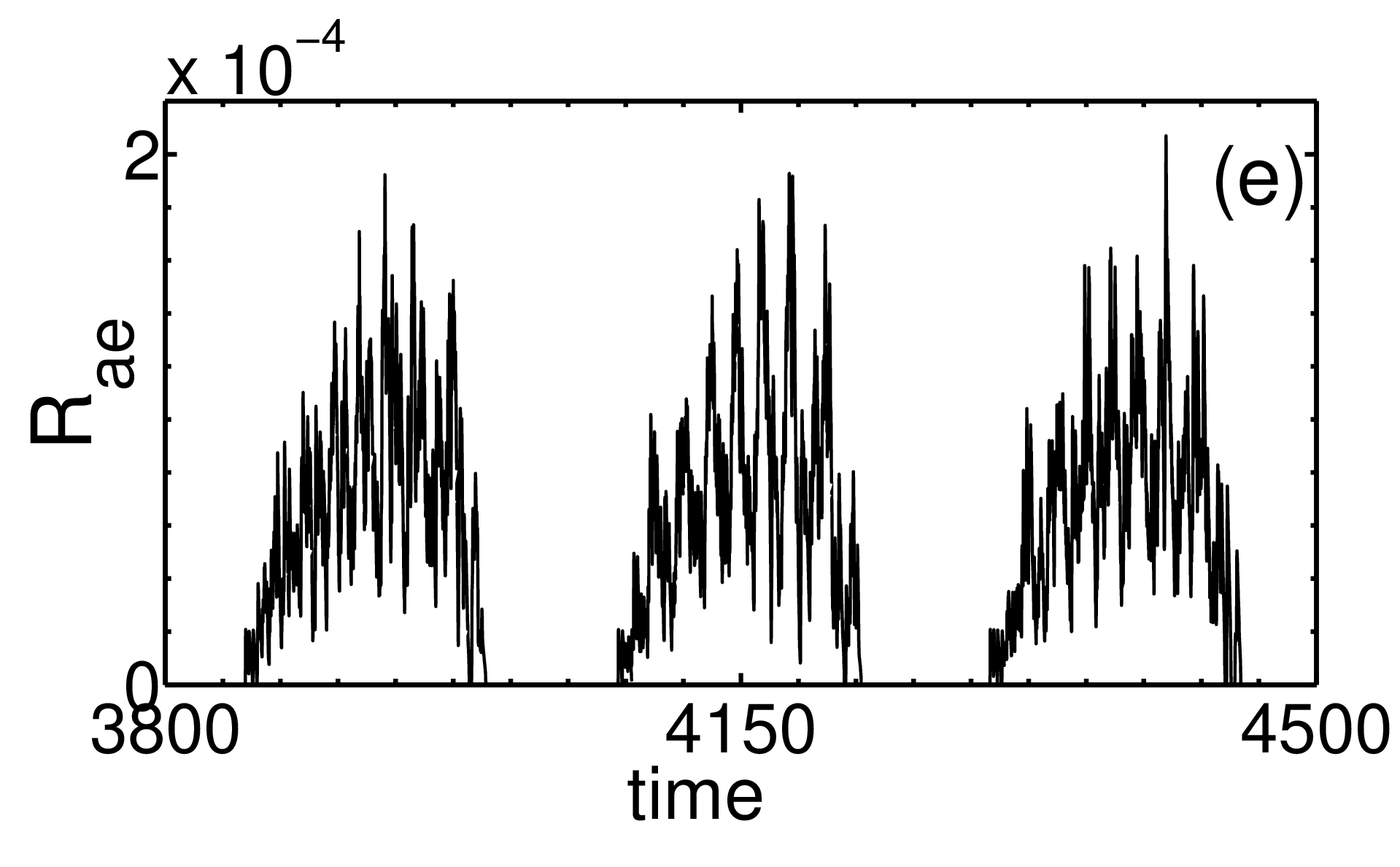}}
\centerline{\includegraphics[height=3.0cm, width=7.5cm]{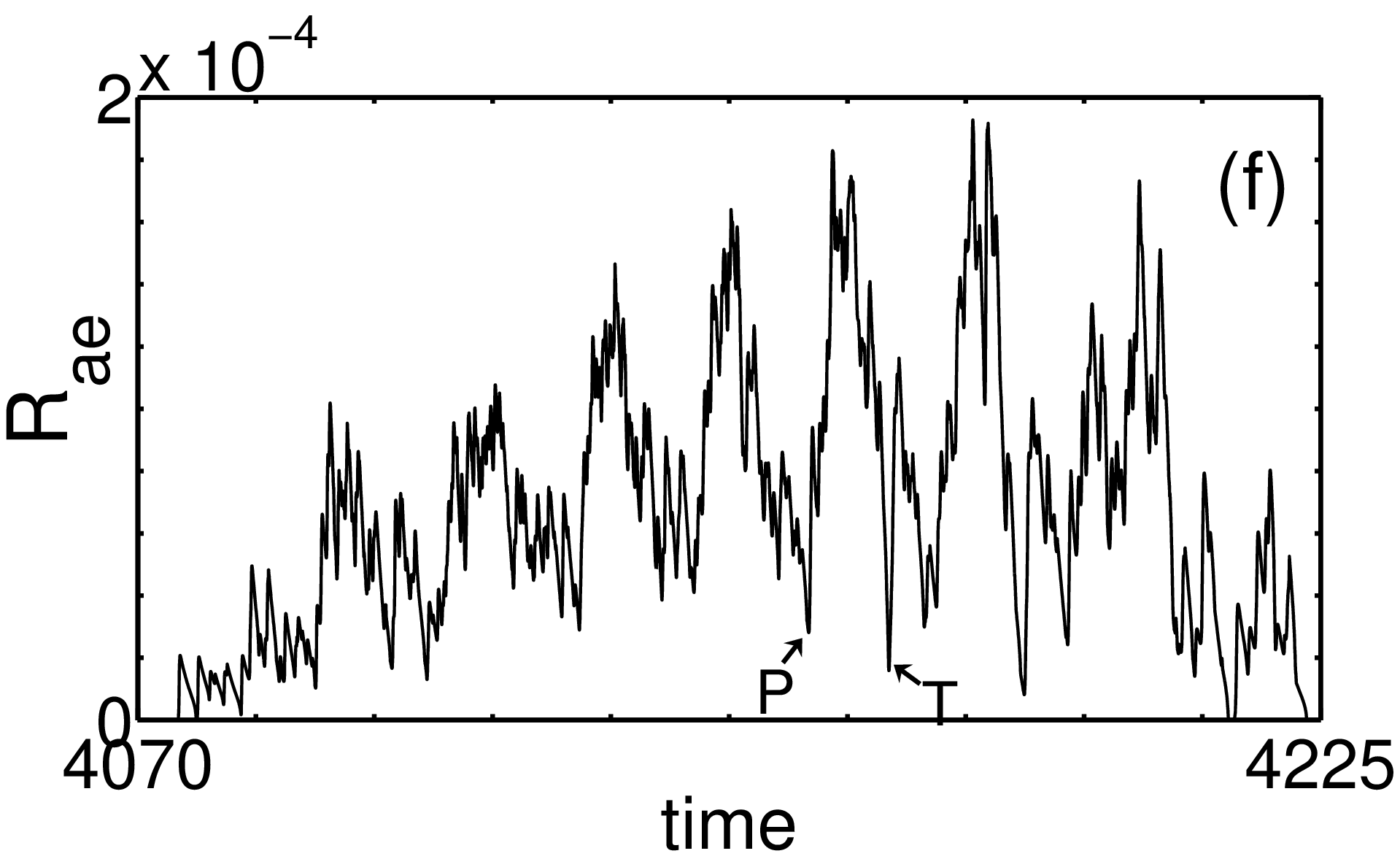}}
}

\caption{(Color online b, c) (a) $R_{ae}$ as a function for $C_f = 0.00788, V^s=1.48$, and  $\gamma_u=1$. (b) a stuck-peeled configuration with more stuck segments  compared to (c) where nearly equal number of stuck and peeled segments  for $C_f = 0.00788, V^s=1.48$, and  $\gamma_u=0.01$. (d) The corresponding phase plot for an arbitrary spatial point along the peeling front. Bold line represents $\phi(v^s)$.   (e) Corresponding $R_{ae}(\tau)$ as a functions of time and (f) substructure of (e).}
\label{I2m3V1gup01}
\end{figure}

\subsubsection{ Case (iii), $C_f =0.00788$, low tape mass and high roller inertia}

For this value of $C_f$, there is just one set of values of tape mass and roller inertia, namely,  $m=0.001$ and $I=0.01$. As the tape mass is low, this also corresponds to the DAE type of solutions for each spatial point. Thus, the velocity jumps between the two branches of the peel force function will always be abrupt with the roller inertia playing a major role in allowing the orbits to jump between the branches of the peel force function as demonstrated earlier \cite{Rumi04}.

\begin{figure}[!b]
\vbox{
\includegraphics[height=3.5cm, width=7.5cm]{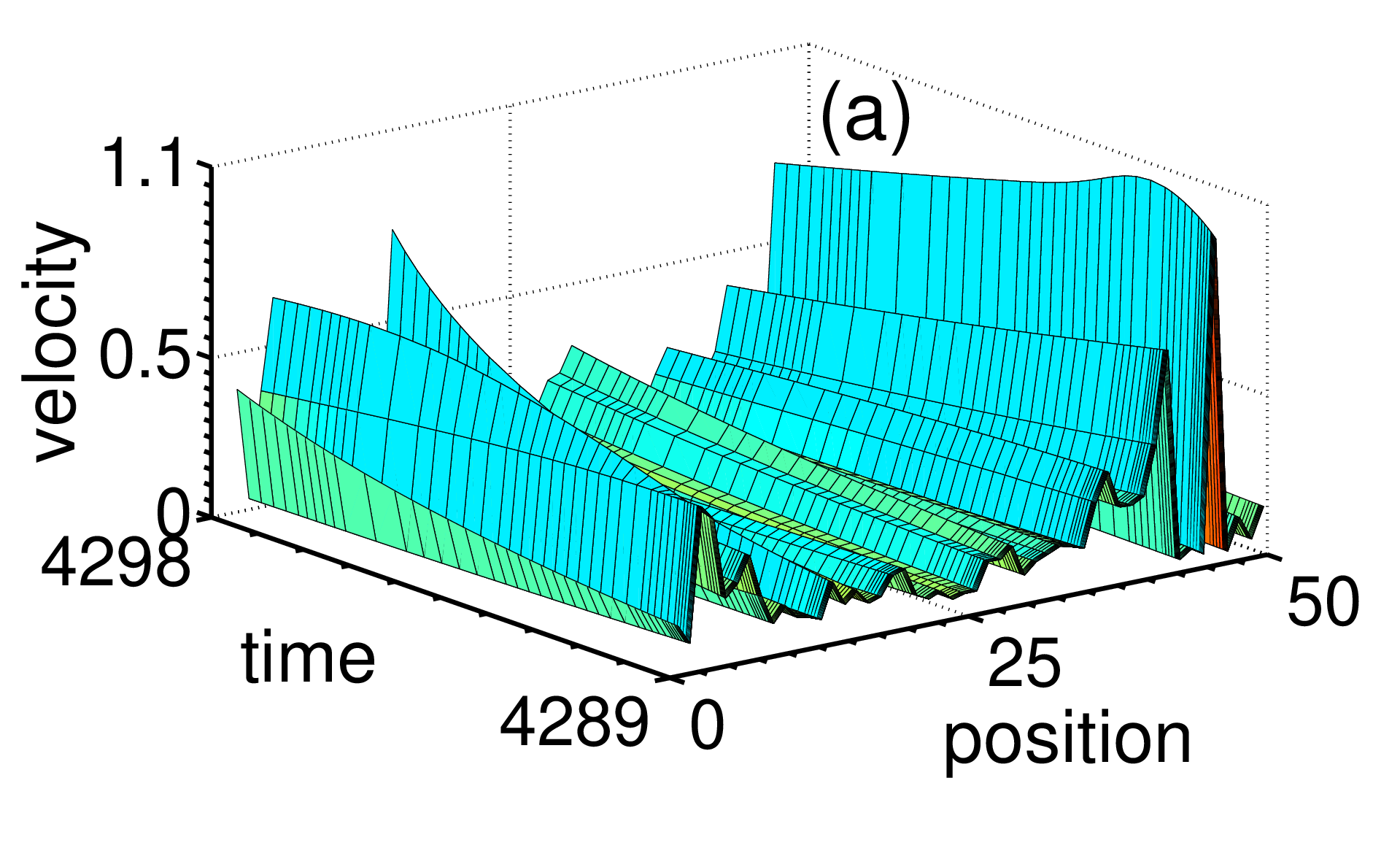}
\includegraphics[height=3.5cm,width=7.5cm]{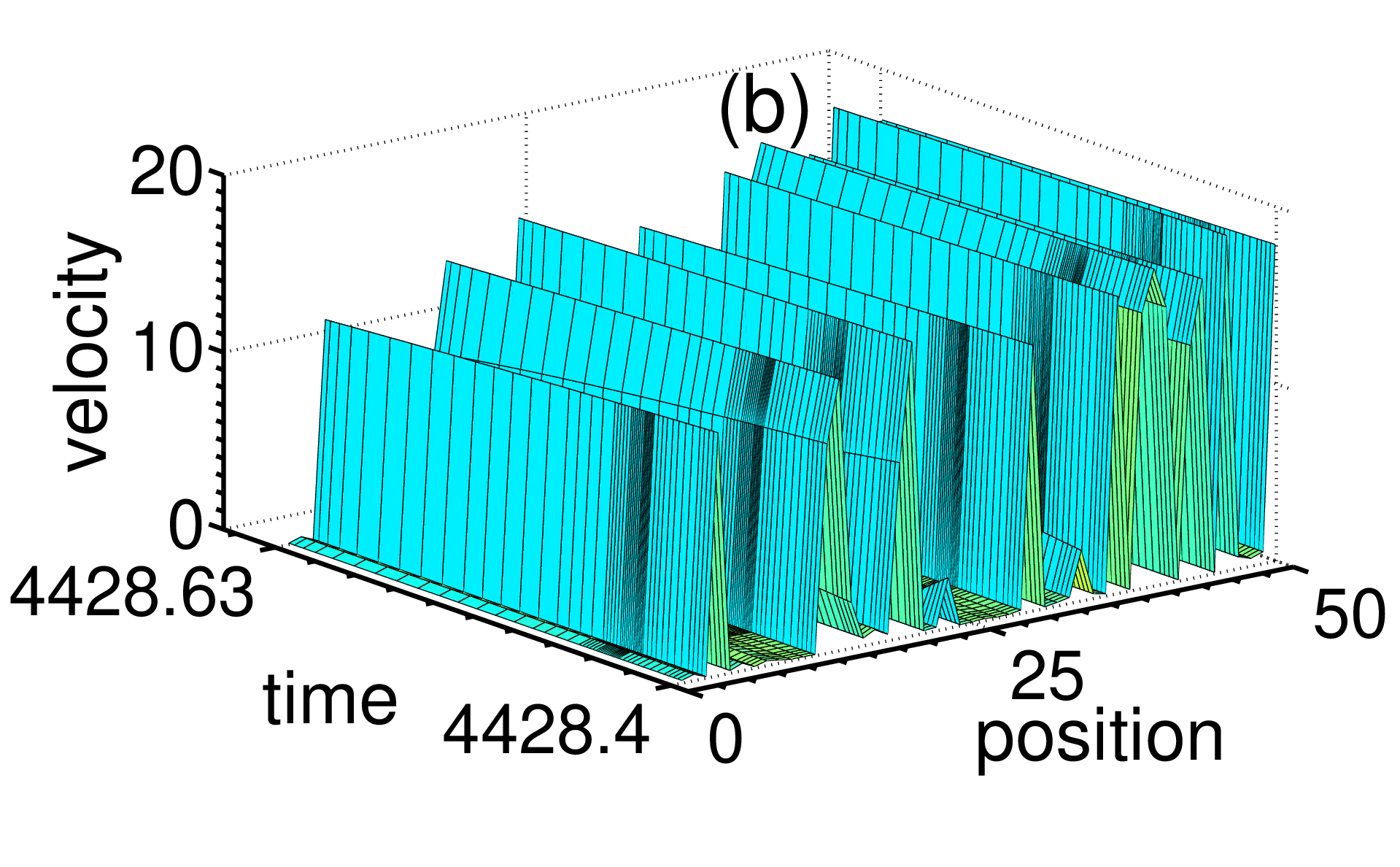}
\includegraphics[height=3.6cm,width=7.5cm]{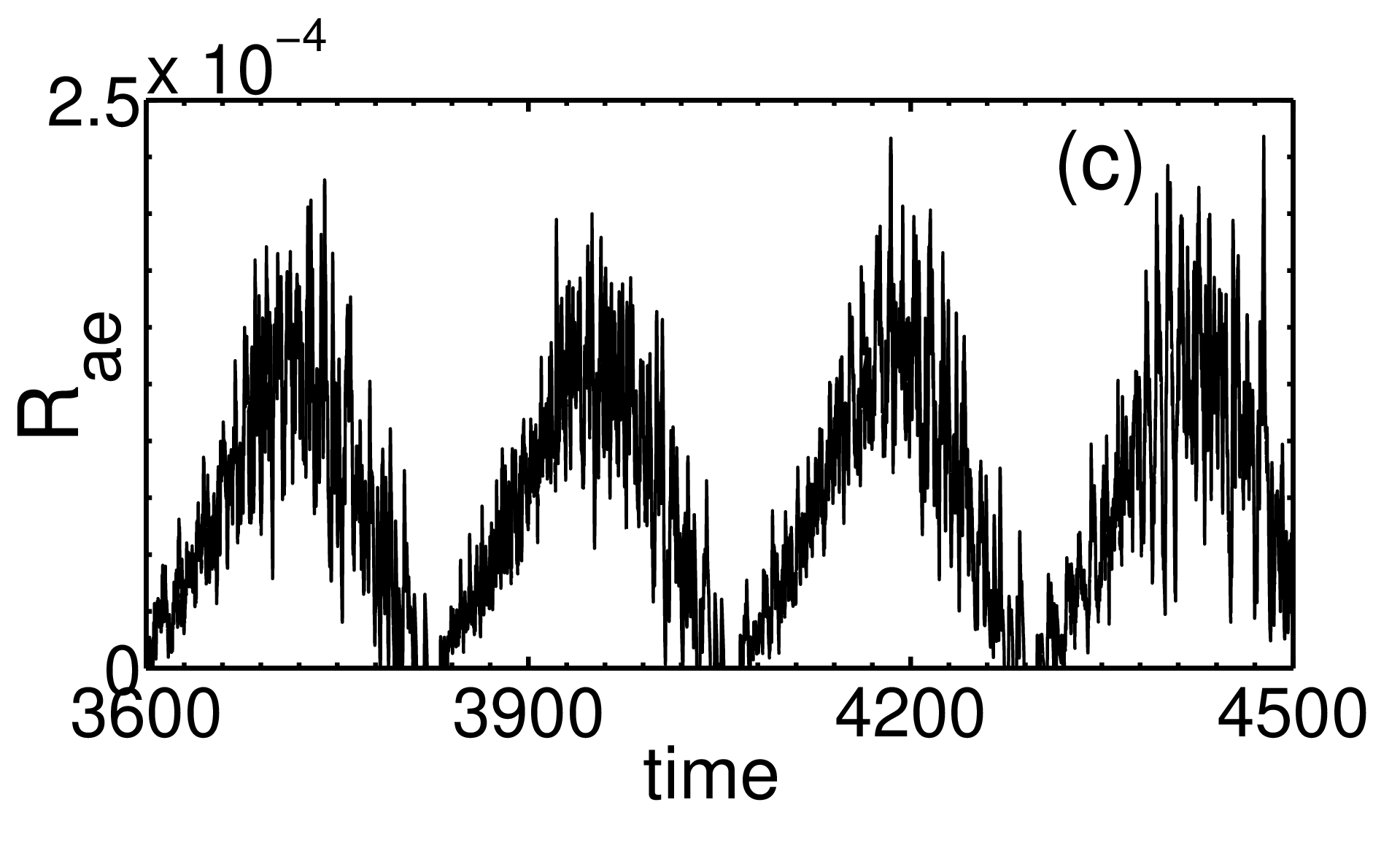}
}
\caption{(Color online a, b) 
(a, b) Snapshots during peel process for $C_f = 0.00788, V^s = 2.48$, and $\gamma_u =0.01$. (c) Corresponding model acoustic energy plot. }
\label{I2m3V2gup0_01}
\end{figure}
\begin{figure}[!t]
\vbox{
\includegraphics[height=3.5cm, width=7.5cm]{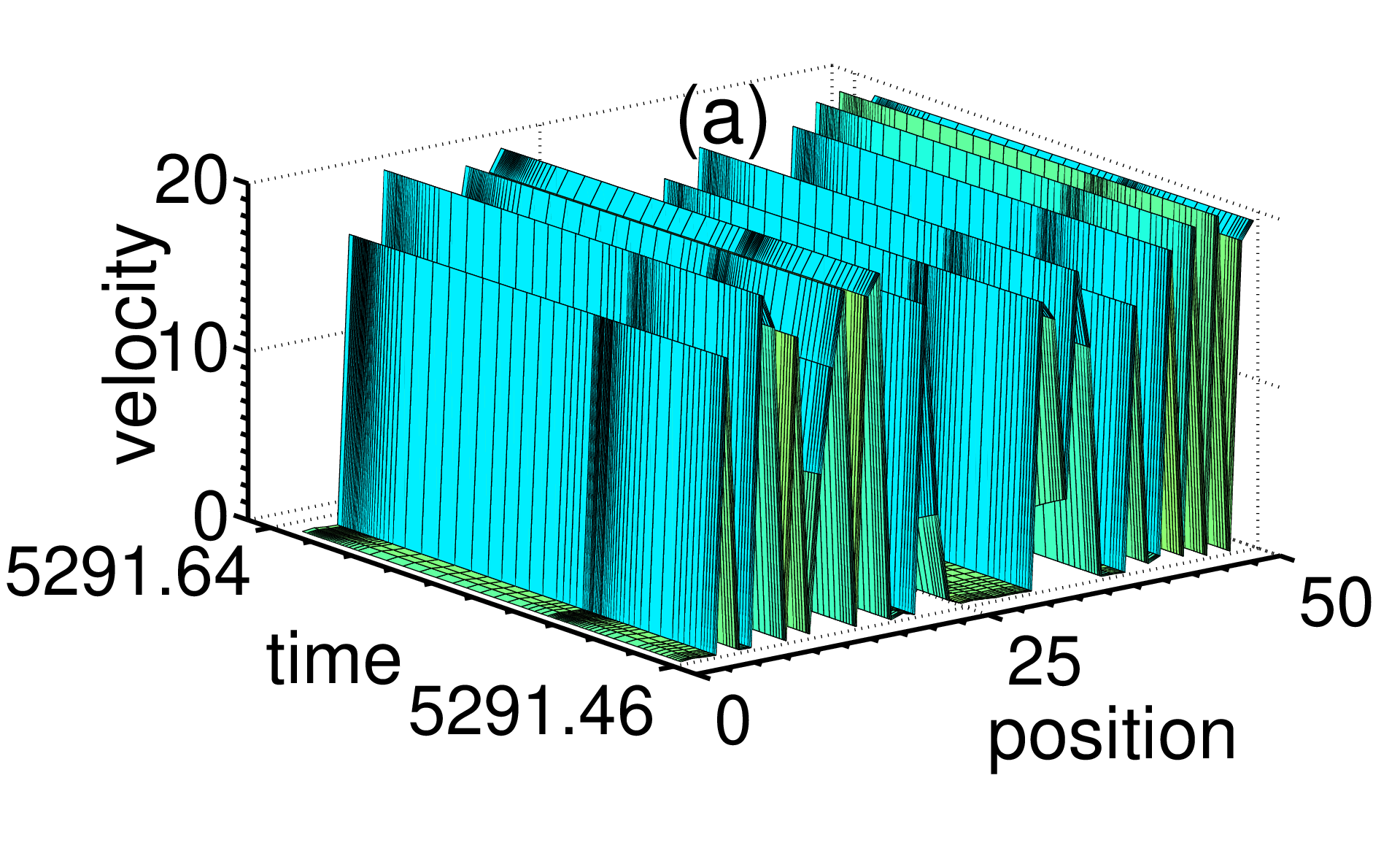}
\includegraphics[height=3.5cm,width=7.5cm]{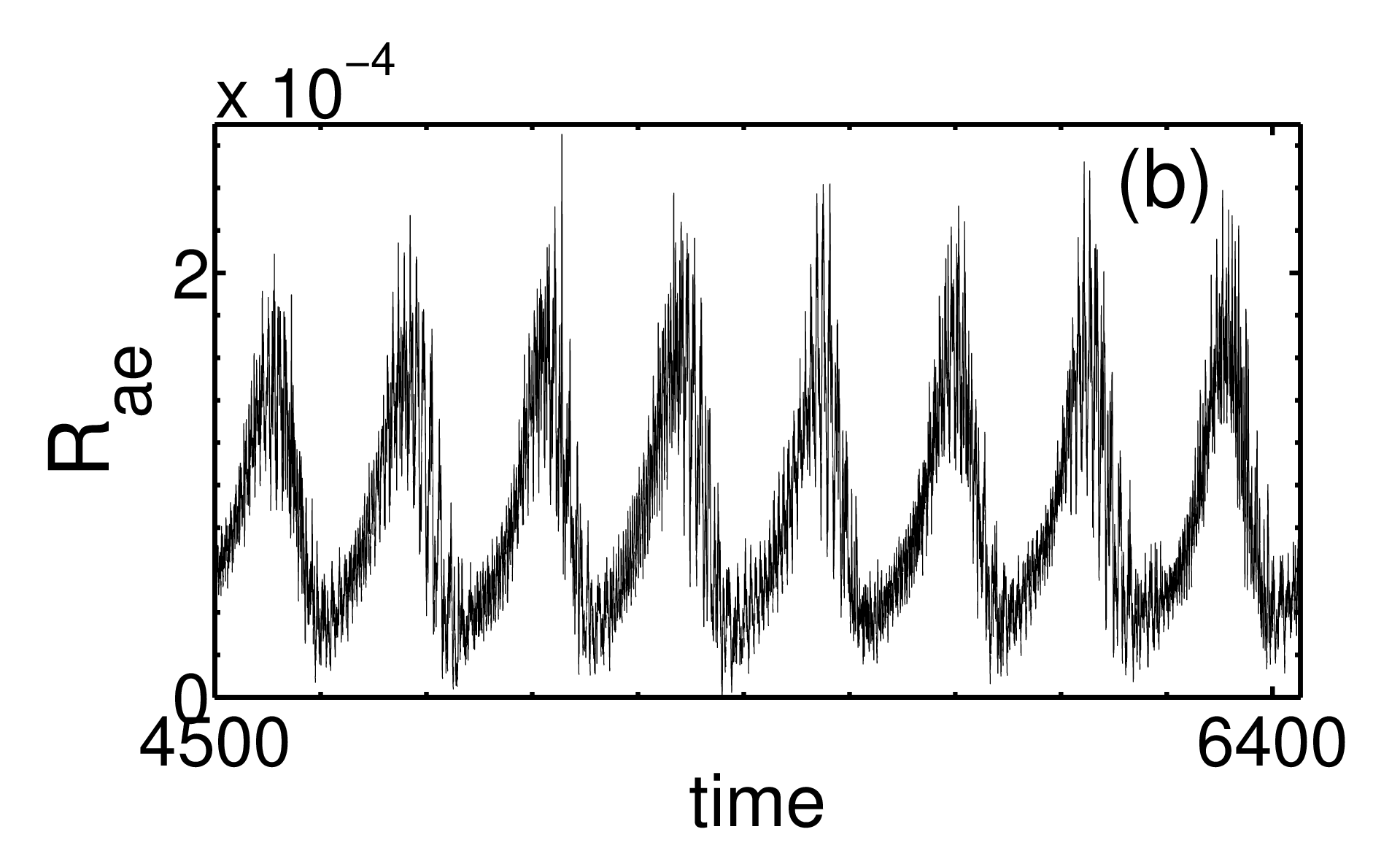}
\includegraphics[height=3.5cm,width=6.2cm]{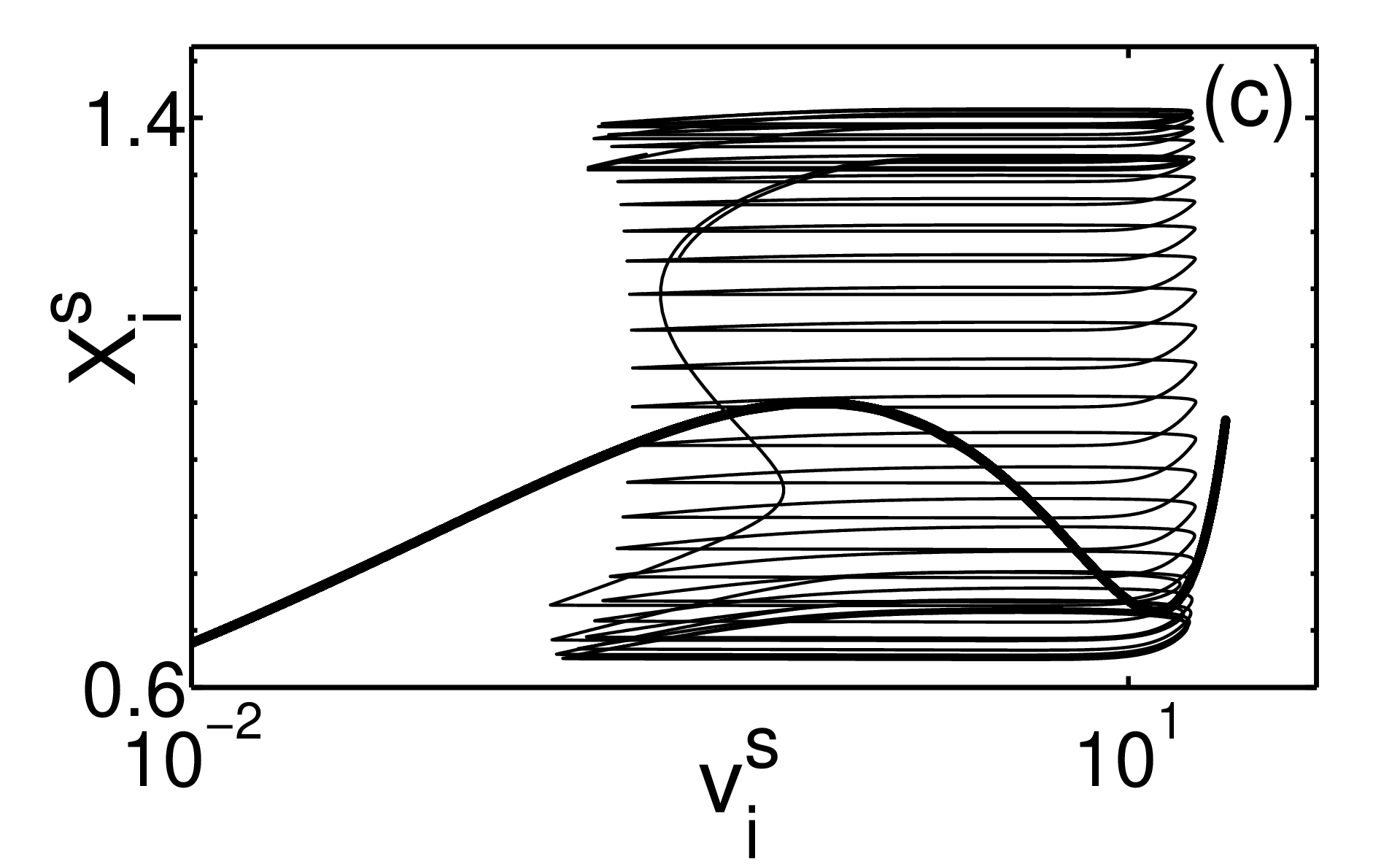}
}
\caption{ (a) (Color online) Snapshot during peel process $C_f = 0.00788, V^s = 4.48$, and $\gamma_u =0.01$. (b) Model acoustic energy plot. (c) Phase plot for an arbitrary spatial point on the  peel front for $C_f = 0.00788 $, $V^s=4.48$, and $\gamma_u=0.01$. Bold line represents $\phi(v^s)$ }
\label{I2m3V4gup0_01}
\end{figure}

Consider the influence of the dissipation parameter $\gamma_u$ keeping  $V^s =1.48$.   For $\gamma_u=1$, peeling is uniform and thus the whole peel front switches between the two branches of the peel force function. The acoustic energy shows a bunch of seven double spikes that appear at regular interval as shown in Fig. \ref{I2m3V1gup01}(a). (The number of spikes is correlated with the number of cascading loops seen in the $X^s-v^s$ phase plot, see below.) As we decrease $\gamma_u$ to 0.01,  the  peel front goes through a cycle of patterns with only few  peeled segments and those with large number of stuck-peeled segments as shown in Figs. \ref{I2m3V1gup01}(b) and (c) respectively. The phase plot in the $X^s-v^s$ plane of an arbitrary point on the peel front  jumps between the $AB$ to $CD$ branches of $\phi(v^s)$.  As shown in Fig. \ref{I2m3V1gup01}(d), in a cycle, the trajectory  starting at the highest value of $\phi(v^s)$ stays on $CD$ for a significantly shorter time compared to that on the left branch. The orbit  then  cascades down through a series of back and forth jumps between the two branches of $\phi(v^s)$. (For $V^s =1.48$, independent of $\gamma_u$ value, the nature of the phase plot is the same with seven loops.)

The corresponding model acoustic energy  consists of rapidly fluctuating time series with an overall convex  envelope of bursts separated by a quiescent state as shown in Fig. \ref{I2m3V1gup01}(e). (Contrast this with Fig. \ref{I2m3V1gup01}(a) for $\gamma_u=1$.)  From the time labels  in Figures. \ref{I2m3V1gup01}(b, c), both configurations belong to the region within the bursts [Fig. \ref{I2m3V1gup01}(e)].  To understand this complex pattern of bursts in $R_{ae}$ we have looked at the fine structure of each of these bursts along with the evolution of the associated configurations. One such plot is shown in Fig. \ref{I2m3V1gup01}(f) which shows  that  fine structure consists of seven bursts within each convex envelope.  These seven bursts can be correlated with the seven loops in the phase plot shown in Fig. \ref{I2m3V1gup01}(d). The time interval marked LM in the phase plot corresponds largely to stuck  configuration (not shown) and hence can be easily identified with the quiescent region  in $R_{ae}$. Following the peel front patterns continuously,  it is possible to identify the sequence of configurations that leads to the substructure shown in $R_{ae}$ [Fig. \ref{I2m3V1gup01}(f)]. For instance, the loop marked PQRST in the $X^s-v^s$ plot corresponds to the burst between P and T  in Fig. \ref{I2m3V1gup01}(f). During this period, the configuration at P is largely in the stuck state (as in Fig. \ref{I2m3V1gup01}(b)) which gradually evolves with more and more segments peeling out [Fig. \ref{I2m3V1gup01}(c)] as the trajectory moves from $P \rightarrow Q \rightarrow R \rightarrow S$. As the number of stuck and peeled segments reaches a maximum, $R_{ae}$ reaches  the peek region. Then, during the interval corresponding to S to T the number of peeled segments decreases abruptly. Thereafter, the next cycle of configurations (corresponding to the next loop in the phases plot) ensues.

As we increase the pull velocity, the peel front is smooth for $\gamma_u =1.0$ as also for 0.1 for the entire range of pull speeds.   However, for $\gamma_u =0.01$, as we increase $V^s$ to 2.48, the peel process goes through a cycle of SP configurations shown in Figs. \ref{I2m3V2gup0_01}(a, b). Note that there is a large dispersion in the jump velocities as is clear from Fig. \ref{I2m3V2gup0_01}(a).  The corresponding $R_{ae}$ shows rapidly fluctuating triangular envelope of  bursts with no quiescent region seen for $V^s=1.48$ case. This is shown in Fig. \ref{I2m3V2gup0_01}(c). The corresponding phase plot is similar to Fig. \ref{I2m3V1gup01}(d) but has twelve loops. In addition, the value of the upper loop extends far beyond that allowed by $\phi(v^s)$. We also see a fine structure similar to that in Fig. \ref{I2m3V1gup01}(e). As in the previous case, it is possible to identify configurations  that correspond to minima and near the maxima of $R_{ae}$. As we increase $V^s$ further to 4.48, only SP configurations are seen. The energy dissipated $R_{ae}$ shows continuous bursts overriding a sawtooth form as shown in Fig. \ref{I2m3V4gup0_01}(b).  The $X^s-v^s$ phase plot shows large excursions way beyond the peel force function values as shown in Fig. \ref{I2m3V4gup0_01}(c). 

A general comment may be relevant regarding large excursions of the trajectory in the phase plot $X^s-v^s$ as we increase the pull velocity. This is easily explained for the low $C_f$ (low tape mass, high roller inertia). It is clear from Eq. (\ref{Meqn}) that $m \rightarrow 0$ we have $F(t) \sim \frac{f(v)}{(1+sin\, \alpha(t))}$. As $\alpha(t)$ can takes on positive and negative values, one can see $F_{max}$ and $F_{min}$ are determined by minimum (negative) and maximum values of $\sin{\alpha}$ as argued in \cite{Rumi04}. It is possible to extend this argument to finite tape mass case. Finally it must be stated that the dynamics is no longer  interesting beyond $V^s=7.48$.

\begin{figure}[!h]
\hbox{
\includegraphics[height=4.5cm,width=8.5cm]{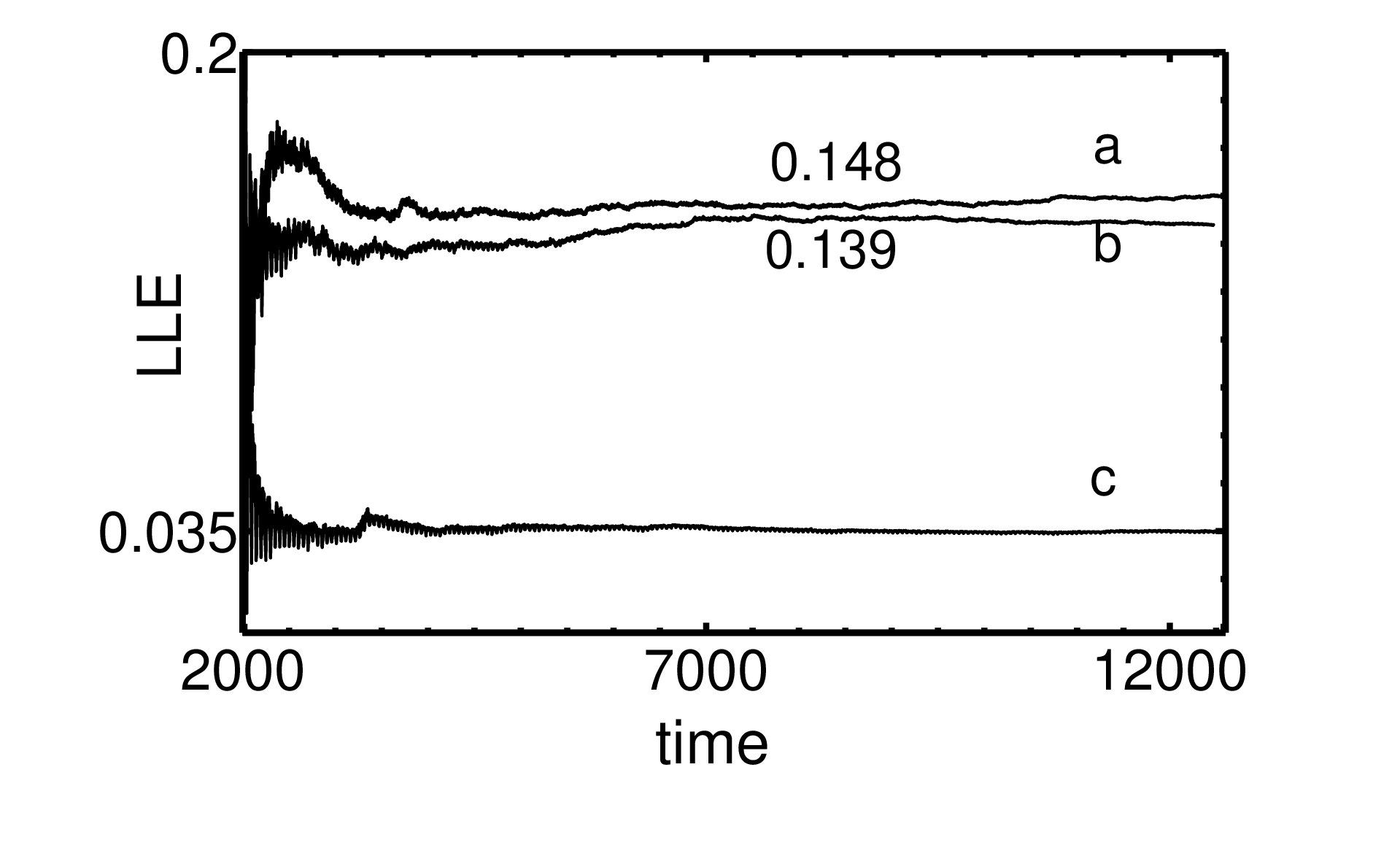}
}
\caption{Largest Lyapunov exponents of the model for $C_f=0.788$ and $\gamma_u=0.01$: (a) $V^s=1.48$ , (b) $V^s=2.48$ and (c) $V^s=4.48$.}
\label{LLE}
\end{figure}
\begin{table}[!t]
\centering
\begin{tabular}{|p{4em}|p{4em}|p{4em}|p{4em}|}  \hline

{$C_f$} &$ V^s$ & $\gamma_u $ &$ LLE$ \\ \hline

$7.88$ & $2.48$ & $0.01$ &$0.110$\\
& $4.48$& $0.01$ &$0.102$ \\ \hline

{$0.788$} & $1.48$ & $1.00$&$0.095$ \\ 
&&$0.10$&$0.120$\\
&&$0.01$&$0.148$\\ \hline
{$0.788$}&$2.48$& $1.00$&$0.068$ \\
&&$0.10$&$0.090$\\
&&$0.01$&$0.139$\\ \hline
{$0.788$}& $4.48$ &  $1.00$ &$0.028$ \\ 
&&$0.10$&$0.030$\\
&&$0.01$&$0.035$\\
\hline

{$0.00788$} & $1.48$   & $0.01$&$0.105$ \\
& $2.48$& $0.01$&$0.180$ \\
& $4.48$ &  $0.01$&$0.224$\\ \hline
\end{tabular}
\caption{Largest Lyapunov exponent for the model for various parameter values. For all $C_f$, the LLE reaches a near value zero for $V^s=5.48$. }
\label{tableLLE}
\end{table}

\subsubsection{Spatiotemporal Chaotic Dynamics}

As discussed above, there are several sets of parameter values for which the phase plots are irregular which may suggest the possibility of spatiotemporal chaotic dynamics. To verify this, we have calculated the largest Lyapunov exponent (LLE) for all the cases using the model equations. (The transient solutions for the first 2000 time units have been ignored for calculating the LLE.)  Figure \ref{LLE} shows a plot of the largest Lyapunov exponent  for $C_f=0.788$ and  $\gamma_u=0.01$ for various pull speeds. The value of LLE for $V^s=5.48$ is close to zero (not displayed in the figure). As can be seen, the LLE is positive being largest for $V^s=1.48$ decreasing to near zero value for $V^s =5.48$. Table \ref{tableLLE} shows the values of LLE for various parameter values for which the spatiotemporal dynamics has been detected. From this we conclude that the dynamics of the peel front is spatiotemporally chaotic for a range of parameters values.  

\section{ Analysis of AE signals}

\subsection{Statistical Analysis of AE signals}
\begin{table}[!b]
\centering
\begin{tabular}{|p{5.5em}|p{3.5em}|p{3.5em}|p{3.5em}|p{3.5em}|p{3.5em}|}  \hline
$V$ cm/s &  $m_{A1}$ & $m_{A2}$  &$\nu$ &$\lambda_1$& $D_{ky}$\\ \hline
{$1.0$} & $...$ & $2.00$ &NC&NC&NC\\ \hline
{$1.6$} & $...$ & $2.15$ &NC&NC&NC\\ \hline
{$3.0$} & $0.31$ & $2.26$ &NC&NC&NC\\ \hline
{$3.8$} & $0.30$ & $2.75$ &$2.80$&$1.70$& $2.94$\\ \hline
{$5.0$} & $0.32$ & $3.00$ &$2.70$&$1.73$&$2.96$\\ \hline
{$6.2$} & $0.27$ & $3.00$ &$2.55$&$1.54$&$2.84$\\ \hline
{$7.4$} & $0.30$ & $2.99$ &NC&NC&NC\\ \hline
\end{tabular}
\caption{Statistical and dynamical invariants for the experimental AE signals for typical traction velocities. The second and third columns show power law exponents $m_{A1}$ and $m_{A2}$ corresponding to small and large amplitudes. When only a single power law is seen, $m_{A2}$ is the exponent value. Fourth to sixth columns list the values of the correlation dimension $\nu$, the largest exponent $\lambda_1$ and Kaplan-Yorke dimension $D_{ky}$ obtained from dynamical analysis of the AE signals. NC corresponds to nonchaotic dynamics where we did not find any convergence of the correlation dimension.}
\label{Exptstat}
\end{table}

\begin{figure}[!t]
\hbox{
\includegraphics[height=3.0cm,width=4.25cm]{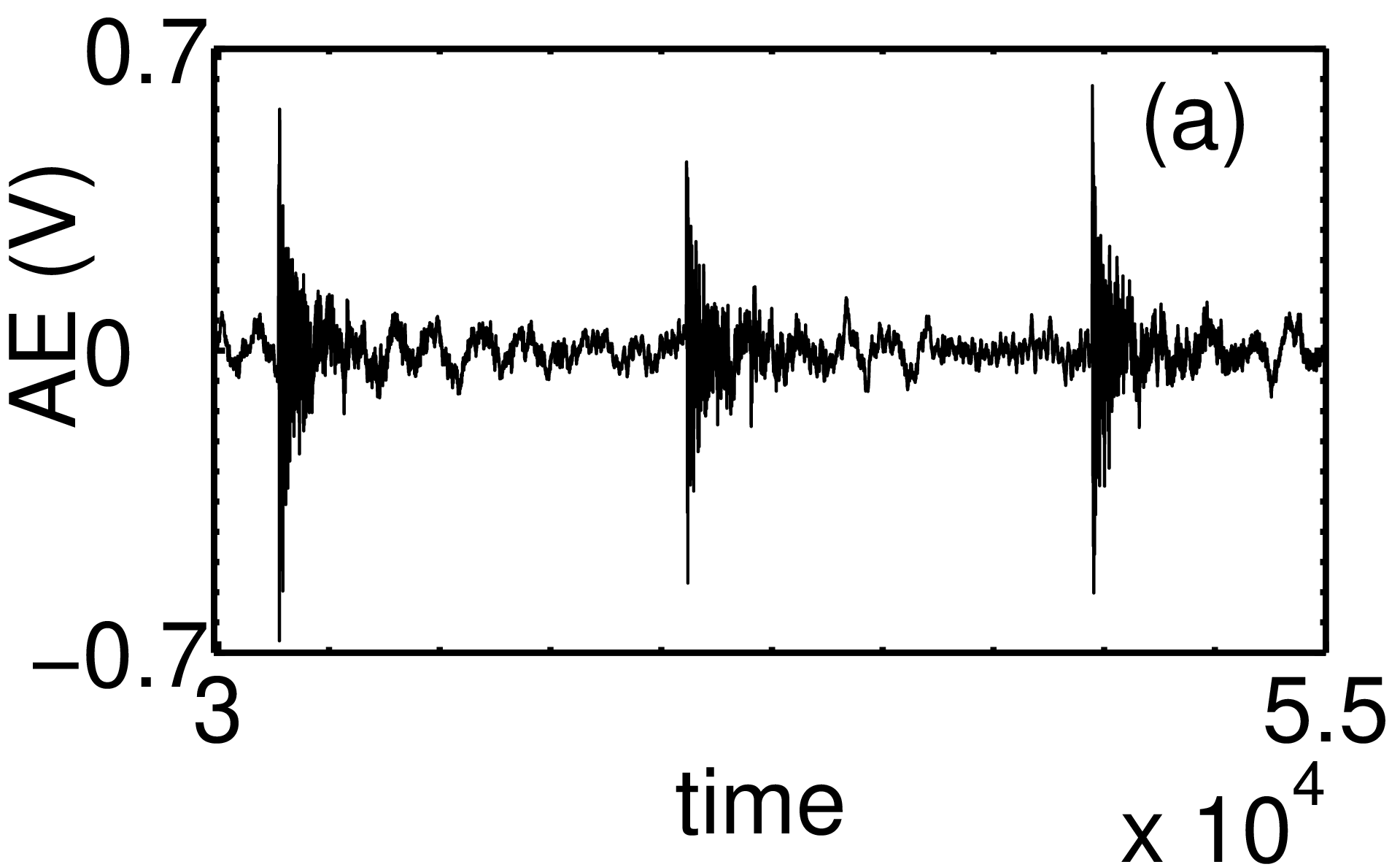}
\includegraphics[height=3.0cm,width=4.25cm]{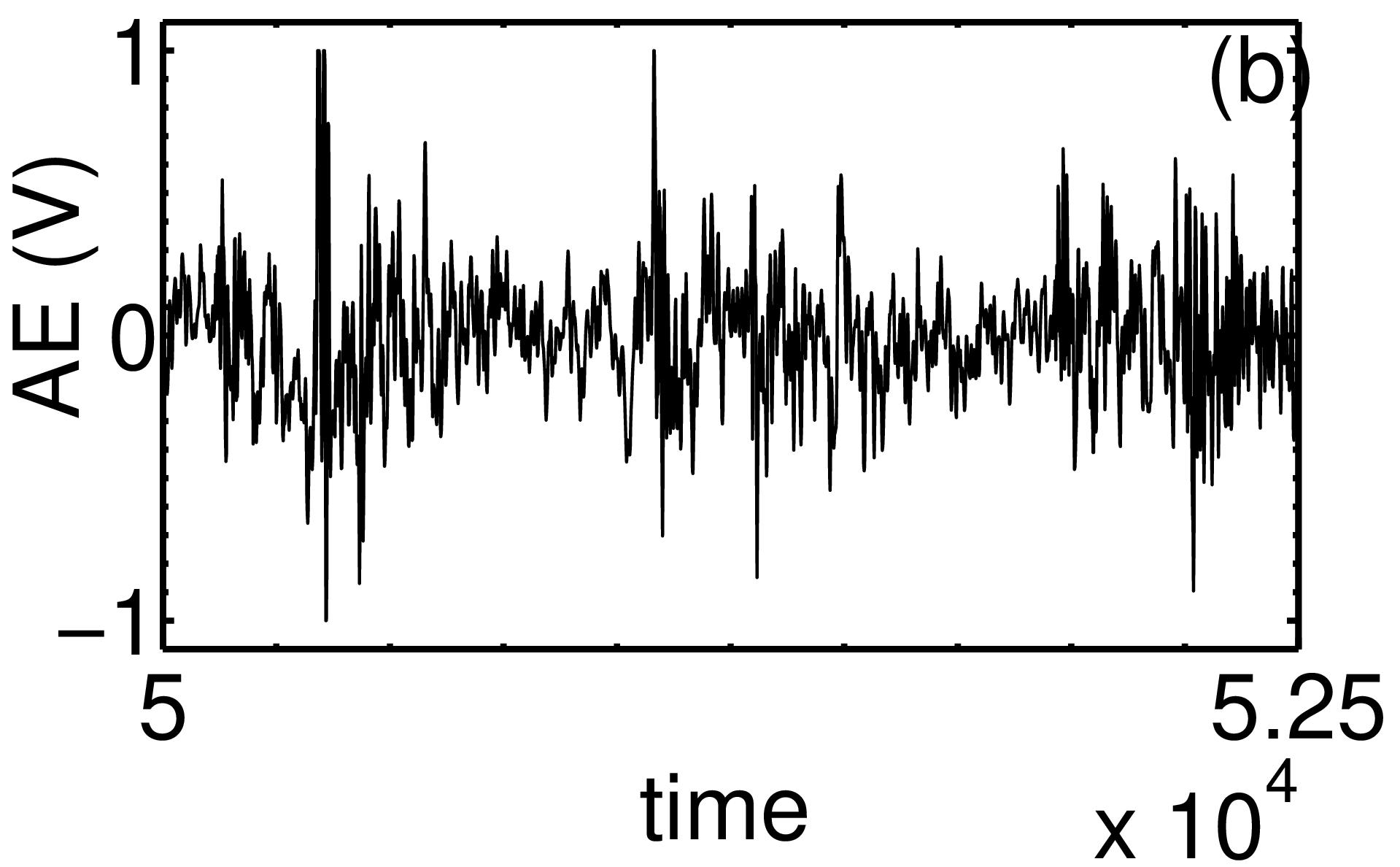}
}
\hbox{
\includegraphics[height=3.0cm,width=4.25cm]{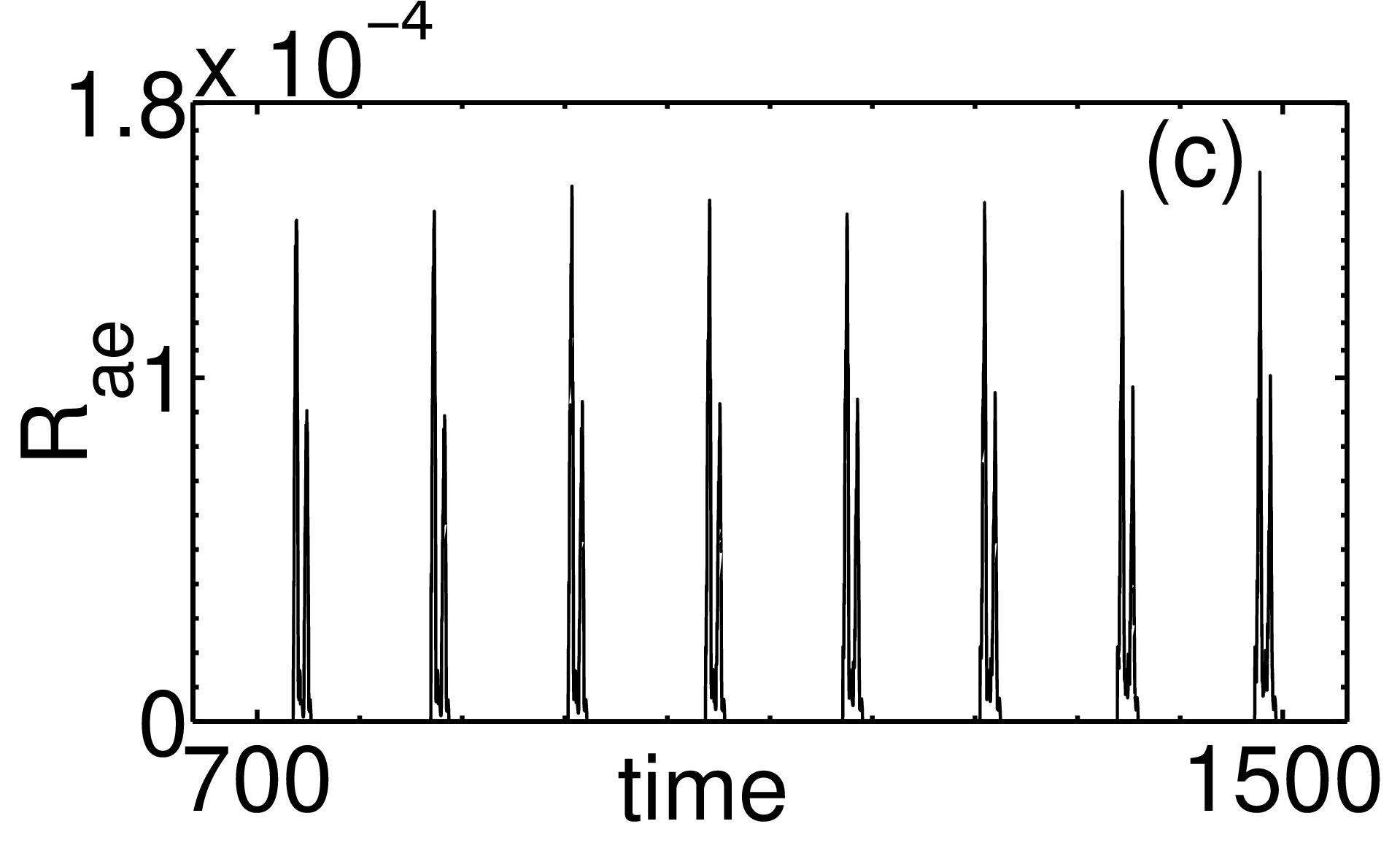}
\includegraphics[height=3.0cm,width=4.25cm]{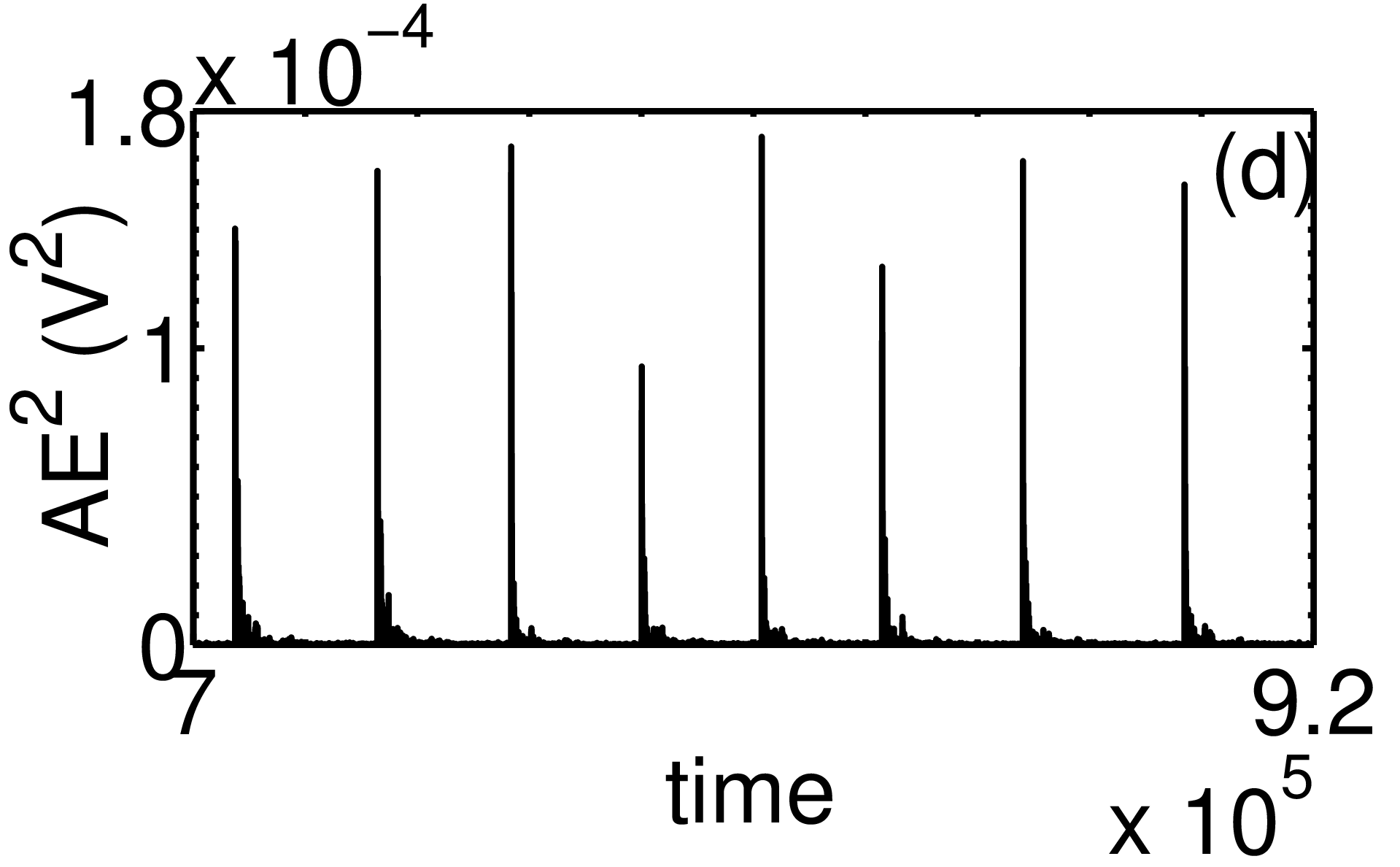}
}
\hbox{
\includegraphics[height=3.0cm,width=4.25cm]{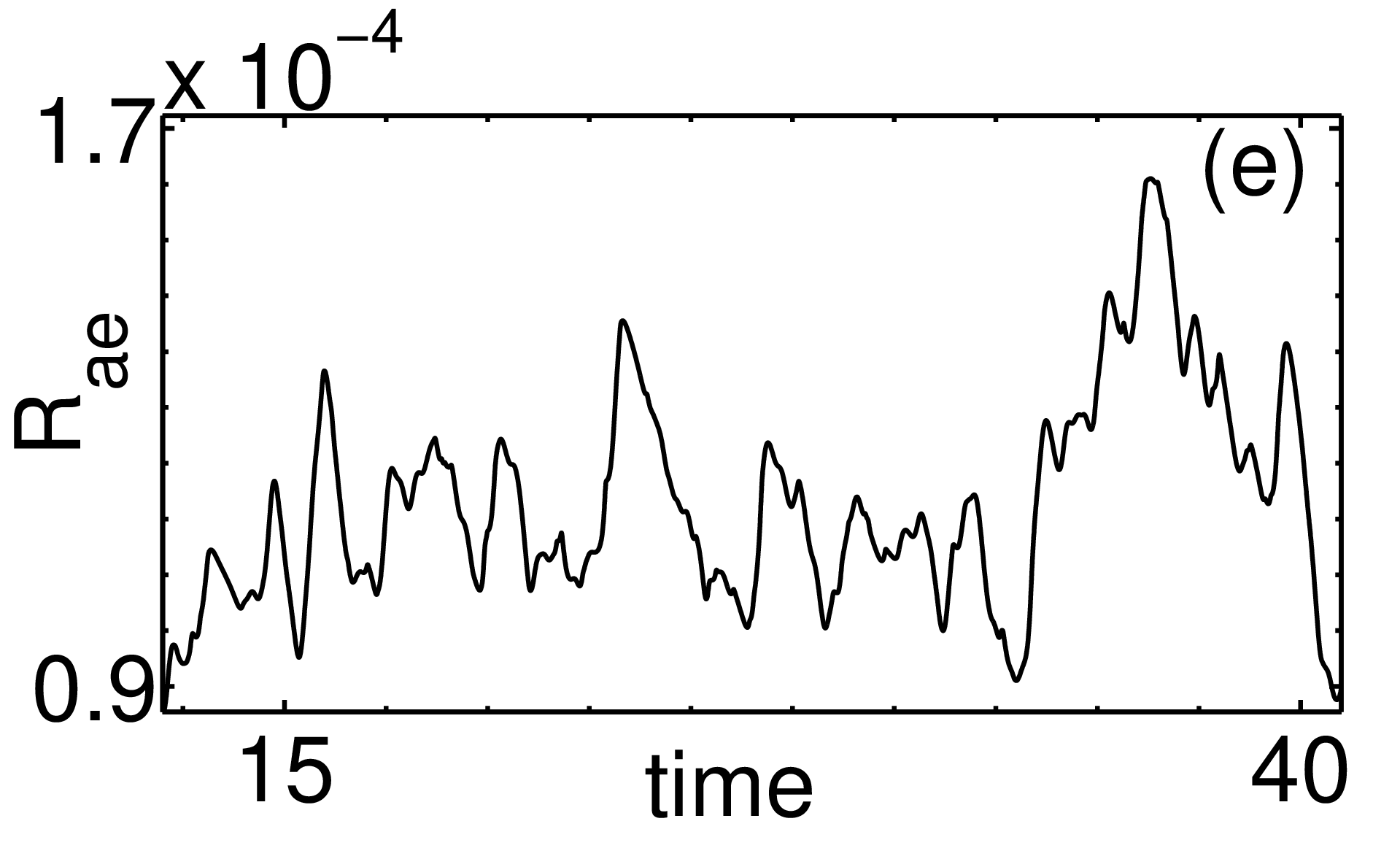}
\includegraphics[height=3.0cm,width=4.25cm]{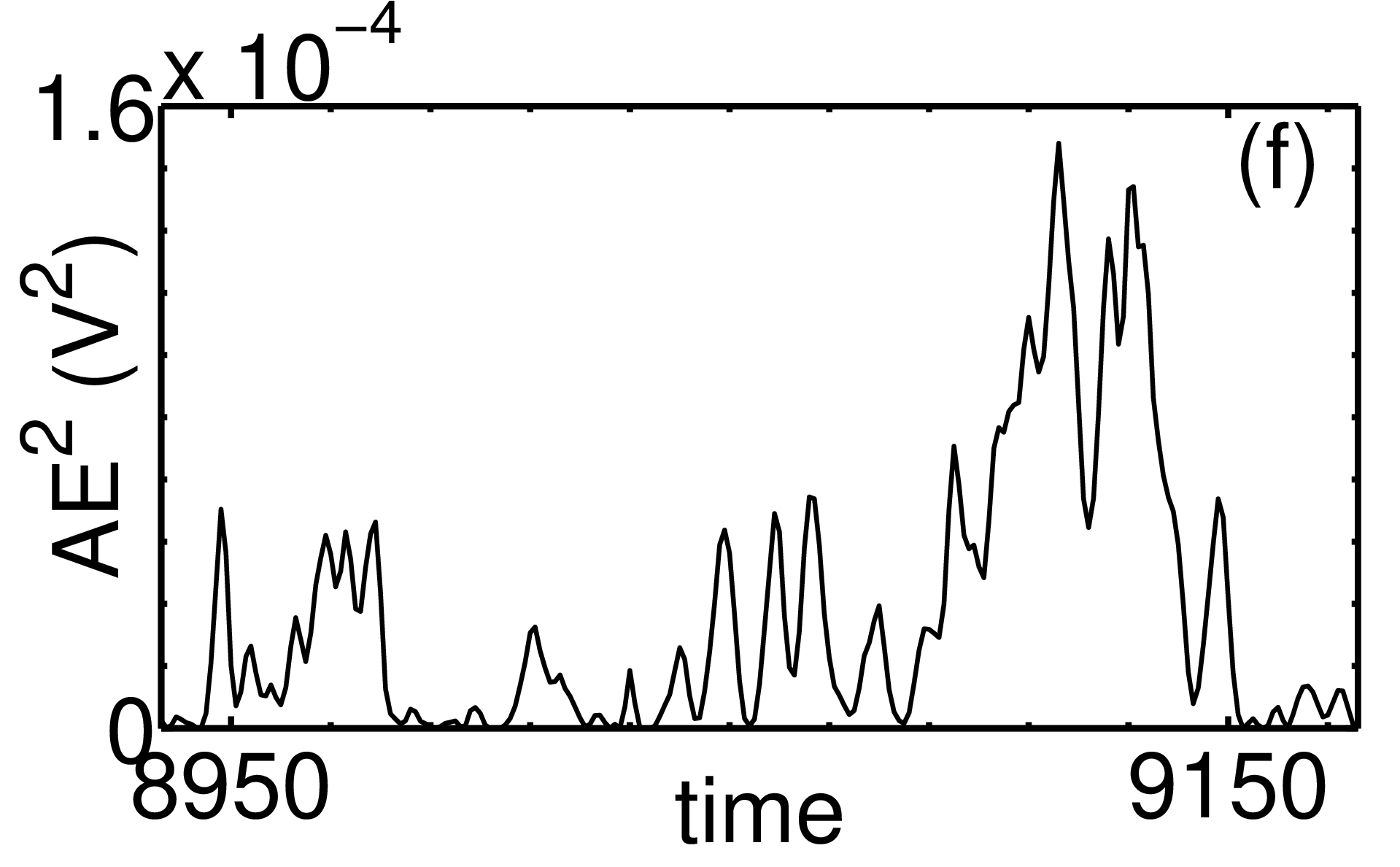}
}
\caption{ (a, b) Raw AE signal  for $V= 1.6$ cm/s and 7.6 cm/s respectively.  (c) Burst like model acoustic energy plot for $C_f=7.88,V^s=1.48$ and $\gamma_u=0.01$. (d) Burst like experimental acoustic energy  for $V= 0.4 $ cm/s. (e) Continuous model acoustic energy plot for $C_f=7.88,V^s=4.48$ and $\gamma_u=0.01$. (f) Continuous experimental acoustic energy  for $V= 6.4 $ cm/s. }
\label{ExptAE}
\end{figure}

The acoustic emission data obtained from experiments  are fluctuating and noisy only within the domain where the peel process is intermittent form $0.2$ to $7.6$ cm/s. As known from early experiments \cite{MB}, force wave forms change as the traction velocity is increased. Correspondingly, the nature of the AE signals also  change with the traction velocity.  At low traction velocities, the AE signals have a burst like character appearing at nearly regular intervals separated by oscillatory decay of the amplitudes. These bursts are correlated with the stick-slip events. With increasing traction velocity, the bursts become increasingly irregular and continuous. Examples of burst and continuous type of AE time series are shown in Figures.  \ref{ExptAE}(a, b). 

As shown in the previous section, the nature of the model AE signal depends on parameter values. In general $R_{ae}$ can be of noisy burst type overriding a periodic component, continuous and irregular, rapidly fluctuating triangular envelope of bursts or  simply a set of spikes. Clearly, interesting cases for comparison with the experimental AE signals are those where $R_{ae}$ is continuous and noisy. Simplest quantity to compare is the nature of the model acoustic signal with the energy of the experimental AE signal (i.e.,  square the amplitude). Figures \ref{ExptAE}(c, d) show a comparison between model acoustic signal for $C_f=7.88,\gamma_u=0.01$ and $V^s= 1.48$ and energy of the experimental signal for $V =0.4$ cm/s. Both show burst type emission. As another example Figs. \ref{ExptAE}(e, f) show respectively the continuous model signal (for $C_f=7.88, \gamma_u=0.01$ and $V^s= 4.48$) and experimental acoustic energy for $V= 6.4$ cm/s.

\begin{figure}[!h]
\vbox{
\hbox{
\includegraphics[height=3.0cm,width=4.25cm]{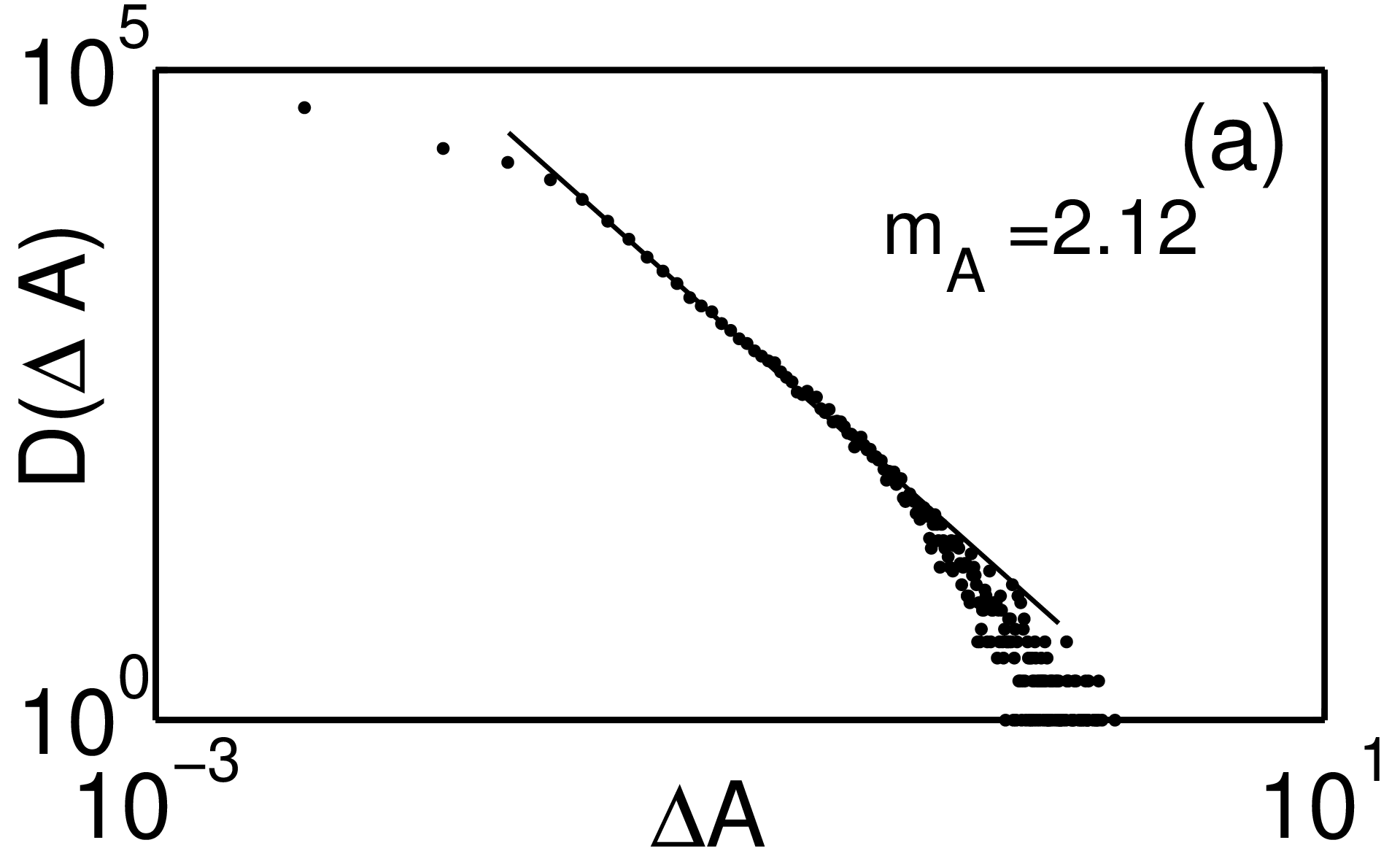}
\includegraphics[height=3.0cm,width=4.25cm]{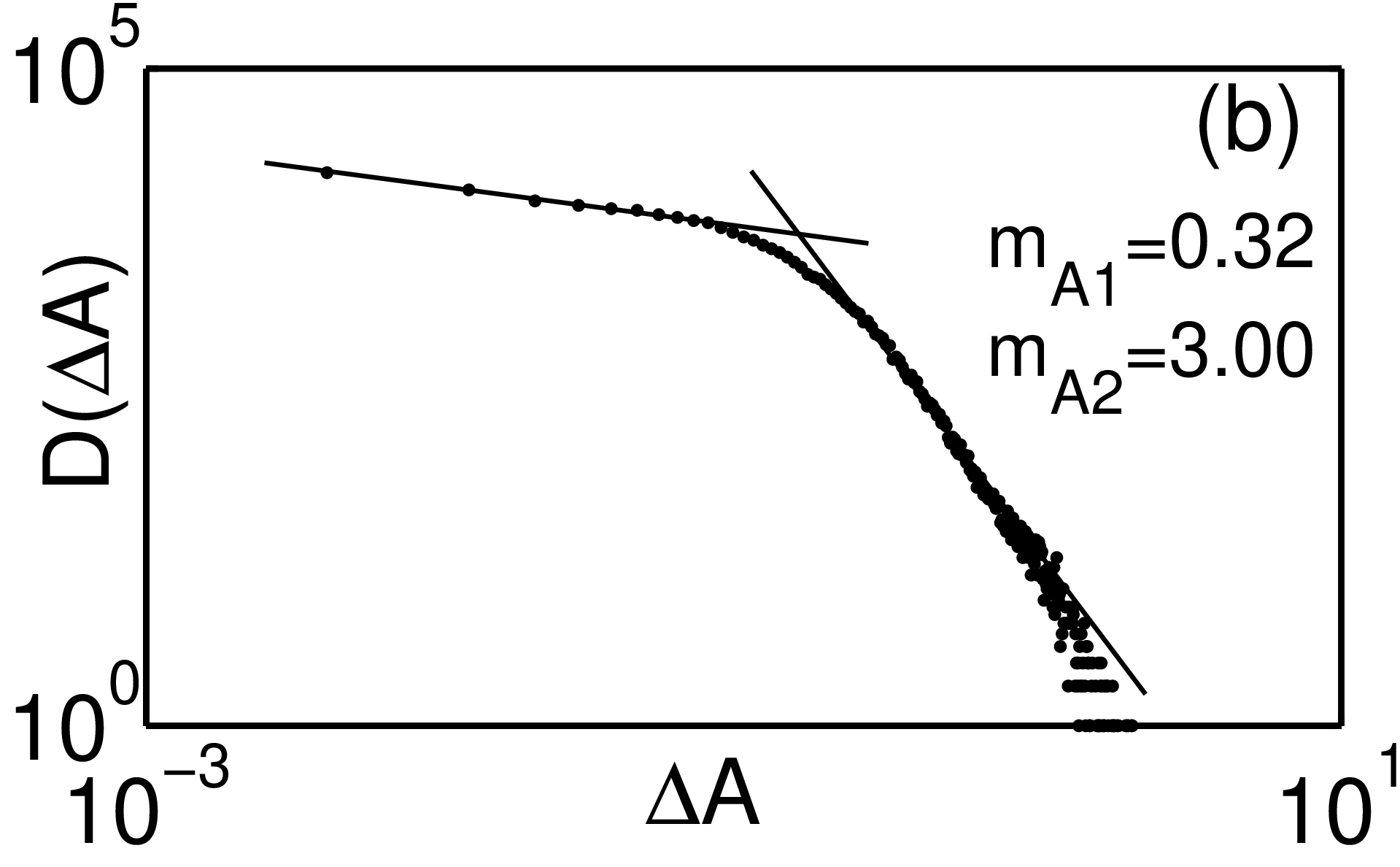}
}
\hbox{
\includegraphics[height=3.0cm,width=4.25cm]{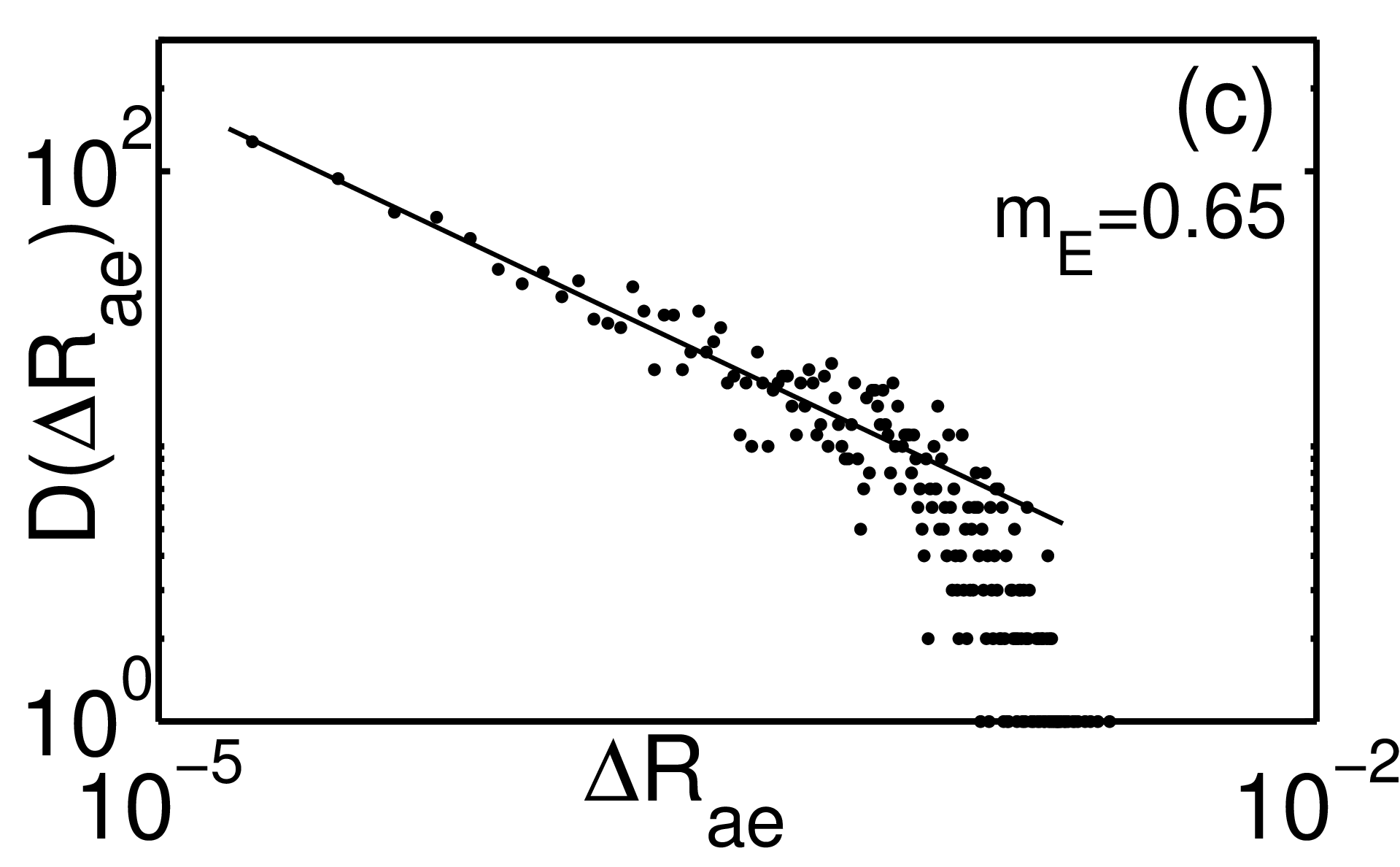}
\includegraphics[height=3.0cm,width=4.25cm]{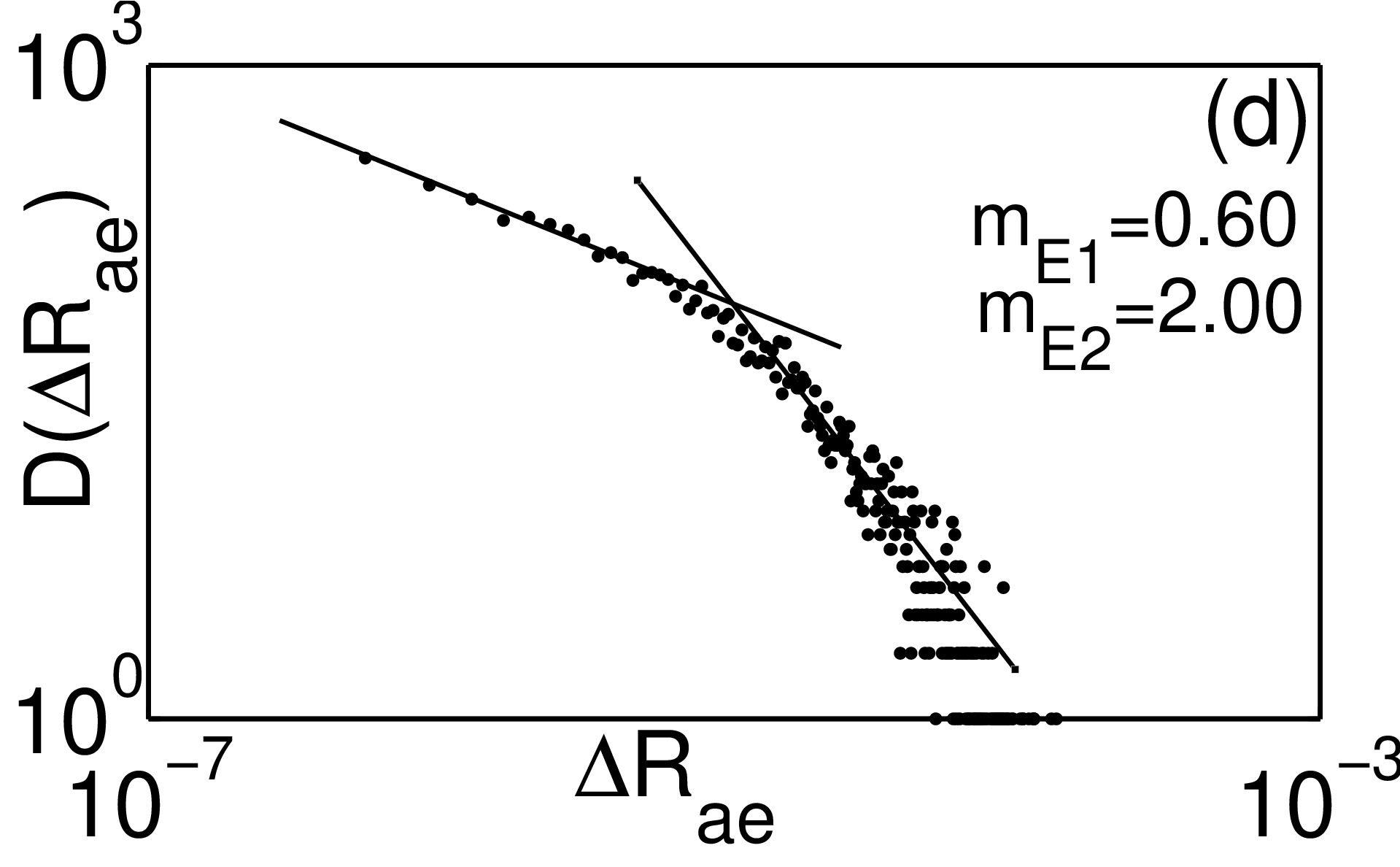}
}
}
\caption{  (a) Plot of a single stage distribution of the amplitudes of the uncured data for V = 1.6 cm/s. (b) Plot of a two stage power law distribution of the amplitudes of the uncured data for V = 5.0 cm/s. (c) A plot of a single stage power law distribution for the magnitudes of acoustic energy dissipated for $C_f=0.788, \gamma_u=0.01$ and $V^s = 1.48$. (d) A plot of a two stage power law distribution for the magnitudes of model acoustic energy for $C_f=0.00788, \gamma_u=0.01$ and $V^s = 2.48$.}
\label{Expt_dist}
\end{figure}

Given an experimental time series, the simplest statistical quantity to compute is the statistics of events. The definition of events depends on the physical situation, which in the case of the AE signal may be the time interval between the bursts of AE, the amplitude of bursts etc. Indeed the former has been computed \cite{CGVB04}. Here we compute the distribution of the amplitudes of the AE signals.  The difference between the maximum and next minimum, denoted by $\Delta A$ can be taken to be a measure of the amplitude of the AE signal. (In experiments, it is measured by setting a cut-off and measuring all amplitudes larger than the cut-off.)  We have computed the distribution of the amplitudes $D(\Delta A)$   for all the 38 data files for pull velocities starting from $0.2$ to $7.6$ cm/s. Surprisingly, we find only power law distributions for all the data files, i.e., $D(\Delta A) \sim \Delta A^{-m_A}$; we do not find peaked distributions.  For small traction velocities, we find a single power law crossing over to a two stage power law for high traction velocities. A typical single power law distribution for $V=1.6$ cm/s is shown in Fig. \ref{Expt_dist}(a) with an exponent $m_A = 2.15$. Figure \ref{Expt_dist}(b) shows a two stage power law for high velocity $V= 5.0$ cm/s. The exponent values are $m_A =0.32$ and $3.0$ respectively for small and large amplitude regimes. The transition from a single to two stage power law distribution occurs with the deviation for small values seen in Fig. \ref{Expt_dist}(b) becoming more dominant with increase in the pull velocity.  A  two stage power law (over one order of magnitude range) is first observed for $V=3.0$ cm/s. The exponent values are functions of the pull velocity. Table \ref{Exptstat} shows the exponent values for a selected set of pull velocities. Even though statistical features are easy to calculate, they are sufficiently discriminating. The analysis will be useful while comparing the  cured data files as also with the statistics of model acoustic energy signals.

These results may be compared with the statistics of the amplitude of the model energy bursts $R_{ae}$, i.e., from the maximum to the next minimum.   Denoting $\Delta R_{ae}$ to be the amplitude of $R_{ae}(\tau)$, let $D(\Delta R_{ae})$ be the distribution  of the amplitude of $R_{ae}$.  For $C_f=0.00788$, only $\gamma_u=0.01$ case is interesting (see Fig. \ref{I2m3V2gup0_01}(d) corresponding to $V^s=2.48$). To determine $D(\Delta R_{ae})$, we use long time series (typically $\sim 10^5$ points in units of the integration step).  For this case, we find a two stage power law as shown in Fig. \ref{Expt_dist}(d) i. e.,  $D(\Delta R_{ae})\sim \Delta R_{ae}^{-m_E}$ with the exponents $m_E \sim 0.60$ and 2.0 for the small and large amplitude regimes respectively. Figure \ref{Expt_dist}(c) shows an example of a single power law for $C_f=0.788, \gamma_u=0.1$ and  $V^s=1.48$. The exponent value is $m_E =0.7$. To compare, we note that the experimental time series refers to  the amplitude of the AE signals while the model signal $R_{ae}$ is the energy. Thus, the two exponents are related through $m_E = (m_A +1)/2$. Using the value of $m_{E}$ shown in Fig. \ref{Expt_dist}(d), we obtain $m_{A1} =0.2$ for the exponent corresponding to small amplitude regime (of the AE signal) and $m_{A2} =3.0$ for large amplitudes. Clearly, the values  are in reasonable agreement with the exponent values for the experimental signals [Fig. \ref{Expt_dist}(b)]. Unlike the experimental signal where scaling regime is good for all  values of the pull velocity, for the model acoustic energy $R_{ae}$, the distributions for $\Delta R_{ae}$ show a  power law statistics (of at least one order of scaling regime)  only for a certain sets of parameter values. Table \ref{RAE2}  lists the exponent values wherever the power law distribution is seen. For high and low values of $C_f$ we find two stage power law distributions. However, for the intermediate $C_f$( 0.788), only a single stage power law is found. It is interesting to note that the power law generated here is purely of dynamical origin. 

\begin{table}[!h]
\centering
\begin{tabular}{|p{3.7em}|p{2.5em}|p{2.5em}|p{2.5em}|p{2.5em}|p{2.5em}|p{2.5em}|p{2.5em}|p{3.0em}|}  \hline
{$C_f$} &{$V^s$} &$ \gamma_u$  &$ m_{E1}$ & $ m_{E2}$ & $\nu$&$\lambda_1$ & $D_{ky}$\\ \hline
{$7.88$} &{$2.48$} & $0.01$ &  $0.45$ & $2.10$ & $2.40$ &$1.05$&$2.90$\\ 
&{$4.48$} & $0.01$ &  $0.65$ &$1.97$ &$2.35$& $0.46$ & $2.74$\\ \hline

{$0.788$} &{$1.48$} & $1.00$ & $0.55$ & $...$&$2.45$&$1.53$&$2.74$\\
&& $0.10$&$0.70$ &$...$&$2.49$&$1.80$ & $2.77$\\
&& $0.01$ &  $0.65$&$...$ &$2.15$ &$1.85$ &$2.45$\\ \hline
{$0.788$}&{$2.48$} & $1.00$ & $1.00$&...&2.45&1.50&2.48 \\
&& $0.10$& $0.90$&...&2.54&1.78&2.59 \\
&& $0.01$ &  $0.67$&...&$2.45$ &$1.59$&$2.86$\\ \hline
{$0.788$}&{$4.48$} & $1.00$ & $0.70$&...&2.55&1.48&2.86 \\
&& $0.10$& $0.60$ &...&2.35&1.57&2.53 \\
&& $0.01$ &  $0.72$&...&$2.50$&$1.50$& $2.52$\\ \hline

{$0.00788$}&{$1.48$} & $0.01$ &  $0.74$ & $2.0$ & NC & NC & NC\\ 
&{$2.48$} & $0.01$ &  $0.60$&$2.0$&$2.20$&$0.32$&$2.40$\\ 
&{$4.48$} & $0.01$ &  $0.75$&$2.0$&$2.70$&$0.13$& $2.76$\\ \hline

\end{tabular}
\caption{Statistical and dynamical quantities for the model acoustic signal. Columns 4 and 5 show the power law exponents. Columns 6 to 8 list the correlation dimension, positive exponent and Lyapunov dimension respectively.}
\label{RAE2}
\end{table}

\subsection{Dynamical Analysis of AE time series}

\begin{figure}[!h]
\vbox{
\hbox{
\includegraphics[height=3.0cm,width=4.25cm]{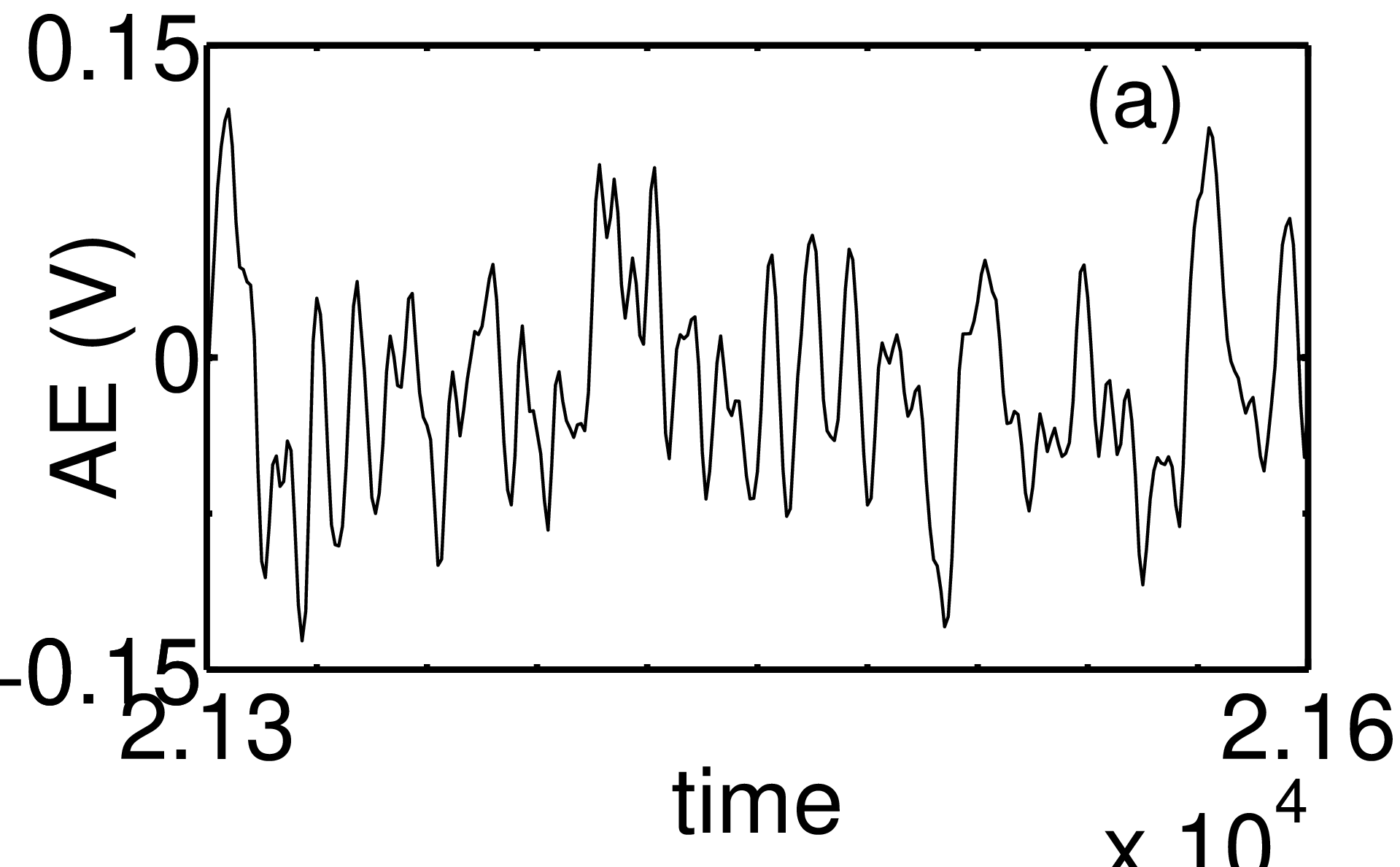}
\includegraphics[height=3.0cm,width=4.25cm]{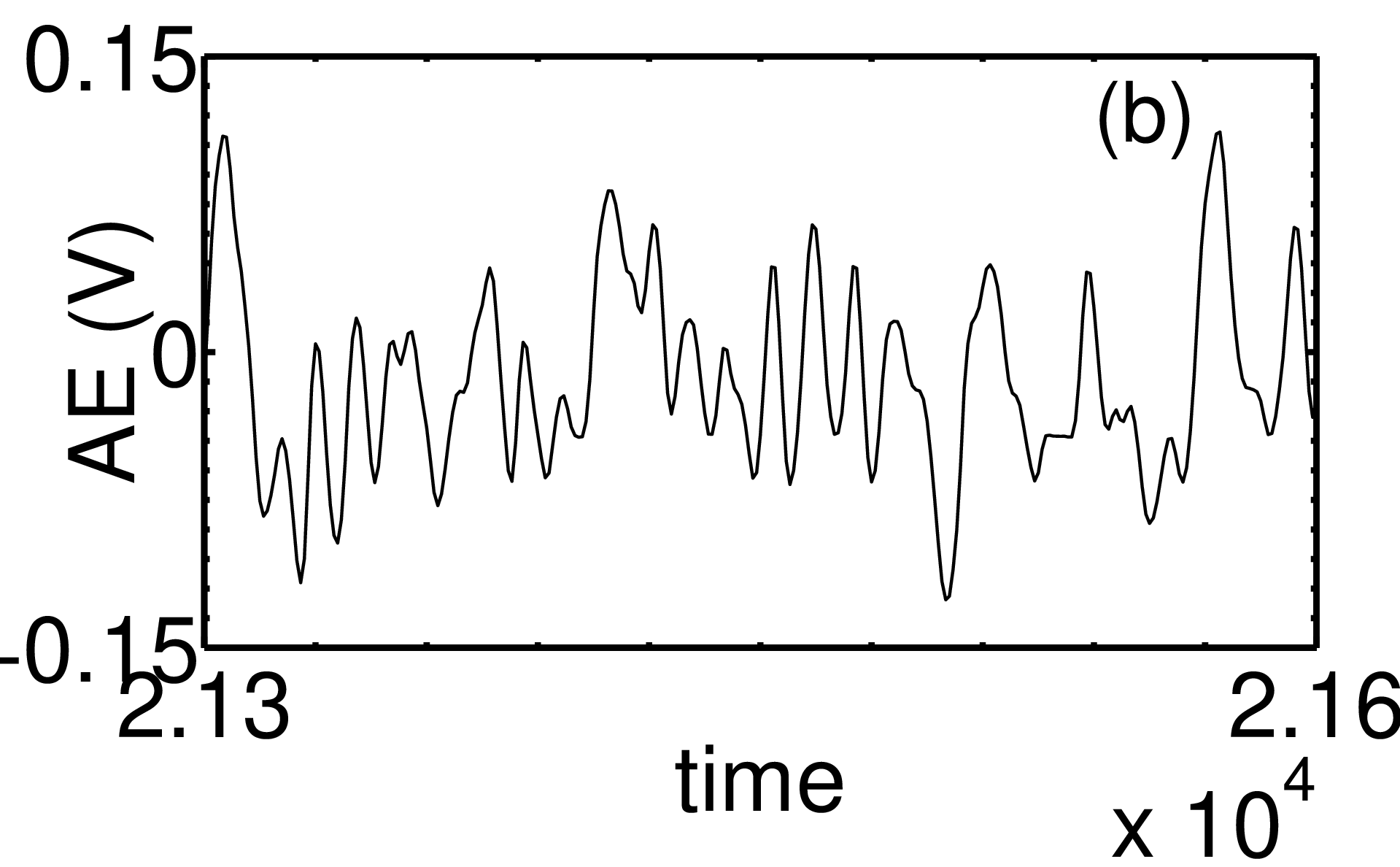}
}
\hbox{
\includegraphics[height=3.0cm,width=4.9cm]{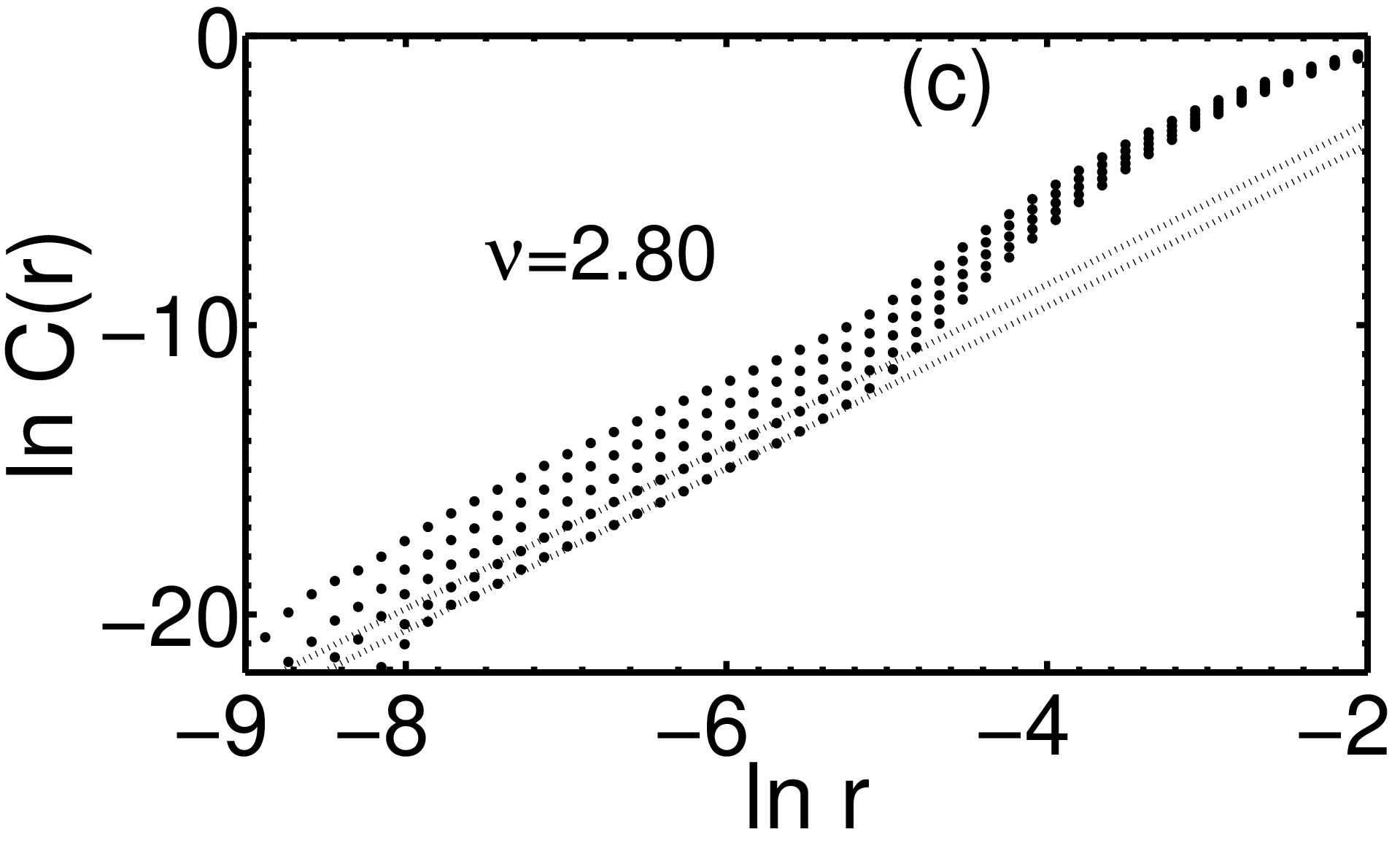}
\includegraphics[height=3.0cm,width=3.6cm]{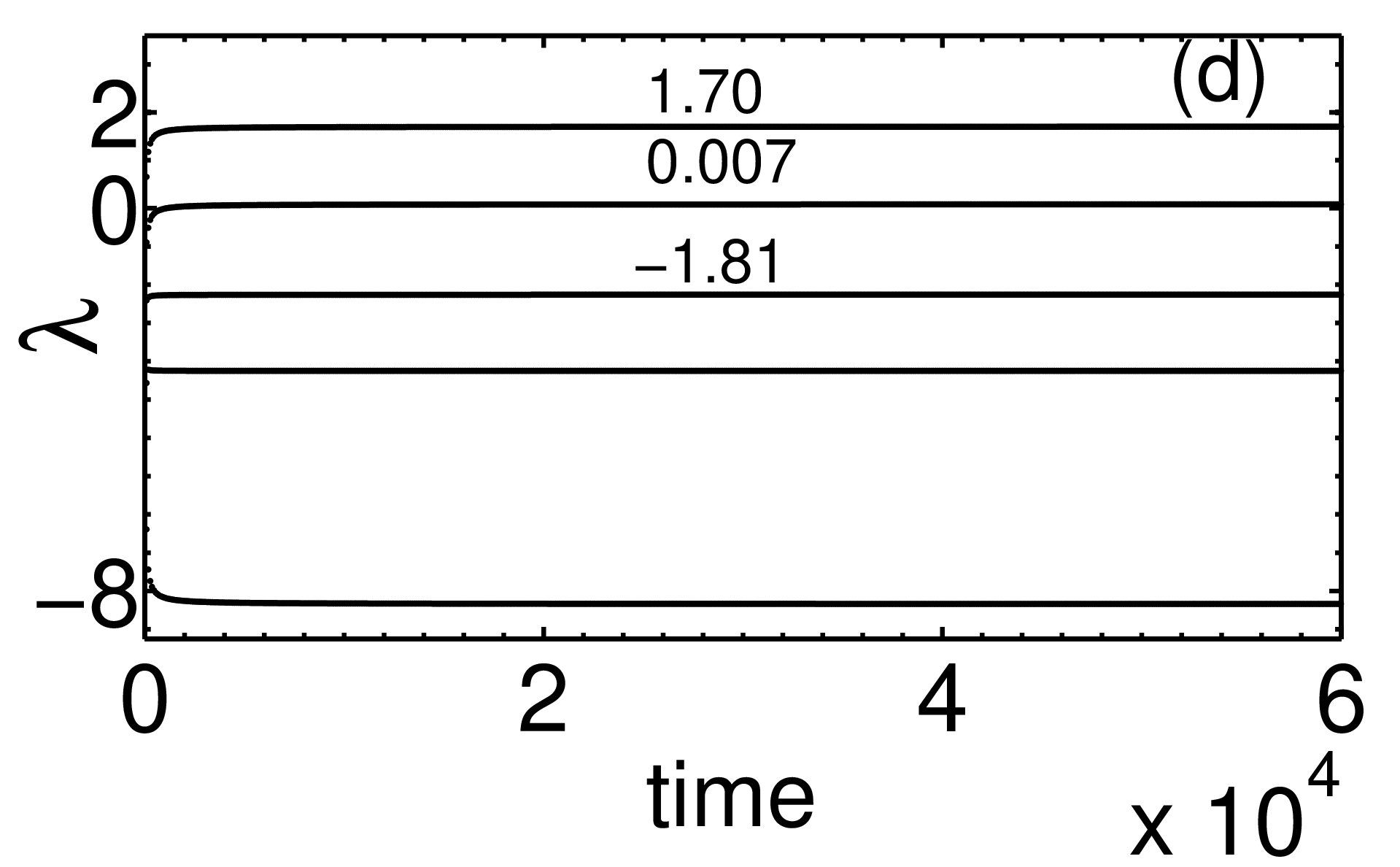}

}
\hbox{
\includegraphics[height=3.0cm,width=4.9cm]{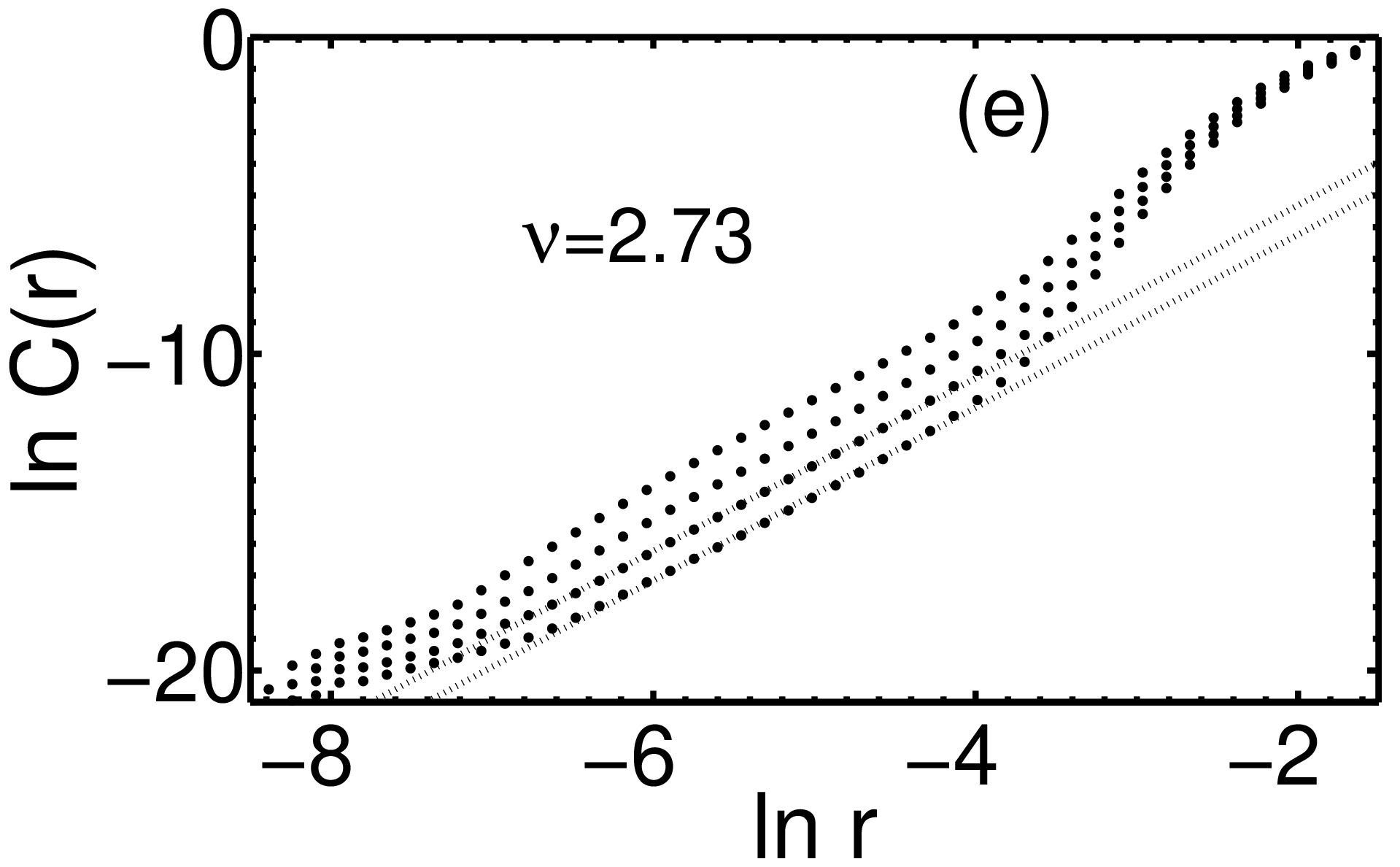}
\includegraphics[height=3.0cm,width=3.6cm]{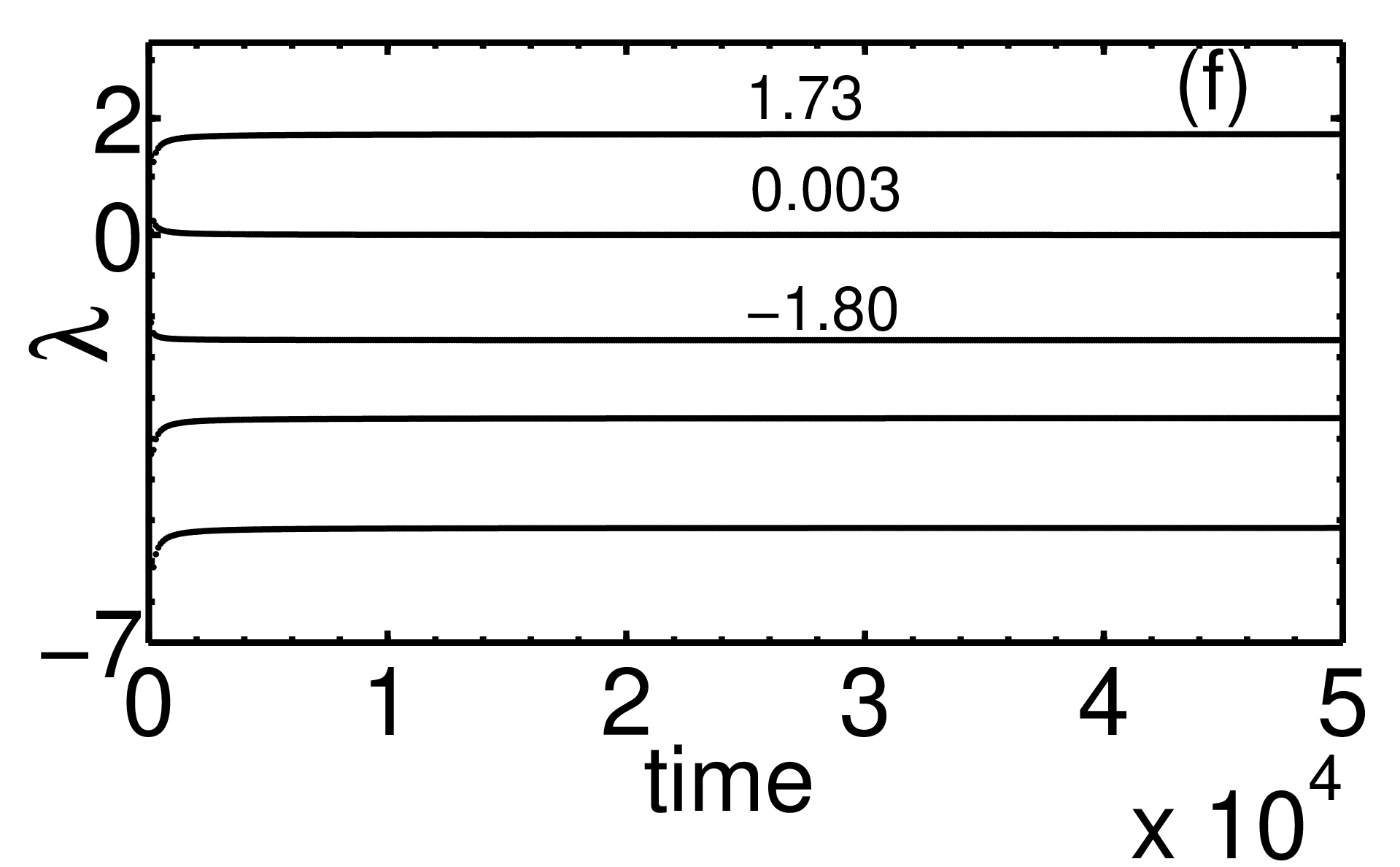}
}
}
\caption{ (a, b) Raw and cured AE signal  respectively for $V= 5.0$ cm/s. (c) Correlation integral for pull velocity $3.8$ cm/s for $d=9$ to $13$. Dashed lines are guide to eye. (d) Lyapunov spectrum for the same data file. (e)   Correlation integral for pull velocity $5.0$ cm/s  from $d=7$ to $10$. Dashed lines are guide to eye. (f) Lyapunov spectrum of the AE signals for traction velocity $5.0$ cm/s.}

\label{DyInv}
\end{figure}

In Section IV, we showed that the  peel front patterns  for several sets of parameters  are spatiotemporally chaotic.   More importantly, the model acoustic energy is quite irregular even as it is of dynamical origin. This  suggests the possibility that the experimental AE signals could be chaotic. However, often time series have undesirable systematic component, which needs to be removed from the original data. For instance, in the PLC effect,  the stress-strain time series has an overall increasing stress  arising from the work hardening component of the stress \cite{Anan99} which needs to be subtracted. In the present case, the experimental data for high pull velocities does show a background variation. A simple way of eliminating this background component is to use a window averaging and subtract this component from the raw data. Moreover, as stated in the introduction, the experimental AE data are quite noisy and therefore it is necessary to cure the data (using standard noise reduction techniques \cite{HKS}) before subjecting them to further analysis.  Simple visual checks for the existence of chaos such as phase plots, power spectrum etc. have been carried out. We have also used singular value decomposition, false neighbor search etc. Figures \ref{DyInv}(a, b) show the raw and cured data respectively for $V= 5.0$ cm/s. Clearly, the dominant features of the time series are retained except that small amplitude fluctuations are reduced or washed out  \cite{HKS}. Statistical features like the distribution function for the amplitude of the AE signals, power spectrum etc. are not altered. For instance, the two stage power law distribution for the amplitude of AE signals for the raw data (shown in Fig. \ref{Expt_dist}(b)) is retained except that the exponent value for the small amplitude regime is reduced from $0.32$ to $0.24$ without altering the exponent corresponding to large amplitudes. This reduction is understandable as small amplitude fluctuations are affected during curing.

The cured data are used to calculate the correlation dimension for all the data files.  However, for calculating the Lyapunov spectrum using our algorithm, raw data is adequate as our algorithm is designed to process noisy data. (In contrast, calculating the Lyapunov spectrum using the TISEAN package requires the cured data). To optimize the computational time, all our calculations are carried out using one fifth of each data set  as each file contains large number of points $\sim  10^6$ points, and there are 38 data sets.

\begin{figure}[!h]
\hbox{
\includegraphics[height=3.1cm,width=4.9cm]{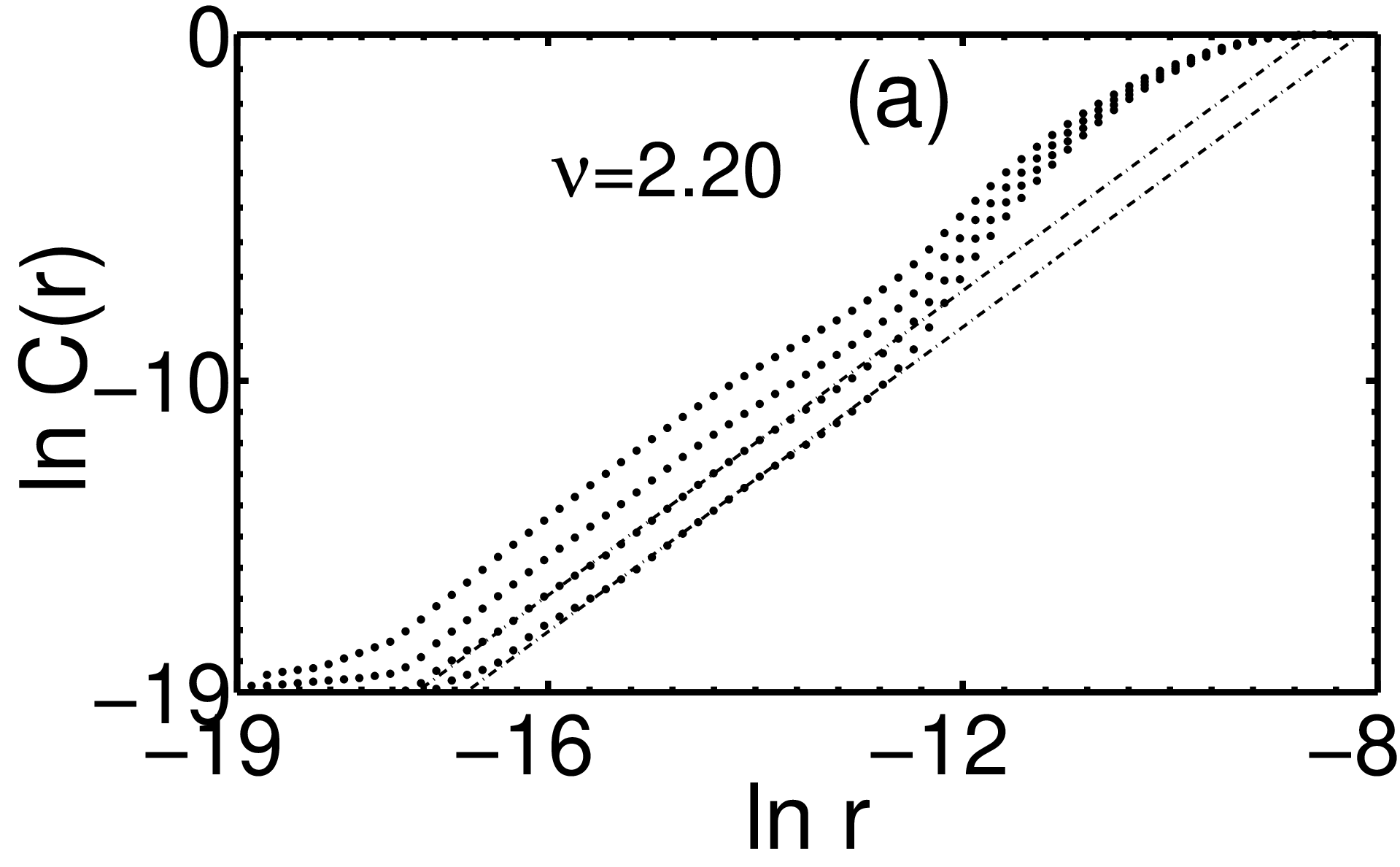}
\includegraphics[height=3.1cm,width=3.6cm]{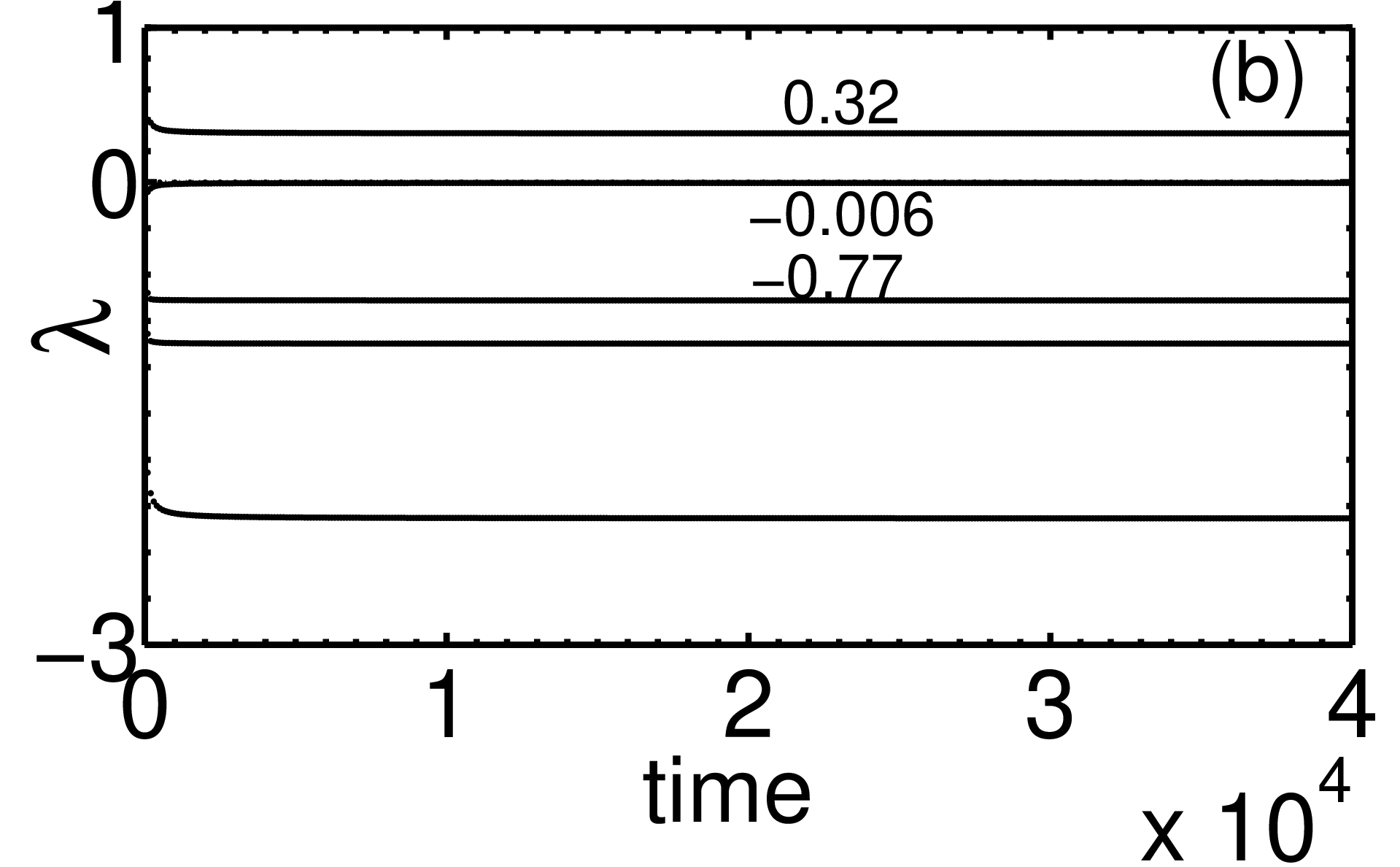}
}
\caption{ (a) Correlation integral of $ R_{ae}(\tau)$ for $C_f=0.00788, \gamma_u=0.01$ and $V^s = 2.48$  with $\nu =2.2 $ ($d=5$ to $8$). Dashed lines are guide to eye.(b) The Corresponding Lyapunov spectrum for $R_{ae}$.}
\label{DyInvModel}
\end{figure}

Typical autocorrelation time is about four units in sampling time. However, using a smaller value of $\tau=1$, we have calculated the correlation integral  $C(r)$ for all the data files. Converged values of correlation dimension are seen only in the region of pull velocities in the subinterval 3.8 to 6.2 cm/s. A log-log plot of $C(r)$  for the pull velocity $3.8$ cm/s is shown in Fig. \ref{DyInv}(c)  for $d=9$ to $13$.   A scaling regime of more than three orders of magnitude is seen with $\nu \sim 2.80 \pm 0.05$.  This is at the beginning of the chaotic window.

We have calculated the Lyapunov spectrum using our algorithm. The Lyapunov spectrum for $V=3.8$ cm/s is shown in Fig. \ref{DyInv}(d). (The outer shell radius is kept at $\epsilon_o=0.065$.) Note that the second exponent is close to zero as should be expected of continuous flow systems. Using the spectrum, we have calculated  the Kaplan-Yorke dimension (also called Lyapunov dimension) $D_{ky}$ using the relation $D_{ky}=j+\frac{\sum_{i=1}^j \lambda_i}{|\lambda_{j+1}|}; \sum_{i=1}^j\lambda_i>0;\sum_{i=1}^{j+1}\lambda_i<0 $. The value so obtained in each case should be consistent with that obtained from the correlation integral. For the case shown in Fig. \ref{DyInv}(d) we get  $D_{ky}= 2+ 1.70/1.81 = 2.94$ consistent  with $\nu=2.80$. (Typical error bars on the first three Lyapunov exponents are $\pm 0.01,\pm 0.005 $ and $\pm0.05$. Thus the errors in $D_{ky}$ values are $\pm 0.05$.) 

As an example of converged value of correlation dimension near the upper end of the chaotic domain, a log-log plot of $C(r)$  for $V^s= 5.0$ cm/s is shown in Fig. \ref{DyInv}(e)  with $\nu =2.73 \pm 0.05$ for $d=7$ to 10. Again, the scaling regime is seen to be nearly three orders of magnitude. The Lyapunov spectrum for the data file is shown in Fig. \ref{DyInv}(f). The  calculated Lyapunov dimension from the  spectrum  is  $D_{ky}= 2+ 1.73/1.80 = 2.96$ which is again consistent with $\nu=2.73$.

The values of $\nu$ for all the files are found to be  in the range $2.55$ to $2.85 \pm 0.05$ as can be seen from Table \ref{Exptstat}.  We have calculated the Lyapunov spectrum for the full range of traction velocities and we find (stable) positive and zero exponents  only in the region 3.8 to 6.2 cm/s, consistent with the range of converged values of $\nu$ as can be seen from Table \ref{Exptstat}. The corresponding values of $D_{ky}$ are in the range of $2.7$ to $3.0$. We have also calculated the Lyapunov spectrum using the TISEAN package using cured files. The $D_{ky}$ values obtained from the TISEAN package  are uniformly closer to the $\nu$ values, typically $D_{ky} = \nu + 0.1$.   Finally, we note that the positive exponent decreases toward  the end of the chaotic domain (6.2 cm/s). 
These results (see Table \ref{Exptstat}) show unambiguously that the underlying dynamics responsible for AE during peeling is chaotic in a mid range of pull speeds.

In order to compare the low dimensional chaotic nature of the experimental AE signals with the model acoustic signal, we have analyzed the low dimensional dynamics of $R_{ae}(\tau)$ using the embedding procedure after subtracting the periodic component.

We have computed  the correlation dimension and Lyapunov spectrum for the entire instability domain. A log-log plot of the $C(r)$ is shown in Fig. \ref{DyInvModel}(a)  for $d=5$ to 8. The convergence over more than three orders of magnitude is clear. The value of $\nu= 2.20 \pm0.05$. For this file, we find stable positive and zero exponents for a range of $\epsilon_o$ values.  A plot of the spectrum for $C_f=0.00788, \gamma_u=0.01$ and $V^s=2.48$ ($\epsilon_o=0.08$) is shown in Fig. \ref{DyInvModel}(b). Using this we get $D_{ky} = 2+ 0.32/0.77 = 2.4$ which is again consistent with $\nu = 2.2 \pm 0.02$.  

We have calculated both correlation dimension and Lyapunov spectrum of $R_{ae}$ for a range of values of the parameters. For each $C_f$, we find converged values of $\nu$ and $D_{ky}$ within a window of pull speeds. Generally, the range of $\nu$ is between 2.15 to 2.70 while $D_{ky}$ is  in the range 2.4 to 2.90.  Table \ref{RAE2} shows the values of correlation dimension and $D_{ky}$ for various sets of parameter values. It is interesting to note that the magnitude of the largest exponent for the model AE signal also decreases as we increase the pull velocity, a feature displayed by the experimental time series as well.

\section{Summary and Conclusions}

In summary, the present investigation is an attempt to understand the origin of the intermittent peeling of an adhesive tape and its connection to acoustic emission. {\it At the  conceptual level, we have established  a relationship between stick-slip dynamics and the acoustic energy}, the latter depends on the local strain rate \cite{Land} which in turn is controlled by the roughness of the peel front.  As the model is fully dynamical, one basic result that emerges is that the model acoustic energy is controlled by the nature of spatiotemporal dynamics  of the peel front.  Further, even as the model acoustic emission is a dynamical quantity,  the nature of $R_{ae}$ turns out to be  quite noisy depending on the possible interplay of different time scales in the model.  Thus, the highly noisy nature of the experimental signals need not necessarily imply stochastic origin of AE signal; instead,  they could  be of deterministic origin. This motivated us to carry out  a detailed analysis of statistical and dynamical features of the experimental AE signals. Despite the high noise content, we have been able to demonstrate the existence of finite correlation dimension and positive Lyapunov exponent for a window of pull speeds. The Kaplan-Yorke dimension (for various traction velocities) calculated from the Lyapunov spectrum is consistent with the value obtained from the correlation integral. Thus, the analysis establishes unambiguously the deterministic chaotic nature of the experimental AE signals.  Interestingly, the largest Lyapunov exponent shows a decreasing trend  toward the end of the chaotic window, a feature displayed by the model acoustic signal as well. The work also addresses the general problem of extracting dynamical information from noisy AE signals. A similar analysis of the model acoustic energy shows that $R_{ae}$ is chaotic for a range of  parameter values. {\it More importantly, several qualitative features of the experimental AE signals such as the statistics of the signals and the change from burst to continuous type with increase in the pull velocity are also displayed by $R_{ae}$.} The observed two stage power law distribution for the experimental AE signals [Fig. \ref{Expt_dist}(b)] is reproduced by the model [Fig. \ref{Expt_dist}(d)]. It must be emphasized that this power law distribution for the amplitudes is completely  of dynamical origin. This result should be of general interest in the context of  dynamical systems  as there are very few models that generate power laws purely from dynamics. The only other example known to the authors is that of the PLC effect where the amplitude of the stress drops shows a power law distribution within the context of the Ananthakrishna model \cite{Anan04}.

The spatiotemporal patterns of the peel front  are indeed rich and depend on the interplay of the three time scales. Although, the nature of spatiotemporal patterns is quite varied, they can be classified as smooth synchronous, rugged, stuck-peeled and even nowhere stuck   patterns. As expected on general consideration of dynamics,  rich patterns are observed for the case when all the three time scales are of similar magnitude (illustrated for $C_f=0.788$). All spatiotemporal patterns, except the smooth synchronous peel front are interesting.  As a function of time, the nature of the peel front can go through a specific sequence of these patterns (depending on the parameter values). The most interesting pattern is the stuck-peeled configuration which is reminiscent of fibrils observed in experiments \cite{Dickinson,Urahama,Yamazaki}. Even among the SP configurations, there are variations, for example, rapidly changing, long lived, edge of peeling etc.  Despite the varied range of patterns, a few general trends of the influence of the parameters on the peel front patterns are worth noting.  First,  in general the number of stuck and peeled segments increases as $\gamma_u$ is decreased. Second, as the pull velocity is increased, the rapidly varying  stuck-peeled configurations observed at low pull velocities become long lived. The dynamical signature of  these two parameters are reflected in the nature of the phase space orbit.  For instance,  given a value of  $C_f$ and $\gamma_u$, the nature of the phase space orbit changes only when $V^s$ is increased which   allows the  orbit to move way beyond the values of  $\phi(v^s)$.

The study of the model shows that while the intermittent peeling is controlled by the peel force function, the dynamics of the peel front is influenced by all the three time scales. This together with the dynamical analysis of the experimental acoustic emission signals establishes  that deterministic dynamics is responsible for AE during peeling. The various sequences of peel front patterns and their time dependences lead to quite varied   model acoustic signals. These can be classified as  bunch of spikes, isolated bursts occurring at near regular intervals, continuous bursts with  an overall envelope separated by  quiescent state, continuous bursts overriding a near periodic triangular form, irregular waveform overriding a periodic component, and continuous irregular type.  {\it Interestingly, our studies show that there is a definite correspondence between the model acoustic energy and the nature of peel front patterns even though $R_{ae}(\tau)$ is the spatial average of the local strain rate.}  Despite this, two distinguishable time scales in $R_{ae}(\tau)$ can be detected, one corresponding to short term fluctuations and another corresponding to overall periodic component. The  short term fluctuations can be readily identified when the model acoustic signal is fluctuating without any background  component (see for example Fig. \ref{I2m1V1gup0_1}(d)). These rapid changes in $R_{ae}$ arise due fast dynamic changes in the SP configurations. The minimum in $R_{ae}$ corresponds to the situation where the average velocity jumps of the SP configurations are smaller compared to that at the preceding maximum.  In contrast,  the  overall periodicity in $R_{ae}(\tau)$ (for instance see Fig. \ref{I2m1V2gup1}(c) among many other cases) can be identified with the changes in the peel front patterns that occur over a cycle in the phase plot.  The minima in the $R_{ae}(\tau)$  corresponds to  the peel patterns where more segments are in stuck state than in the peeled state while the maxima corresponds to more stuck and peeled segments (see Figs. \ref{I2m1V2gup1}(a), (b)). The corresponding phase plot usually goes through a cycle of visits between the low and high velocity branches.

Often, however, the nature of the model acoustic energy signal can be complicated as in the case of low tape mass ($C_f =0.00788$). Even in such cases, some insight is possible. This is aided by the analysis of the corresponding phase plot. For example for the low $C_f$ case  where the roller inertia plays an important role in the dynamics, the rapidly fluctuating acoustic energy has an overall triangular envelope [Fig. \ref{I2m3V1gup01}(e)].  On an  expanded scale, the convex envelope consists of seven local peaks [Fig. \ref{I2m3V1gup01}(f)]. Each of these  is  generated when the various peel front segments make abrupt jumps between the two branches of the peel force function. Note that the phase space orbit has  seven loops in this case [Fig. \ref{I2m3V1gup01}(d)]. The general identification of the minima in $R_{ae}$ with patterns that have more stuck segments than peeled segments still holds. Similarly, the maxima in $R_{ae}$ usually correspond to the presence of large number of stuck-peeled configurations. 

The above correspondence  between the model acoustic energy and the peel front patterns provides insight into the transition from burst to continuous type of AE seen in experiments as a similar transition from burst type to continuous type is also seen in  the model acoustic energy (for large $C_f=7.88$ ). At low pull velocities, the peel front goes through a cycle of patterns where most segments of the peel front (or the entire peel front) spends substantial time in the stuck state switching to  stuck-peeled configuration. As the duration of the SP configuration is short and velocity bursts are large, $R_{ae}(\tau)$ is of burst type [Fig. \ref{ExptAE}(c)]. With increasing pull velocity, only dynamic stuck-peeled configurations are seen  which in turn leads to continuous AE signals [Figures. \ref{I5m3V2gu0_1_01}(a, b)].  This coupled with time series analysis of the model acoustic signal shows that the associated positive Lyapunov exponent  decreases  with increase in the traction velocity. This is precisely the trend observed for experimental  signals as well. Thus, the decreasing trend of the largest Lyapunov exponent can  be attributed to the peel front breaking up into large number of small segments providing insight into stick-slip dynamics and its connection to the AE process. 

The present study has relevance to the general area of stick-slip dynamics. As mentioned earlier, models for stick-slip dynamics use negative force-drive rate relation. In such models, the phase space orbit generally sticks to the slow manifold (stable branches) of the force-drive rate function. This leads to clearly identifiable stick and slip phases, the former lasting much longer than the latter. However, recent work on imaging the peel point dynamics  \cite{Cortet} shows that the ratio of the stick phase to the slip phase, is about two or even  less than unity for high  peel velocities. While all the known models of the peel process predict that the duration of the stick phase is longer than that of the slip phase, our model displays the experimentally observed feature. This feature emerges in the model due to the interplay of the three time scales aided by incomplete relaxation of the relevant modes. Our  studies  show that {\it only} for low pull velocity and high $C_f$ do we observe the stick phase lasting  much longer than the slip phase. {\it As the pull velocity is increased, and for all other parameter values, we find that the duration of the slip phase} (peel velocity being larger than unity) {\it is nearly  the same as or less than that of the  stick phase} (peel velocity less than unity). Further,  the present model provides an example of the richness of spatiotemporal dynamics arising when  more than two time scales are involved. In this context, we emphasize that the introduction of the Rayleigh dissipation functional to model the acoustic energy is crucial for the richness of the spatiotemporal peel front patterns. It is important to note that this kind of dissipative term is specific to spatially extended systems as it represents relaxation of neighboring points on the peel front.

The present  study  has relevance to time dependent issues of adhesion. For instance, apart from the fact that the time series analysis addresses the general problem of extracting dynamical information from noisy AE signals, it may have relevance to failure of adhesive joints and composites that are subject to fluctuating loads. The failure time can be estimated by calculating the Lyapunov spectrum for the AE signals. If the largest Lyapunov exponent is positive, the inverse of the exponent should give an estimate of the time scale over which the failure can occur and hence  could prove to be useful in predicting failure of joints. The present study should also help to optimize  production schedules in peeling tapes.

Finally, several features of the present study are common to the  PLC effect even though the underlying mechanism is very different. In this case the repeated occurrence of stress drops during constant strain rate deformation \cite{PLC,GA07,GA07a}, are associated with the formation and possible propagation of dislocation bands that are visible to the naked eye. The phenomenon occurs only in a window of applied strain rates. The instability is attributed to the pinning and unpinning of dislocations from solute atmosphere, yet, the  dominant feature underlying the instability is the negative strain rate sensitivity of the flow stress that has two stable branches  separated by an unstable branch. Clearly, these features are similar to the occurrence of the peel instability within a window of pull velocities and the existence of unstable branch in the peel force function. Further, the Ananthakrishna (AK) model for the PLC instability predicts that the stress drops should be chaotic in a subinterval of the instability domain \cite{Anan82}. This prediction has been verified subsequently through the analysis of experimental stress-strain curves obtained from single and polycrystals \cite{Anan93,Anan95,Anan99,Bhaprl}. This feature is again similar to the existence of chaotic dynamics observed in a mid range of pull velocities in the peeling  problem, both in  experiment and in the model.  In the case of the PLC effect, one finds that the positive Lyapunov exponent characterizing the stress-time series decreases toward the end of chaotic window, both in experiments and in the AK model.  
Again this feature is also seen in the present peel model as also in experimental AE signals. Dynamically, in the case of the AK model for the PLC effect the  decreasing trend of the positive Lyapunov exponent  has been shown to be a result of a forward Hopf bifurcation (HB) followed by a reverse HB \cite{Anan04}. In the case of peeling problem as well, the instability begins with a forward HB followed by a reverse HB. Finally, in the PLC effect (both in experiments and the AK model), as in the peeling problem, the duration of the slip  phase can be longer than that of the  stick phase   with increasing drive rate. As many of these features are common to two different systems, it is likely that these are general features in other stick-slip situations with multiple time scales that are limited to a window of  drive rates with multiple participating time scales.  

A few comments may be in order about the  model, in particular about the parameters that are  crucial for the dynamics. While the agreement of several statistical and dynamical features of the model ( for several  sets of parameter values) with the experimental AE series is encouraging, it would be interesting to verify model results for other sets of parameters. For instance, it is clear that the roller inertia and the inertia of the tape mass are experimentally assessable parameters. Thus, the influence of these two  inertial time scales can in principal be studied in experiments. However, conventional experiments have been performed keeping these parameters fixed, presumably, as there has been no suggestion that the dynamics can be sensitive to these variables.  It would be interesting to verify the predicted dynamical changes in the AE signals as a function of these two parameters. As for the influence of the dissipation parameter $\gamma_u$, the range of physically reasonable values of  $\gamma_u$ is expected to be small ($10^{-4}$ to $10^{-3}$) as argued. Interestingly, the region of low $\gamma_u$ is indeed the region where both statistics and dynamical features compare well with that of the experiments. However, within the scope of the model, the visco-elastic properties of the adhesive have been modeled using an effective spring constant. (This kind of assumption is common to studies in adhesion and tackiness etc  \cite{Gay}.) However, it is possible to include this feature as well.  Finally, it must be stated that  features that critically depend on thickness of the film and its visco-elastic properties such the shape of peel front are beyond the scope of the present model.

Acknowledgment: The authors wish to thank Professor M. Ciccotti for providing the AE data. GA  acknowledges support from BRNS Grant No. $2005/37/16/BRNS, 2007/36/62-BRNS/2564$, and Raja Ramanna Fellowship.


\begin{thebibliography}{99}

\bibitem{Kendall00} K. Kendall, {\it Molecular Adhesion and its Applications} (Kluwer Academic, New York, 2001).

\bibitem{Urbakh04} M. Urbakh {\it et al.,} Nature (London) {\bf 430}, 525 (2004).
\textit{}

\bibitem{Persson} B. N. J. Persson, {\it Sliding Friction: Physical Principles and Applications}, 2nd ed. (Springer, Heidelberg, 2000).

\bibitem{Jagota07} N. J. Glassmaker {\it et al}., Proc. Natl. Acad. Sci. USA {\bf 104}, 10786 (2007), and references therein.

\bibitem{Rumi07} Rumi De, A. Zemel, and S. A. Safran, Nat. Phys. {\bf 3}, 655 (2007).

\bibitem{MB} D. Maugis and M. Barquins, in {\it Adhesion 12}, edited by K. W. Allen (Elsevier, London, 1988), p. 205.

\bibitem{CGVB04} M. Ciccotti, B. Giorgini, D. Villet, and M. Barquins, Int. J. Adhes. Adhes. {\bf 24}, 143 (2004).

\bibitem{BC97} M. Barquins and M. Ciccotti, Int. J. Adhes. Adhes. {\bf 17}, 65 (1997).

\bibitem{Cortet} P-P Cortet, M. Ciccotti, and L. Vanel, J. Stat. Mech., P03005 (2007). 

\bibitem{Heslot} F. Heslot, T. Baumberger, B. Perrin, B. Caroli, and C. Caroli, Phys. Rev. E {\bf 49}, 4973(1994).

\bibitem{soc04} A. Socoliuc, R. Bennewitz, E. Gnecco, and E. Meyer, Phys. Rev. lett. {\bf 92}, 134301 (2004).

\bibitem{BK} P. Burridge and L. Knopoff, Bull. Seissmol. Soc. Am. {\bf
57}, 341 (1967).


\bibitem{PLC} A. Portevin and F. Le Chatelier, C. R. Acad. Sci. Paris {\bf 176},
507 (1923); F. Le Chatelier, Re. de Metallurgie {\bf 6}, 914 (1909).

\bibitem{GA07} G. Ananthakrishna, Phys. Rep. {\bf 440}, 113 (2007).

\bibitem{Anan04} G. Ananthakrishna and M. S. Bharathi, Phys. Rev. E {\bf 70}, 026111 (2004).

\bibitem{BKC} See Modelling Critical and Catastrophic Phenomena in Geosciences: a Statistical Physics Approach, Eds. P. Bhattacharyya and B. K. Chakrabarti, Lecture Notes in Physics 705, (Springer, Berlin 2006).

\bibitem{Anan82} G. Ananthakrishna and M. C. Valsakumar, J. Phys. D {\bf 15}, L171 (1982).

\bibitem{Rajesh} S. Rajesh and G. Ananthakrishna, Phys. Rev. E {\bf 61}, 03664 (2000).

\bibitem{HY1} D. C. Hong and S. Yue, Phys. Rev. Lett. {\bf 74}, 254 (1995).

\bibitem{HY2} D. C. Hong (private communication).

\bibitem{CGB} M. Ciccotti, B. Giorgini, and M. Barquins, Int. J. Adhes. and Adhes. {\bf 18}, 35 (1998).

\bibitem{Rumi04} Rumi De, Anil Maybhate, and G. Ananthakrishna, Phys. Rev. E {\bf 70}, 046223 (2004).

\bibitem{HLR} E. Hairer, C. Lubich, and M. Roche, {\it Numerical Solutions of Differential-algebraic Systems by Runge-Kutta Methods}, (Springer-Verlag, Berlin, 1989).

\bibitem{Rumi05} Rumi De and G. Ananthakrishna, Phys. Rev. E {\bf 71}, 055201(R) (2005).
\bibitem{Rumiprl} Rumi De and G. Anantahakrishna, Phys. Rev. Lett. {\bf 97}, 165503 (2006).

\bibitem{Jagdish08} Jagadish Kumar, M. Ciccotti, and G. Ananthakrishna, Phys. Rev. E {\bf 77}, 045202(R) (2008).

\bibitem{Scholz68a} C. H. Scholz, J. Geophys. Res. {\bf 73}, 1417 (1968).

\bibitem{Lockner96} A. Lockner, J. Acous.  Emission {\bf 14}, S88 (1996).

\bibitem{Sam92} P. R. Sammonds, P. G. Meredith, and I. G.  Main, Nature {\bf 359}, 228 (1992).

\bibitem{vives} E. Vives, J. Ort\'in, L. Ma\~nosa, I. R\'afols, R. P\'erez-Magran\'e, and A. Planes Phys. Rev. Lett. {\bf 72 }, 1694 (1994)

\bibitem{rajeevprl} R. Ahluwalia and G. Ananthakrishna, Phys. Rev. Lett. {\bf 86}, 4076 (2001).

\bibitem{kalaprl} S. Sreekala and G. Ananthakrishna, Phys. Rev. Lett. {\bf 90}, 135501 (2003).

\bibitem{Petri94} A. Petri, G. Paparo, A. Vespignani, A. Alippi, and M. Costantini, Phys. Rev. Lett. {\bf 73}, 3423 (1994).

\bibitem{Diodati} P. Diodati, F. Marchesoni, and S. Piazza, Phys. Rev. Lett. {\bf 67}, 2239 (1991).

\bibitem{Miguel} M. C. Miguel {\it et al.,} Nature (London) {\bf 410}, 667 (2001).

\bibitem{Weiss} J. Weiss and D. Marsan, Science {\bf 299}, 89 (2003).

\bibitem{Bak} P. Bak, C. Tang, and K. Wiesenfeld, Phys. Rev. Lett. {\bf 59}, 381 (1987).

\bibitem{GP} P. Grassberger and  I. Procaccia, Physica D {\bf 9}, 189 (1983).

\bibitem{HKS}  R. Hegger, H. Kantz and T. Schreiber, CHAOS {\bf 9}, 413 (1999).

\bibitem{KS} H. Kantz and T. Schreiber, {\it Nonlinear Time Series Analysis}, Cambridge Unviersity Press (UK, 1997).

\bibitem{Land}L. D. Landau and E. M. Lifschitz, {\it Theory of Elasticity} (Pergamon, Oxford, 1986).

\bibitem{Rumiepl} Rumi De and G. Ananthakrishna, Europhys. Lett. {\bf 66}, 715 (2004).

\bibitem{ECG} E. C. G. Sudarshan and N. Mukunda, {\it Classical Dynamics: A Modern Perspective} (John Wiley and Sons, New York, 1974).

\bibitem{Ciccotti07} The data has been kindly supplied by Professor M. Ciccotti of Universit\'e de Montpellier II, France.

\bibitem{Packard} N. H. Packard, J. P. Crutchfield, J. D. Framer and R. S. Shaw, Phys. Rev. Lett. {\bf 45}, 712 (1980).

\bibitem{KS1} E. Kostelich and T. Schreiber, Phys. Rev. E {\bf 48}, 1752 (1993).

\bibitem{GHKSS} P. Grassberger, R. Hegger, H. Kantz, C. Schaffrath and T. Schreiber, CHAOS {\bf 3}, 127 (1993).

\bibitem{CH} R. Cawley and G-H Hsu, Phys. Rev. A {\bf 46}, 3057 (1992).

\bibitem{Ding} M. Ding, C. Grebogi, E. Ott, T. Sauer, J. A Yorke, Phys. Rev. Lett. {\bf 70}, 3872 (1993).

\bibitem{Anan99} G. Ananthakrishna, S. J. Noronha, C. Fressengeas, and L. P. Kubin, Phys. Rev. E {\bf 60}, 5455 (1999).

\bibitem{Noro01} S. Noronha, G. Ananthakrishna and C. Fressengeas, in {\it Nonlinear Dyanmics, Integrability and Chaos,} (Norosa, New Delhi, 2000), p. 235.

\bibitem{Eckmann} J. P. Eckmann, S.O Kamphorst, D. Rulle and S. Ciliberto, Phys. Rev. A {\bf 34}, 4971 (1986).

\bibitem{SF85} R. B Schwarz and L. L Funk, Acta metall. {\bf 33}, 295 (1985).

\bibitem{Dickinson} L. Scudiero, I. T. Dickenson, L. C. Jensen and S. C. Langford,  J. Adhes. Sci. Technol. {\bf 9}, 27 (1995).

\bibitem{Yamazaki} Y. Yamazaki and A. Toda, Physica D {\bf 214}, 120 (2006).

\bibitem{Canard} M. Diener, Math. Intell. {\bf 6}, 38 (1984).

\bibitem{Urahama} Y. Urahama, J. Adhes. {\bf 31}, 47 (1989).

\bibitem{GA07a} G. Ananthakrishna, in Dislocations in Solids, Eds. F. R. N. Nabarrow and J. P. Hirth, Vol. 13  P 81-223 (2007).

\bibitem{Anan93} G. Ananthakrishna, Scipta Met. {\bf 29}, 1183  (1993).

\bibitem{Anan95} G. Ananthakrishna {\it et al.,} Scripta Met. {\bf 32}, 1731(1995).

\bibitem{Bhaprl} M. S. Bharathi, M. Lebyodkin, G. Ananthakrishna, C. Fressengeas, and L. P. Kubin Phys. Rev. Lett. {\bf 87}, 165508 (2001).  

\bibitem{Gay} C. Gay and L. Leibler, Phys. Today {\bf 52}, No. 11, 48 (1999).


\end{thebibliography}
\end{document}